\documentclass[12pt,preprint]{aastex}
\usepackage{graphicx}
\usepackage{natbib}
\usepackage{changepage} 
\def\batse{BATSE\/}
\def\swift{{\em Swift\/}}
\def\sax{{\em BeppoSAX\/}}
\def\gro{{\em Compton Gamma Ray Observatory\/}}   
\def\fermi{{\em Fermi\/}}

\def\ep{E_\mathrm{p}}
\def\kte{kT_\mathrm{e}}

\makeatletter
\def\@cite#1#2{(#1\if@tempswa , #2\fi)}
\def\preprint{preprint}   \newif\ifPreprintMode
\ifx\preprint\revtex@genre\PreprintModetrue\else\PreprintModefalse\fi
\makeatother

\begin{document}

\title{Comptonization signatures in the prompt emission of \\ Gamma Ray Bursts}
\author{F.~Frontera\altaffilmark{1,2},
L.~Amati\altaffilmark{2},
R.~Farinelli\altaffilmark{1,3},
S.~Dichiara\altaffilmark{1},
C.~Guidorzi\altaffilmark{1},
R.~Landi\altaffilmark{2},
L.~Titarchuk\altaffilmark{1}}
\altaffiltext{1}{Universit\`a di Ferrara, Dipartimento di Fisica e Scienze della Terra, Via Saragat 1, 
44100 Ferrara, Italy; email: frontera@fe.infn.it}
\altaffiltext{2}{INAF, Istituto di Astrofisica Spaziale e Fisica Cosmica,
Bologna, Via Gobetti 101, 40129 Bologna, Italy}
\altaffiltext{3}{ISDC Data Center for Astrophysics, Universit\'e de Gen\`eve, chemin d'\'Ecogia 16, 1290 Versoix, Switzerland
}
%
%

%
\begin{abstract}
We report results of a systematic study of the broad--band (2--2000~keV) 
time--resolved prompt emission spectra of 
a sample of Gamma-Ray Bursts (GRBs) detected with both the Wide Field Cameras 
(WFCs) aboard the \sax\ satellite and the BATSE experiment aboard CGRO. The main 
goal of the paper is to test spectral models of the GRB prompt emission that have recently been proposed. 
In particular, we test the photospheric model proposed by Ryde and Pe'er (2009), i.e., blackbody plus power--law, the addition of a blackbody emission to the Band function in the cases in which this function does not fit the data, 
and the Comptonization model developed by Titarchuk et al. (2012). 
By considering the few spectra for which the simple Band function does not provide a fully acceptable fit to the data (Frontera et al. 2012), only in one case we find a statistically significant better fit by adding a blackbody to this function. 
We confirm the results found fitting the BATSE spectra alone with a blackbody plus a power law. Instead when the BATSE GRB spectra are joined to those obtained with 
WFCs (2--28 keV), this model becomes unacceptable in most of time intervals in which 
we subdivide the GRB light curves. We find instead that the Comptonization model 
is always acceptable, even in the few cases in which the Band function is inconsistent with the data. We discuss the implications of these results.
\end{abstract}


\maketitle

\section{Introduction}
\label{s:intro}

In spite of the huge advances in the knowledge of the GRB afterglow properties mainly with \swift, 
the GRB phenomenon is still poorly understood \citep[e.g.,][]{Lyutikov09,Zhang11a}. It is recognized to be of crucial importance 
the study of the prompt emission, which is more directly connected with the original explosion. 
One of the still open issues is the radiation emission mechanism at work. 
Most of the GRB properties derived thus far come from the time--averaged spectra, that are 
mainly described with empirical functions. The function that has been found to better describe
the prompt emission spectra from 15 keV up to at least 10 MeV  is a 
smoothly broken power--law proposed by \citet{Band93}({\em Band function}, {\sc bf}).
On the basis of the spectral data obtained with the {\em Burst and Transient Source Experiment} (BATSE), 
aboard the \gro\ satellite (CGRO) and with other satellite data \citep[e.g.,][]{Guidorzi11}, for
long GRBs ($>$2 s), the mean value of the low energy photon index $\alpha$ of the {\sc bf} is 
about $-1$, while that of the high energy photon index $\beta$ is about $-2.3$ \citep{Kaneko06}. 
As a consequence of this result, the received power per unit logarithmic 
energy interval  $EF(E)$ shows a peak value, that in the BATSE era seemed to show a sharp 
Gaussian distribution around 200 keV. Actually, with the discovery of the X--ray flashes with 
\sax, later also found with HETE-2, \swift, and, now, with the \fermi\ Gamma-Ray Burst Monitor (GBM), 
this distribution results to be much flatter \citep[e.g.,][]{Kippen03,Sakamoto05}.  In the cases 
in which $\beta$ cannot be constrained, a power--law model with
a high energy exponential cutoff ({\sc cpl}) gives a good fit to the data, and, 
in some cases, even a simple power law ({\sc pl}) can describe the GRB time--averaged
spectra up to several MeV photon energies.

Besides fitting with an empirical function, different radiative emission models have been developed to interpret the GRB
spectra. Given their non thermal shape, the first model proposed was a
synchrotron emission model by non thermal electrons
in strong magnetic fields \citep{Rees94,Katz94,Tavani96}. Indeed, the time--averaged spectra of many GRBs
are consistent with an optically thin synchrotron shock model 
\citep[e.g.,][]{Tavani96,Amati01}. However, there is a significant number of GRBs for which
this model does not work. Indeed, while for an optically thin synchrotron
spectrum, the expected power-law index of the $E F(E)$ spectrum below the
peak energy $E_p$ cannot be steeper than 4/3 (ideal case of an instantaneous
spectrum in which the electron cooling is not taken into account), in many cases \citep[e.g.,][]{Preece98,Preece00})
the measured spectra, even those time--resolved \citep{Crider97,Frontera00a}, 
are inconsistent with these expectations. 

To overcome these difficulties, either modifications of the above synchrotron scenario
\citep[e.g.,][]{Lloyd00a}, or other radiative models, have been suggested. Among them,
we mention the synchrotron self-Compton model \citep{Meszaros00,Stern04}, the Compton up-scattering 
of low energy photons by a quasi--static plasma \citep{Liang97}, the superposition of 
blackbody spectra \citep{Blinnikov99}, the Compton drag emission model \citep{Lazzati00}, 
thermal emission plus a possible non-thermal tail model \citep{Lazzati09}. Each of these models 
interprets some of  the emission features, but fails to interpret others.  

One of the GRB spectral properties that the emission models should interpret
is a correlation between the intrinsic peak energy $E_{p,i}$ of 
the $E F(E)$ function and either the GRB released energy $E_{iso}$ \citep{Amati02} or the peak bolometric luminosity
$L_{p,iso}$ \citep{Yonetoku04}. Both correlations (Amati relation and Yonetoku relation)
have been derived from the time integrated spectra assuming  isotropic emission.
Actually, the Amati relation ($E_{p,i} = a E_{iso}^m$, with $a = 98\pm 7$~keV when $E_{iso}$ is given in units of $10^{52}$~erg, and $m = 0.54\pm 0.03$, \citet{Amati08})
has been questioned by various authors \citep{Band05,Butler07b,Butler09,Shahmoradi11}, 
but it is a matter of fact that this relation 
is confirmed by all GRBs (more than 150) with known redshift $z$ and determined $E_{p,i}$ discovered thus far, 
except GRB\,980425 \citep[but see][]{Yamazaki03,Ghisellini06}, the nearest and less energetic event ever observed ($z= 0.0085$).
Many models have been suggested to explain the Amati relation (see discussion in 
\citet{Ghirlanda10}). 
We should list some of them. Namely, they are: the model proposed by \citet{Zhang02} in the context of the standard
synchrotron shock scenario, which is based on the assumption that a considerable fraction of the
prompt emission flux is due to blackbody \citep{Rees05,Thompson06,Thompson07}, the model proposed
by \citet{Giannios07}, which is based on the magnetic reconnection mechanism, in which the flow 
that powers a GRB is initially Poynting flux dominated, the model proposed by \citet{Panaitescu09},
in which the prompt emission is synchrotron radiation from the external shock. The tests already performed for
 these models show that either the theoretical expectations are not found in the data \citep[e.g.,][]{Ghirlanda07} 
or the constraints imposed by the models could not be verified with the observations. 

Given the significant evolution of the GRB spectra and the fact that many models can predict the shape of the instantaneous spectra, studies of the measured time--resolved spectra have also been performed by several
authors. These studies are strongly expected  to go deeper in the issue of the radiative mechanisms
at work during the prompt emission. Thus far, fits of 
time--resolved spectra 
with physical and/or semi--empirical models have already been performed. One of these models is a simple 
blackbody ({\sc bb}) \citep{Ghirlanda03}, that was found to be suitable to describe
the first few seconds of GRB prompt emission. Another model is the sum of a {\sc bb} plus a 
power--law component \citep{Ryde04,Ryde05} with time-dependent spectral parameters. This spectral model was found
to  describe the time--resolved spectra of the prompt emission of a sample of BATSE GRBs. 
Later, \citet{Ryde09} have confirmed the validity of this blackbody plus 
power-law model using a sample of 56 bright BATSE GRBs. 
The authors stress the fact that the existence of a 
thermal component is a natural outcome of the fireball model. 
In fact, in the fireball scenario, in addition to an internal shock emission region, it 
is expected to see  a thermal emission from the region where the fireball becomes 
optically thin (optical depth $\tau = 1$) during its expansion (fireball photosphere). 
Indeed, detailed calculations performed by \citet{Fan10} show that a thermal component should be apparent
for reasonable values of the GRB parameters (isotropic-equivalent outflow luminosity $L_{iso} \approx 10^{52}$~erg~s$^{-1}$,
final bulk Lorentz factor $\Gamma \approx 10^3$ and initial radius of the outflow 
$ R_0 \approx 10^6$ cm).

However, observational results show a more complex phenomenology. In a 
well--studied case of the \fermi\ GRB\,080916C \citep{Zhang09,Fan10}, in which outflow parameters 
have been constrained and a significant thermal emission should have been visible, no evidence 
of a thermal component was found \citep{Abdo09a}. 
A multicolor blackbody plus a power--law  model is the best description of the prompt emission spectrum of GRB\,090902B given by \citet{Ryde10}, but see also \citet{Zhang11,Peer12}.  In some other cases, like GRB\,100724B, the spectrum shows, in addition to the Band function, a component which is consistent with a blackbody with a temperature of about 40~keV \citep{Guiriec11}. Also GRB110721A, a burst dominated by a single emission episode, shows a time--resolved spectrum that has been described with a Band function plus a blackbody component \citep{Axelsson12}. 
But, in general, the Band function, with a possible cutoff at very high energies \citep{Ackermann12}, fits the data. The absence of a thermal component in many GRBs
could be due to the fact that the outflow properties change from burst to burst.
For example, for a relativistic outflow dominated by a Poynting flux component \citep{Lyutikov03}, 
a very weak thermal photospheric emission is expected. Alternatively, its absence could be, e.g., due to the limited passband 
of the used detector,  for example, a threshold passband  higher than that would be required for the 
detection of the thermal component. In the case of BATSE, the lower energy threshold is about 25 keV, while
in the case of the {\em Fermi}/GBM instrument it is $\sim$10 keV. If the thermal emission 
is characterized by a lower temperature (a $kT_{bb}\sim$1 keV was found by \citet{Frontera01}), it could escape 
most of the observations of the GRB prompt emission performed thus far. Actually, thermal components have been observed in the early afterglow of a number of \swift\ GRBs \citep{Starling12,Sparre12,Friis13}, thanks to the low
energy threshold (0.5~keV) of the X--ray Telescope (XRT) aboard.  

The Wide Field Cameras (WFCs) aboard the \sax\ satellite were among the few instruments 
that detected GRB prompt emission down to 2 keV \citep{Frontera09,Guidorzi11,Frontera12a,Frontera12b}. Some of these GRBs were
simultaneously observed with BATSE. In this work, using the time--resolved spectra of 
GRBs detected with both WFCs and BATSE (among which GRB~990123, that is part of the \citet{Ryde09} sample), 
we report on  the results obtained from the test of the {\sc bb} plus {\sc pl} model adopted by \citet{Ryde09} but also suggested
by \citet{Lazzati09}, the search of a {\sc bb} component in addition to {\sc bf}, and the test of the X--ray spectral model ({\sc grbcomp}) recently proposed by some of us \citep[][hereafter T12]{Titarchuk12}.

{\sc grbcomp} is essentially a photospheric model for the prompt emission
of GRBs. Its main ingredients are a thermal bath of soft seed photons
which are Comptonized by a subrelativistically expanding outflow characterized by a Maxwellian electron plasma with temperature $kT_e$ and Thomson optical depth $\tau$.
A non--relativistic expansion at the beginning of the explosion is also expected in the electromagnetic 
outflow model by \citet{Lyutikov03}. 
The outgoing emerging spectrum at least up to the peak of the $EF(E)$ diagram 
is the result of  multiple Compton scatterings of the seed photons in an hot environment having $\tau > 1$.
Under these conditions, the Comptonization parameter $Y \propto \Theta \tau^2 \gg 1$ where $\Theta= kT_e/m_e c^2$, and quasi-saturated spectra are produced.
The immediate consequence of this  result is that the peak energy $E_p$ of the $E F(E)$
spectrum mostly depends on the electron temperature $kT_e$, with a modification induced by the fact that the plasma is not static, but subrelativistically moving outwards (T12).
The  high-energy power-law tail above the energy peak of the model is instead phenomenologically obtained by the convolution of the subrelativistic Comptonized spectrum with a
broken powerlaw upscattering Green's function.
The reason for this pure mathematical treatment of the second part of the spectrum resides on the fact that, whatever its origin, the  underlying process giving rise
to tails extending in some case up the GeV energies cannot be treated using classical
Comptonization equations (e.g. Fokker-Planck approximation), but presumably will require a fully relativistic treatment of the photon-electron interaction. Possible physical interpretations of the last convolution are discussed by T12.
A possible scenario is an explosion inside the progenitor star with the production of a sea of seed photons plus, in a small region, likely around the star spin axis, a hot Compton cloud. Due to the radiation pression, a hot outflow  is activated, whose velocity is sub-relativistic until the sonic point, where it becomes relativistic. In the relativistic jet region the optical depth is less than 1, thus only a small fraction of the photons Comptonized during the non relativistic phase are further up scattered. Most of the photons freely pass through the jet. The small fraction of up scattered  in the jet is strictly directed along the jet direction, while the photons Comptonized in the subrelativistic stage of the outflow are isotropically emitted. The photons up-scattered by the jet only modify the high energy tail of the emitted spectrum. 
In this case, the convolution can be interpreted as an Inverse Compton of the relativistic plasma outflow with the Comptonized photons. 

Another possibility is that the electron distribution of the Compton cloud is not fully Maxwellian, but is characterized by an hybrid distribution with a suprathermal powerlaw component. In this case, the two-step approach of the radiative transfer equation of {\sc grbcomp} -- solution of subrelativistic Fokker-Planck equation and its convolution  with the up-scattering  Green's function -- would exactly mimic
the multiple scattering photon interaction off a hybrid thermal/non-thermal electron population.

The {\sc grbcomp} model somewhat resembles the scenario proposed by \citet{Lazzati00} for their Compton-drag model, but with two main differences. In {\sc grbcomp} the outflow is subrelativistic and multiple scatterings are assumed, while in  \citet{Lazzati00} the spectrum is formed from single-scattering Inverse Compton of the thermal photons off a relativistic outflow.
From the numerical simulations performed by T12, it results
that the photon peak energy of the $E F(E)$ spectrum is mainly related to $kT_e$ and, at lower level,
to the bulk parameter $\delta$ (see below), while the released energy depends on both the electron and seed photon temperatures.

%

\section{GRB sample and spectral analysis}
\label{s:models}

The entire number of GRBs simultaneously detected with WFCs and BATSE is 20: 960720, 970111, 970508, 971206B, 971214, 971227, 980109, 980326,
980329, 980425, 980519, 990123, 990510, 990625, 990806, 990907, 991014, 991030, 991105, and 000424.
%
However, only four of them (970111, 980329, 990123, and 990510) were sufficiently bright to allow a 
fine time--resolved spectroscopy (see Table~\ref{t:grb-intervals}) in both instruments. These GRBs are the same used to derive the 
time--resolved
$E_p$--$L_{iso}$ correlation \citep{Frontera12a,Frontera12b}. 

The instrumentation that detected these bursts is widely described in the literature.  
For the BATSE experiment see, e.g., \citet{Fishman94}, for WFC, see \citet{Jager97}. 
The BATSE spectra were taken from the Large Area Detectors (LADs), whose
typical passband is 25--2000 keV. The LADs provide various types of data products.
The data used for this analysis are the High Energy Resolution Burst (HERB) data, that
provide 128 energy channels with a minimum integration time of 64 ms. For details on the
BATSE spectral data products and detector response matrix, see \citet{Kaneko06} 
and references therein.

The WFCs consisted of two coded aperture cameras, each with a field
of view of 40$^\circ \times 40^\circ$ (full width at zero response) and
an angular resolution of 5 arcmin. They operated in normal mode with 31 energy 
channels in 2--28~keV and 0.5~ms time resolution.

\begin{figure}
\begin{center}
  \includegraphics[width=.4\textwidth]{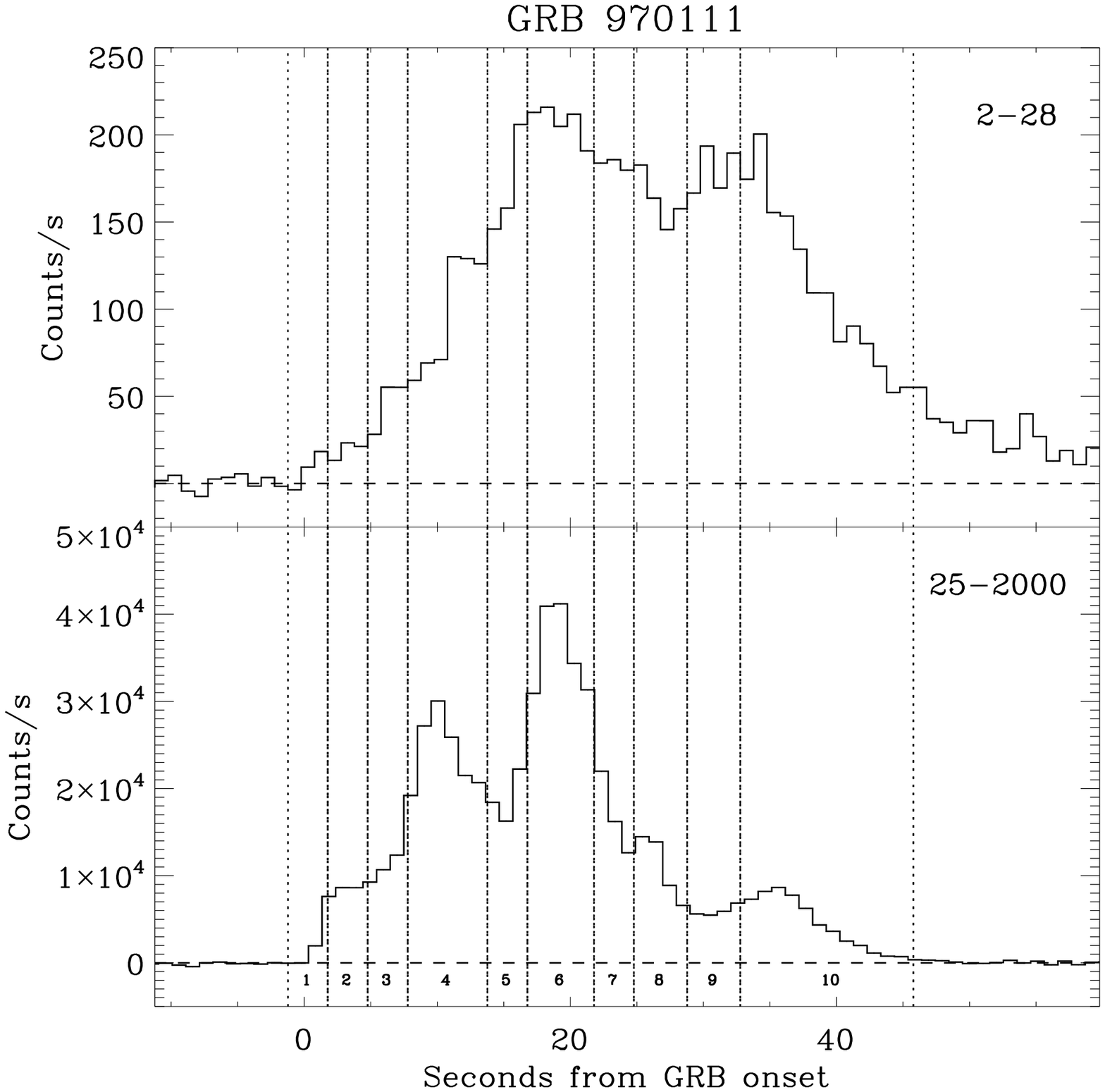}
  \includegraphics[width=.4\textwidth]{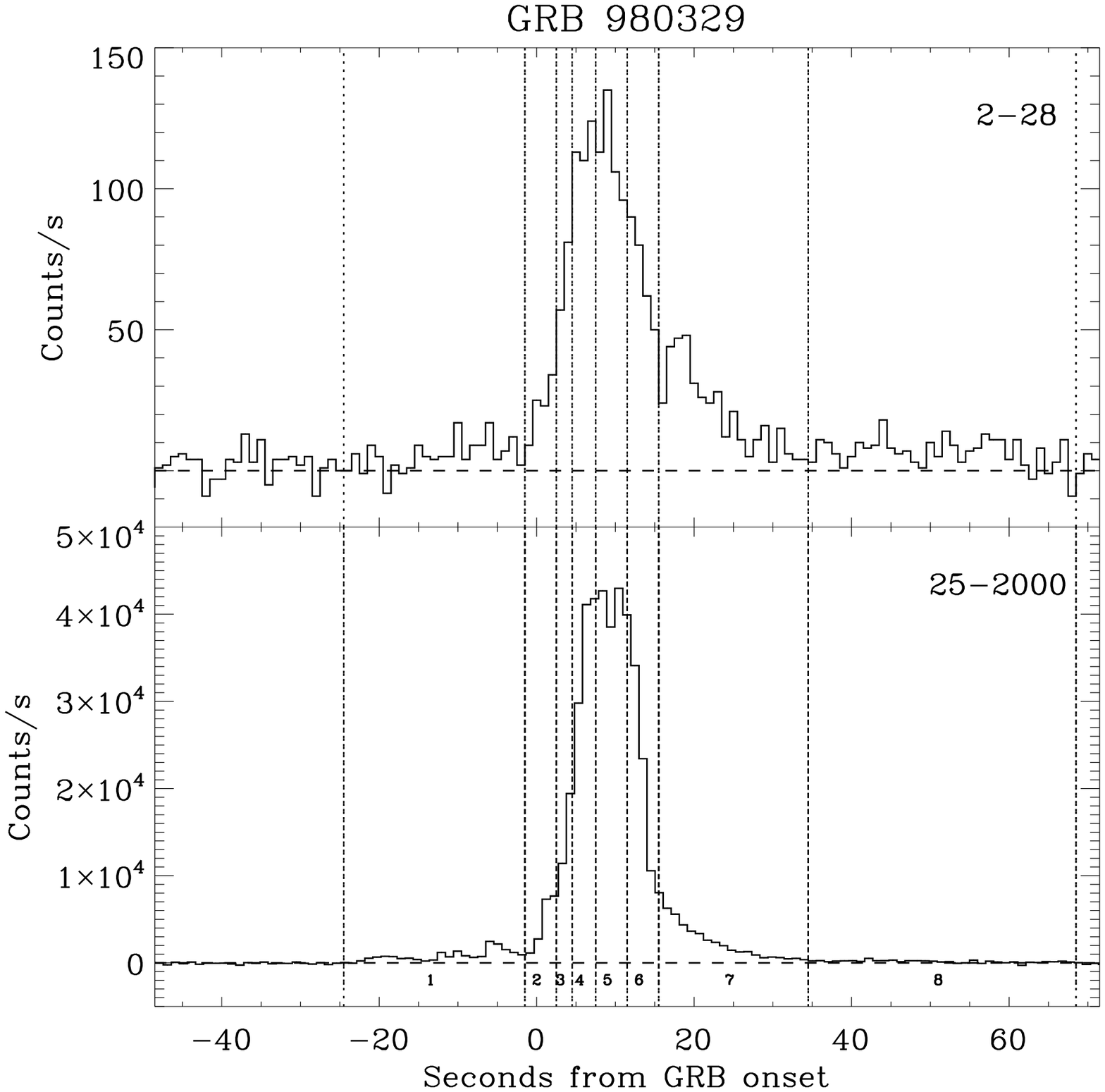}

\vspace{0.5in}

  \includegraphics[width=.4\textwidth]{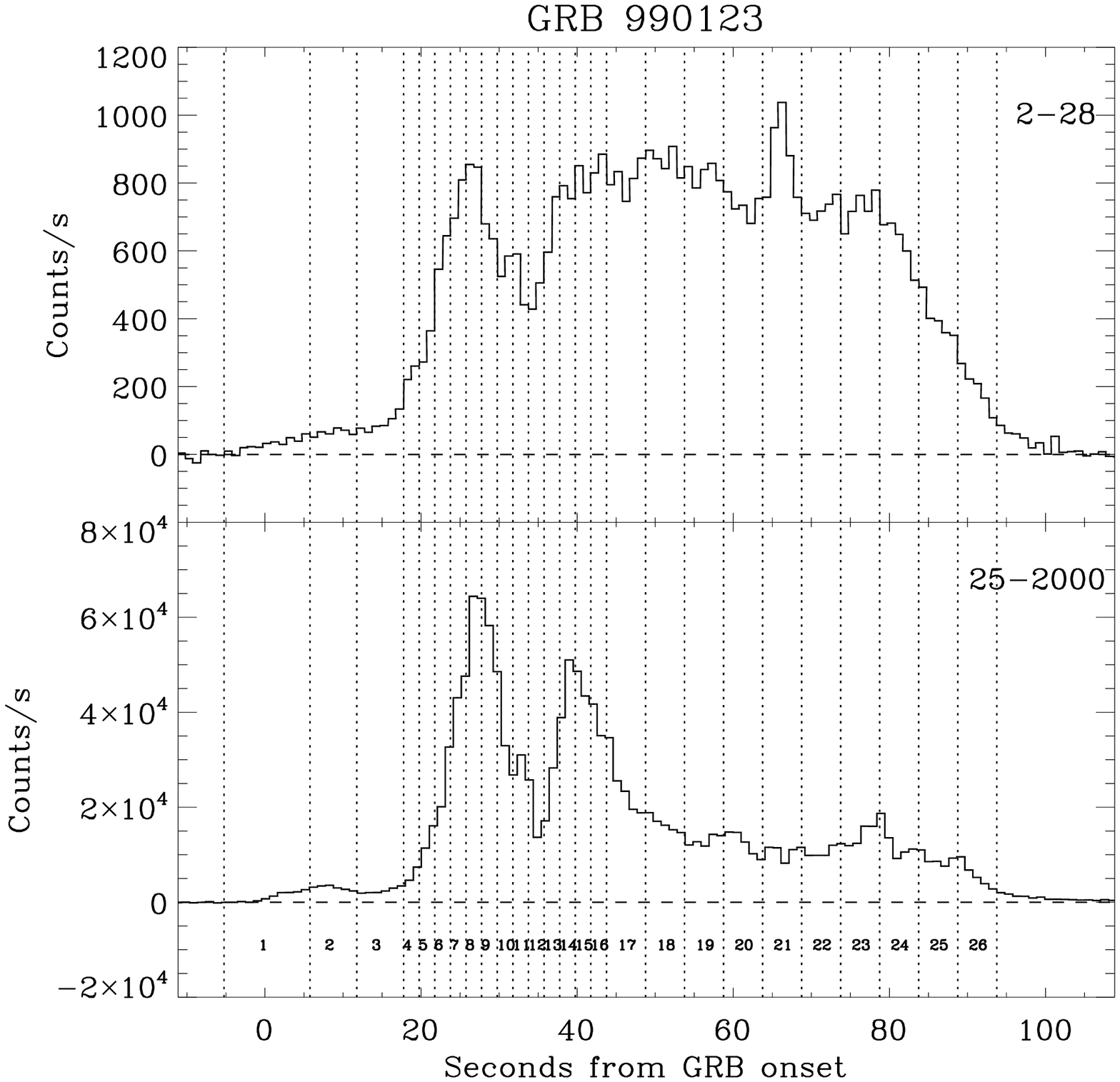}
  \includegraphics[width=.4\textwidth]{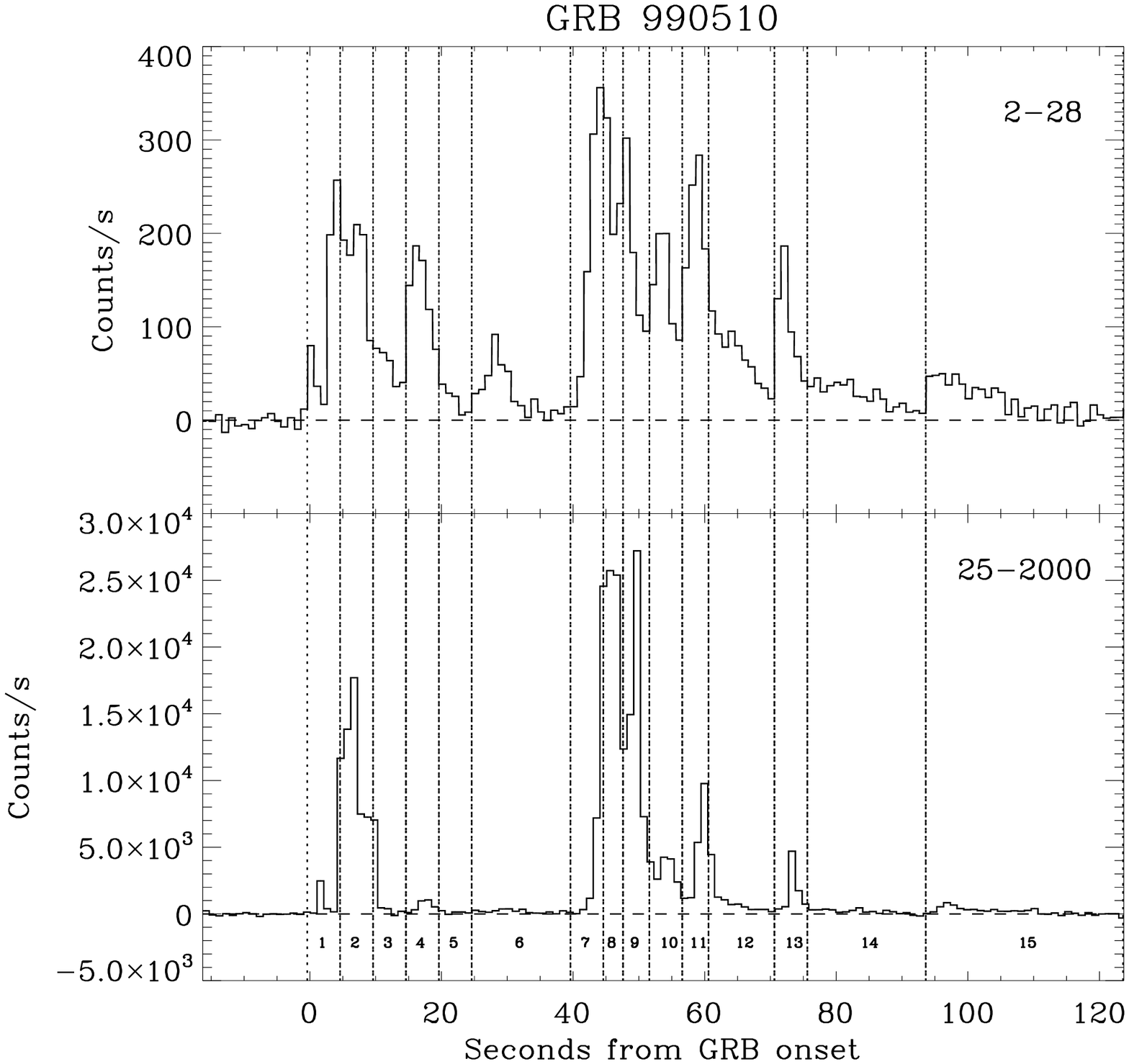}
\end{center}
\vspace{0.5cm}
\caption{Light curves of GRBs 970111, 980329, 990123, 990510, detected with
\sax\ WFC (2--28 keV) and BATSE (25--2000 keV). The intervals, in which the 
time--resolved spectra were derived, are also shown.}
\label{f:lc}
\end{figure}

The background--subtracted light curves of the 4 strongest GRBs in our sample 
detected with both BATSE and WFCs are 
shown in Fig.~\ref{f:lc}. For the BATSE data, the background level was estimated using the count
rates immediately before and after the GRBs. Given that the background is
variable during the GRB, it was  estimated by means of a parabolic interpolation, channel
by channel, between the background measured before the event and that after the event.
Instead, for WFC spectra, the background level was estimated 
using an equivalent section of the detector area not illuminated by the
burst or by other known X-ray sources. We also checked the consistency
of this background level with that obtained by using the data before and
after the burst. 

We subdivided the time profile of each GRB into a number of time slices (see 
Table~\ref{t:grb-intervals}) taking into account the GRB profile (visible pulses, their rise, peak, and decay) as observed 
with 1 s time resolution (in the case of BATSE the real integration time was 1.024~s) and 
the count statistics (see Figure~\ref{f:lc}). We performed 
the spectral analysis in all of the time slices (most of them) in which it was possible 
to constrain the model parameters.  In the case of GRB\,990123, we excluded from the 
analysis the time intervals from 21 to 26, given that this part of the event was 
observed by WFC through the Earth's atmosphere. The number of time intervals in which 
we subdivided the time profile of each event, the number of selected time slices in 
which the spectral analysis was performed, the GRB fluence, and its 
redshift when known, are given in Table~\ref{t:grb-intervals}. 
 
We used as input models {\sc bb$+$pl}, {\sc bb$+$bf}, and {\sc grbcomp}. In all fits, a normalization 
factor 
between \batse\ and WFC data was included and left free to vary in the range 0.8--1.2, 
to account for a possible intercalibration error. Actually, we found that, 
for all analyzed GRB spectra, this parameter was consistent with 1 within one standard deviation, except in two cases (one time resolved of GRB\,970111 and another of GRB\,990123), in which it is consistent with 1 at 90\% confidence level.  
The systematic error used by the
\batse\ team to take into account the uncertainty in the background subtraction and 
the uncertainty in the instrument response function (see, e.g., \citet{Kaneko06} for details)
was included in the fit.

\begin{figure}
\begin{center}
  \includegraphics[angle=-90,width=.8\textwidth]{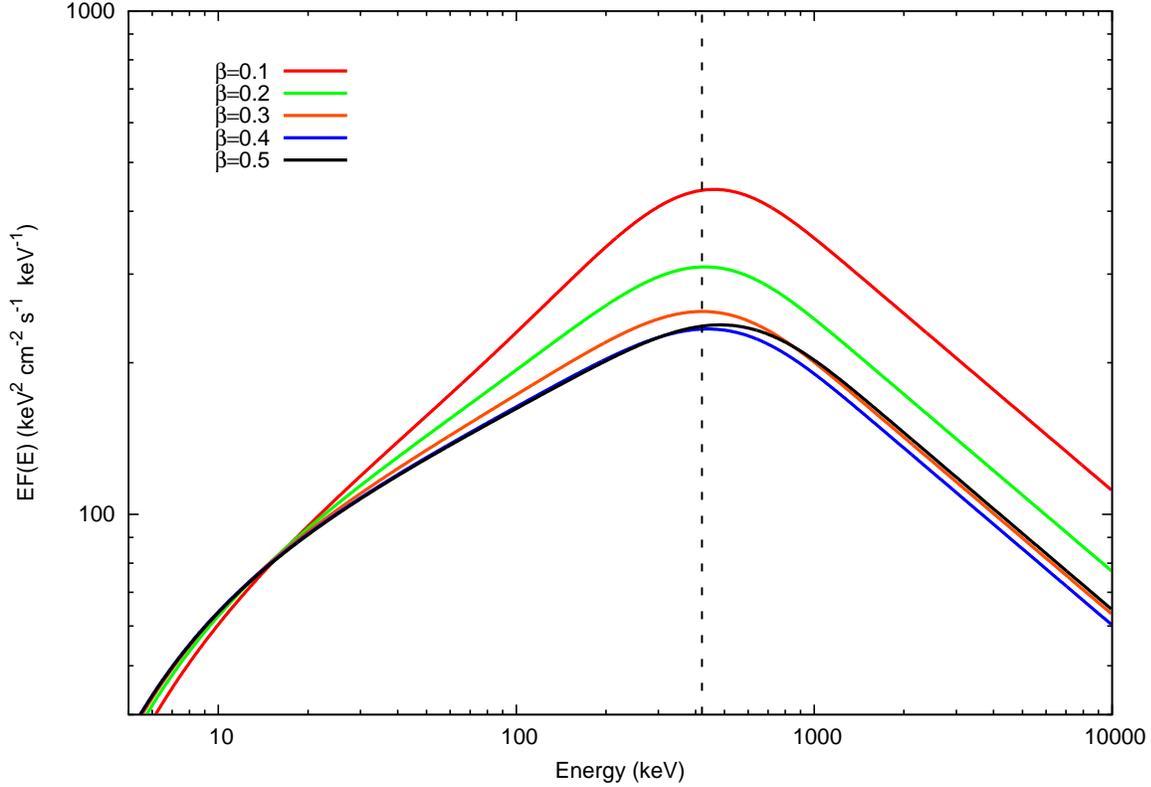}
\end{center}
  \caption{Spectral behaviour of the GRBCOMP model in EF(E) units as a function of the 
outflow velocity $\beta$. The fixed parameters are $kT_{bb}=1$~ keV, $kT_{e,i} = 100$~keV, $\tau=$5 and $\alpha_{boost}= 1.5$.
The peak energy $E_p$ slightly decreases for increasing values of $\beta$ 
because of the Doppler effect for subrelativistic outflow  \citep[see][]{Laurent07} 
due to first order Fermi effect, which is proportional to $\beta$.
 When $\beta=0.5$ or higher, the second order Fermi effect (proportional to 
$\beta^2$) starts to be important and essentially adds its contribution to the 
thermal Comptonization process. This effect is dictated by the parameter
$f_b=1+\beta^{2/3\theta}$ in the thermal Comptonization term of the
Fokker-Planck operator (see Eqs. 1 and 4 in T12).
Of course, the relative contribution of the $\beta$ and $\beta^2$ terms depends on the electron temperature $kT_e$.} 
\label{f:spectrum-vs-beta}
\end{figure}

The fit results with the {\sc bf} function alone have  already been published \citep{Frontera12a,Frontera12b}. 
In the case of the first model, in addition to the 
normalization constants for {\sc bb} and {\sc pl}, the other free parameters are the {\sc bb} temperature 
$kT_{bb}$ (in keV) and the power--law photon index $\Gamma$.
In the case of the {\sc bb$+$bf}, in addition to the normalization constants for {\sc bb} and {\sc bf}, the other free parameters are the {\sc bb} temperature 
$kT_{bb}$ (in keV), the photon indices $\alpha$ and $\beta$, and the peak energy in the observer frame $E_{p,o}$.  In the case of the {\sc grbmcomp} model, in
addition to its normalization constant $N = R_9^2/D^2_{Mpc}$, (where $R_9$ is the apparent photospheric radius $R_{ph}$ in units of $10^9$~cm, and $D^2_{Mpc}$ is the source distance in Mpc), 
the other free parameters, in the rest frame, are the temperature of the seed photons $kT_{s,i}$ (in keV), 
the plasma electron temperature $kT_{e,i}$ (in keV), the effective optical depth $\tau_{eff}$ of the plasma outflow, 
the plasma outflow bulk velocity $v$, and the power--law photon index $\alpha_{boost}$ of the component above the peak energy. From the model best--fit parameters, it is also possible to derive the bulk parameter $\delta$, which, for the case of a constant outflow velocity, is defined as:
\begin{equation}
 \delta = 2 \beta/(3 \tau_{eff} \Theta)
\label{e:delta}
\end{equation}
where $\beta = v/c$, and  $\Theta = kT_{e,i}/ m_e c^2$ and $\tau_{eff}$ is an effective optical depth such that  $\tau_{eff} \la \tau$. 
The definition of an effective optical depth $\tau_{eff}$ was introduced to separate the space and energy operators in the radiative transfer Fokker-Planck equation (see Eq.~(6) in T12), which provides a much faster way for getting numerical solutions. In {\sc grbcomp} it is assumed $\tau_{eff}=0.5 \tau$, where $\tau$ is 
the actual free parameter of the model.
Finally, despite being in principle a free parameter, the outflow subrelativistic velocity $\beta$
is kept frozen in the fitting procedure, to avoid parameter degeneracy or too large
uncertainties. In our analysis we assumed $\beta=0.2$. This value is the median value of those consistent with {\sc grbcomp}
(see Fig.~\ref{f:spectrum-vs-beta}).

Each model was assumed to be photoelectrically absorbed (WABS model in {\sc xspec}).  
Given that the absorption column density $N_{\rm H}$ could not be constrained, a Galactic absorption 
along the GRB direction \citep{Kalberla05} was assumed. 
While the {\sc bb}$+${\sc pl} model was separately fit to  the BATSE spectra alone and to the joint 
WFC plus BATSE spectra, the {\sc grbcomp} was only fit to the joint WFC plus BATSE spectra. 
To deconvolve the count rate spectra, we adopted the {\sc xspec} ({\em v. 12.5}) software package \citep{Arnaud96}.  
If not explicitly stated, the quoted uncertainties are single parameter errors at 90\% confidence level.

\section{Results}

The fit results of the  {\sc bb$+$pl} and {\sc grbcomp}  models to the joint WFC$+$BATSE time--resolved spectra are reported in Table~\ref{t:results}, while the time behavior of the {\sc grbcomp} parameters and of the Null Hypothesis Probability ($NHP$) for all joint fits and when the {\sc bb$+$pl} model is fit to the time--resolved BATSE spectra alone, are reported in the four panels of  Fig.~\ref{f:time_evol}. 
In Figure~\ref{f:NHP}, the cumulative distribution of $NHP$ displays the fraction of intervals with a good fit for each model.

\subsection{Fit with {\sc bb$+$pl} and {\sc bf$+$bb}}

 As it can be seen from Table~\ref{t:results}  and from behavior of $NHP$ ( see Fig.~\ref{f:time_evol} and Fig.~\ref{f:NHP}), the fit with {\sc bb$+$pl} is acceptable only when  we fit this model to the time--resolved BATSE spectra alone, as found by \citet{Ryde09}.  When the joint WFC$+$BATSE spectra are considered, for the majority of the time--resolved 
spectra this model  cannot be accepted. Examples of these fits are shown in Fig.~\ref{f:bbpl_fit}. The reason for the unsuccessful fit of a BB+PL model appears to reside in the fact that
the data show a steepening at lower energies below the {\sc bb}  peak. Indeed, as it can be clearly seen in Fig.~\ref{f:bbpl_fit},  the residuals  between the data and the model show that below 10 keV the data are systematically under the model continuum. Thus the slopes of the observed GRB spectra below and above
the {\sc bb} component are different, and cannot be simultaneously fitted by a single power--law index.
This behavior cannot be observed with BATSE because of its low energy threshold around 20 keV. This result shows the importance of having a spectral coverage as large as possible, especially down 1-2 keV, for constraining the prompt emission models.

The fit with {\sc bf$+$bb} was performed only for those GRB time--resolved spectra, for which the fit with the {\sc bf} alone provides a $\chi^2/{\rm dof}$ significantly higher than that expected in the case of a good fit \citep[see results in][]{Frontera12a,Frontera12b}. The list of these spectra and of the results obtained by fitting them with
{\sc bf$+$bb} are reported in Table~\ref{t:bb-plus-bf}. For each time interval not well fit with a Band function, mindful of the Protassov et al. (2002) warnings, we report, in addition to the results of the "F-test/add" for the addition of a further component ({\sc bb}) to a fitting function ({\sc bf}), 
the results of the "F-test/plain" for testing the discrepancy between an assumed fit function and its parent function .  This test (see Bevington 1969, p. 195\nocite{Bevington69}) makes use of the probability distribution
of the ratio $F_{12} = \chi^2_{\nu_1}/\chi^2_{\nu_2}$ (or $F_{21} = 1/F_{12}$) where $\chi^2_{\nu_1} = \chi^2/\nu_1$ is obtained from the
fit of the function 1 (in our case {\sc bf}) to the data,
and $\chi^2_{\nu_2} = \chi^2/\nu_2$ is obtained from the fit of the function 2 (in our case, {\sc bf$+$bb}). (For an application of the latter test see \citet{Frontera04}).

As it can be seen from Table~\ref{t:bb-plus-bf}, we do not find any positive result using the F-test/plain, while, using the F-test/add, in one case (interval No.~7 of  GRB\,970111 light curve, see Fig.~\ref{f:lc}) we find a significant decrease of the $\chi^2$ found with {\sc bf}, with a chance probability of $6.7\times 10^{-3}$ that the $\chi^2$ reduction is due to chance when a {\sc bb} is added to {\sc bf}. In the other cases, also reported in Table~\ref{t:bb-plus-bf}, the chance probability  obtained with the F-test/add is not less than 1\%.

In the case of the interval No. 7 of GRB\,970111, using a {\sc bf$+$bb} model, the best--fit spectral parameters, in the observer frame, are the following: $\alpha = -0.43_{-0.14}^{+0.19}$,  $\beta = -3.7_{-0.8}^{+0.4}$, $E_{p,o} = 99\pm11$~keV, and $kT_{bb} = 8.9\pm 2.2$~keV. The {\sc bf} parameter values are slightly different from those obtained when the spectrum was fit with a {\sc bf} alone: $\alpha = -0.58_{-0.04}^{+0.04}$,  $\beta = -3.46_{-0.21}^{+0.15}$, $E_{p,o} = 89_{-3}^{+4}$~keV \citep{Frontera12a,Frontera12b}. In  Fig.~\ref{f:970111-7} (top panels), we show the best--fit curve of the {\sc bf} and  {\sc bf$+$bb} models to the observed spectrum, while in the bottom panel (left side), we show the $E F(E)$ spectrum when the fit is performed with {\sc bf$+$bb}. Unlike that found by, e.g., \citet{Guiriec11},  we find (see Table~\ref{t:bb-plus-bf}) low {\sc bb} temperatures ($\sim 9 $~keV for GRB\,970111, interval 7), except in two cases (GRB\,990123, intervals No. 9 and 12). 
Apart from these few spectra (14\% of the total), no additional component to the {\sc bf} is required by the data \citep{Frontera12a,Frontera12b}.

\subsection{Fit  with {\sc grbcomp}}

\begin{figure}
\begin{center}
  \includegraphics[width=.35\textwidth]{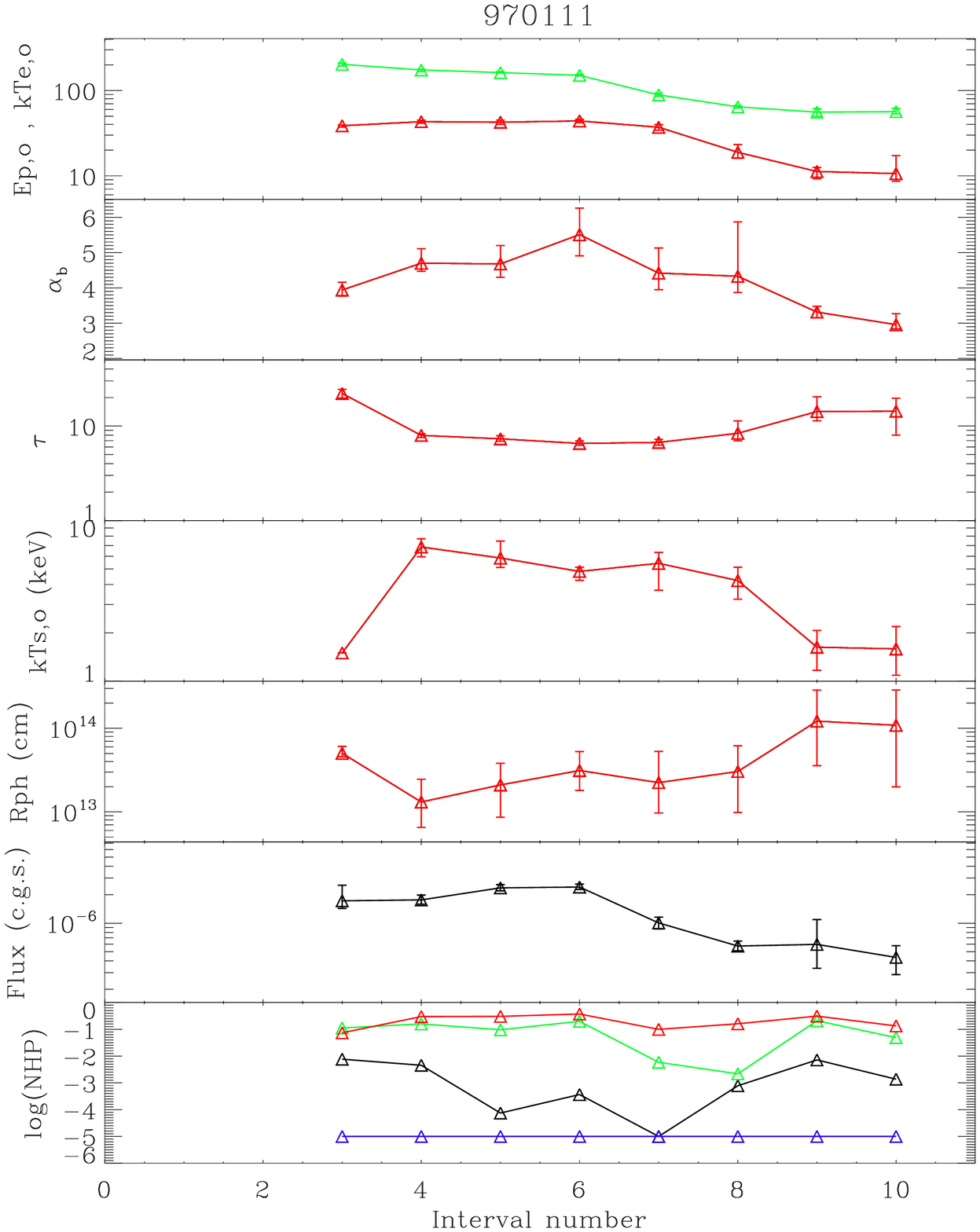}
  \includegraphics[width=.35\textwidth]{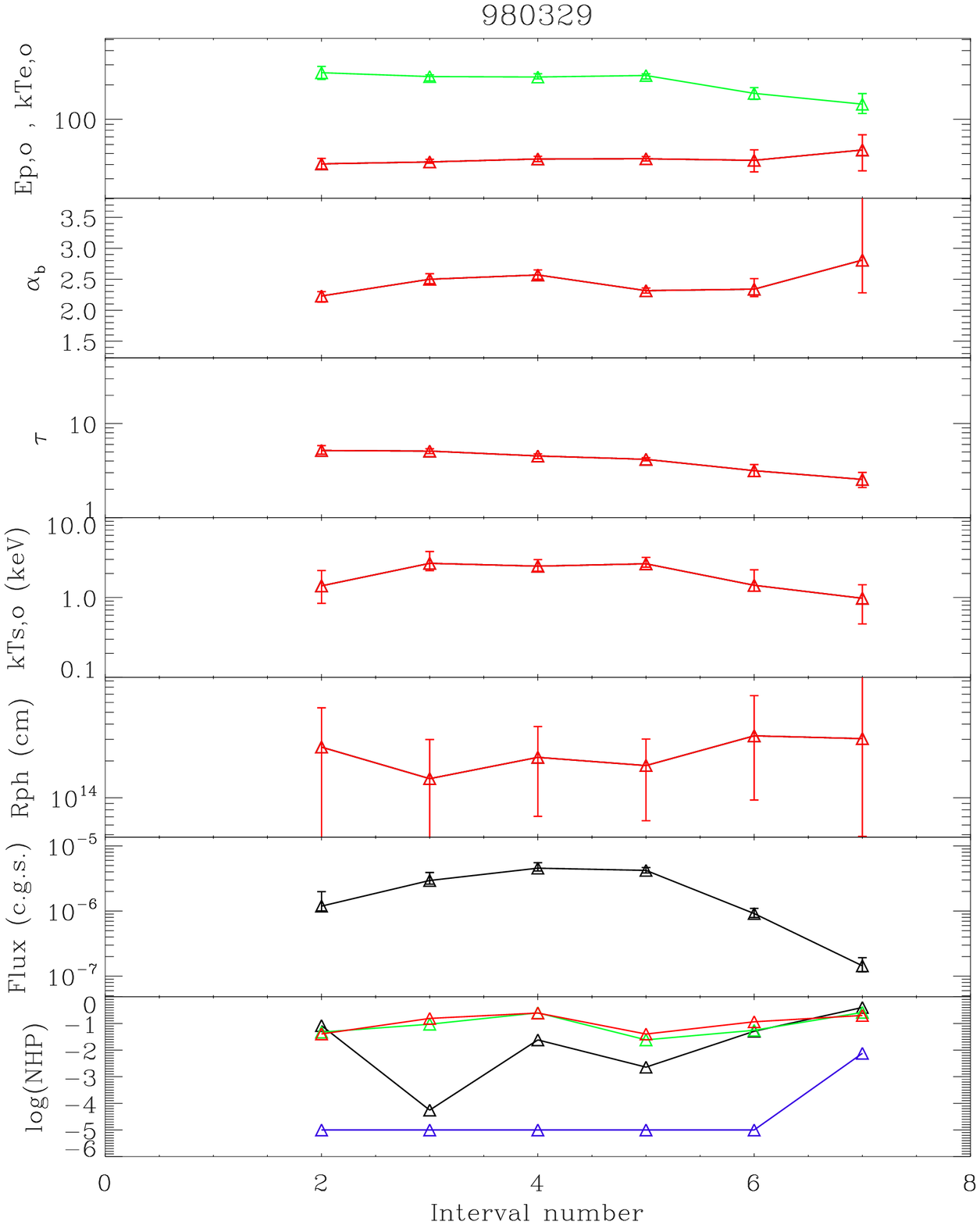}

\vspace{0.5in}

  \includegraphics[width=.35\textwidth]{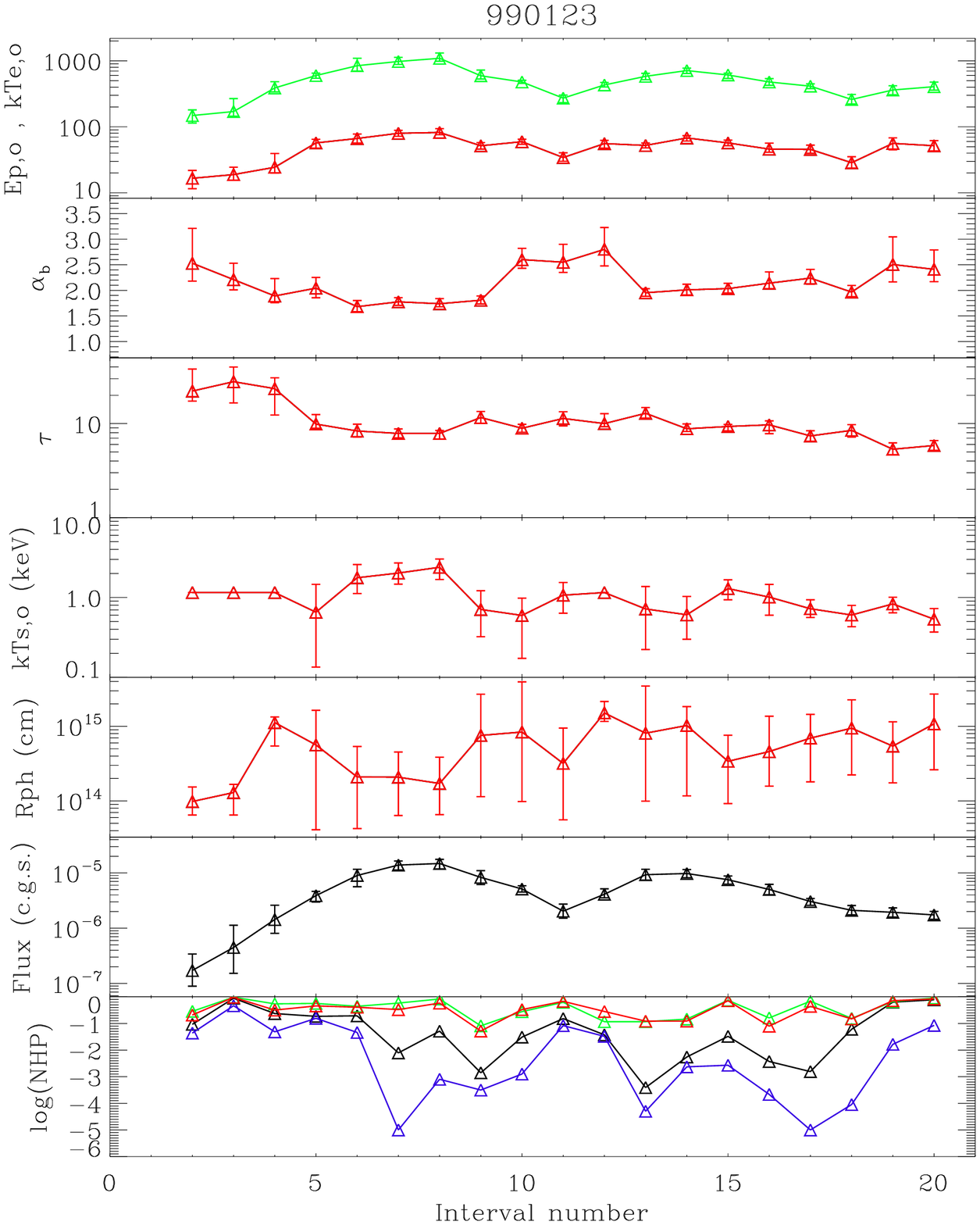}
  \includegraphics[width=.35\textwidth]{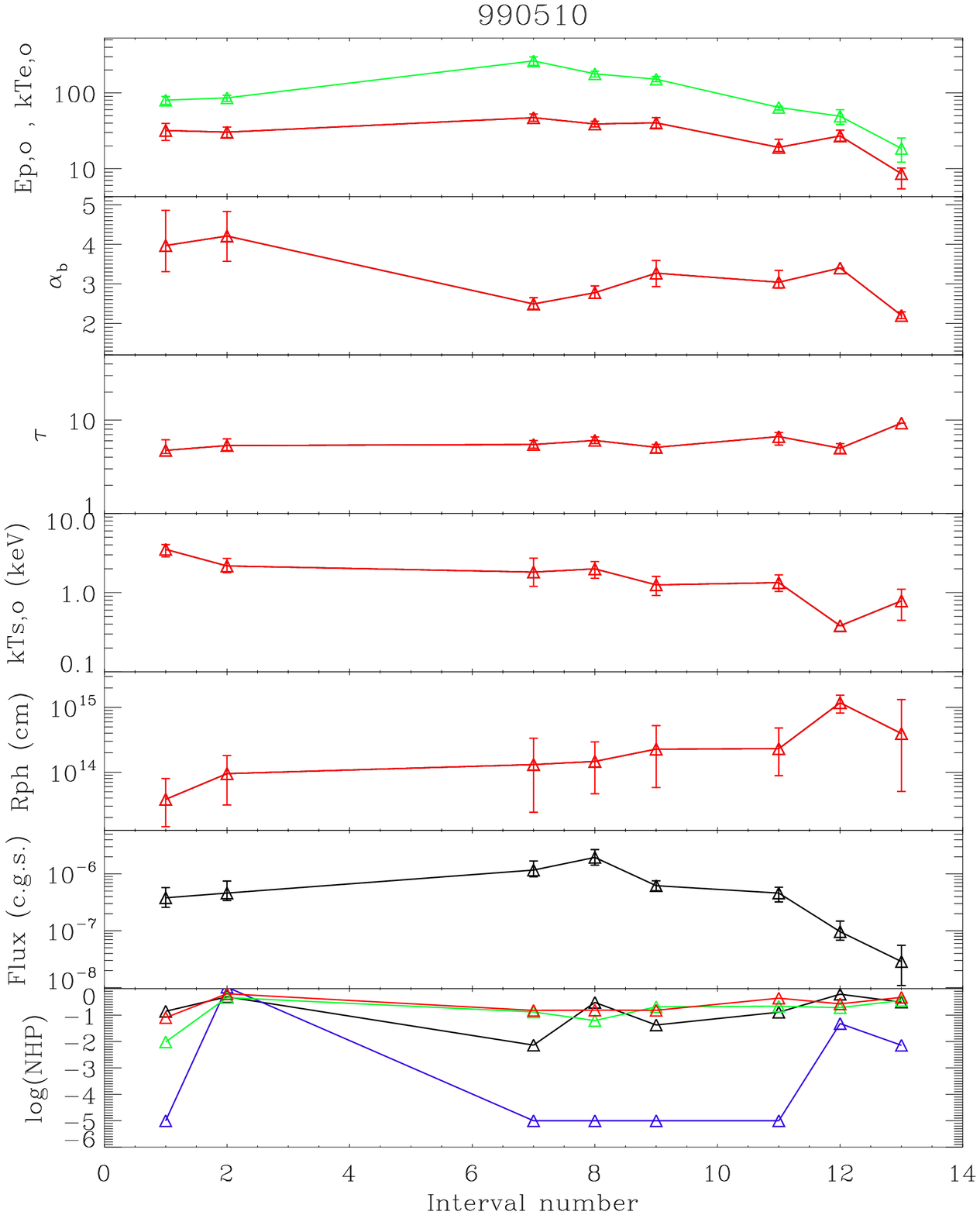}
\end{center}
  \caption{{\em In red}: Temporal evolution of the {\sc grbcomp} best--fit parameters, 2--2000 keV flux and Null Hypothesis Probability ($NHP$). The best--fit parameters are given in the observer frame. For GRB\,970111, a redshift $z=1$ was assumed. In the order from top to bottom: electron temperature $kT_e$, power--law photon index $\alpha_{boost}$, effective optical depth $\tau_{eff}$, seed photon temperature $kT_s$, photospheric radius $R_{ph}$, 2--2000 keV flux, and $NHP$, for each of the GRBs 970111, 980329, 990123, and 990510. For comparison with the other tested models, at the top of each panel it is shown the time behavior of the best--fit peak energy $E_{p,o}$ of the $E F(E)$ spectrum, when {\sc bf} as input model is adopted, while   at the bottom of each panel it is shown the $NHP$ behaviour obtained for {\sc bf} ({\em in green}), for {\sc bb $+$ pl} ({\em in blue}), when only BATSE spectra of the GRBs in our sample are fit with {\sc bb $+$ pl} ({\em in black}). 
$R_{ph}$ is  obtained from the normalization constant of the {\sc grbcomp} model.
}
\label{f:time_evol}
\end{figure}

\begin{figure}
\begin{center}
  \includegraphics[width=.4\textwidth]{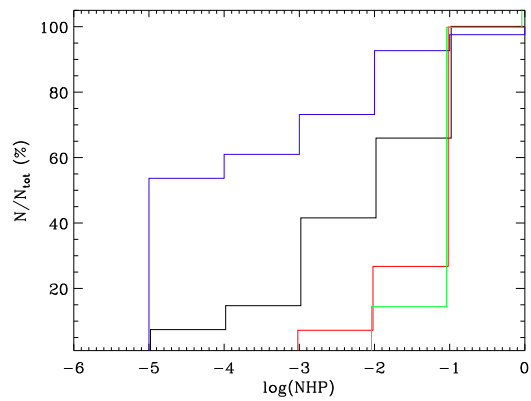}
\end{center}
  \caption{Cumulative distribution of the Null Hypothesis Probability ($NHP$) derived from the best fit of the 
tested models to all time--resolved spectra. In ordinate, the fraction of spectra that show a $NHP$ less than the value reported in the x--axis. For joint BATSE$+$WFC spectra, {\em green:}
Band function; {\em red:} {\sc grbcomp}; {\em blue:} {\sc bb $+$ pl}. For  best fits of {\sc bb $+$ pl} 
to BATSE spectra alone: {\em black}.} 
\label{f:NHP}
\end{figure}

\begin{figure}
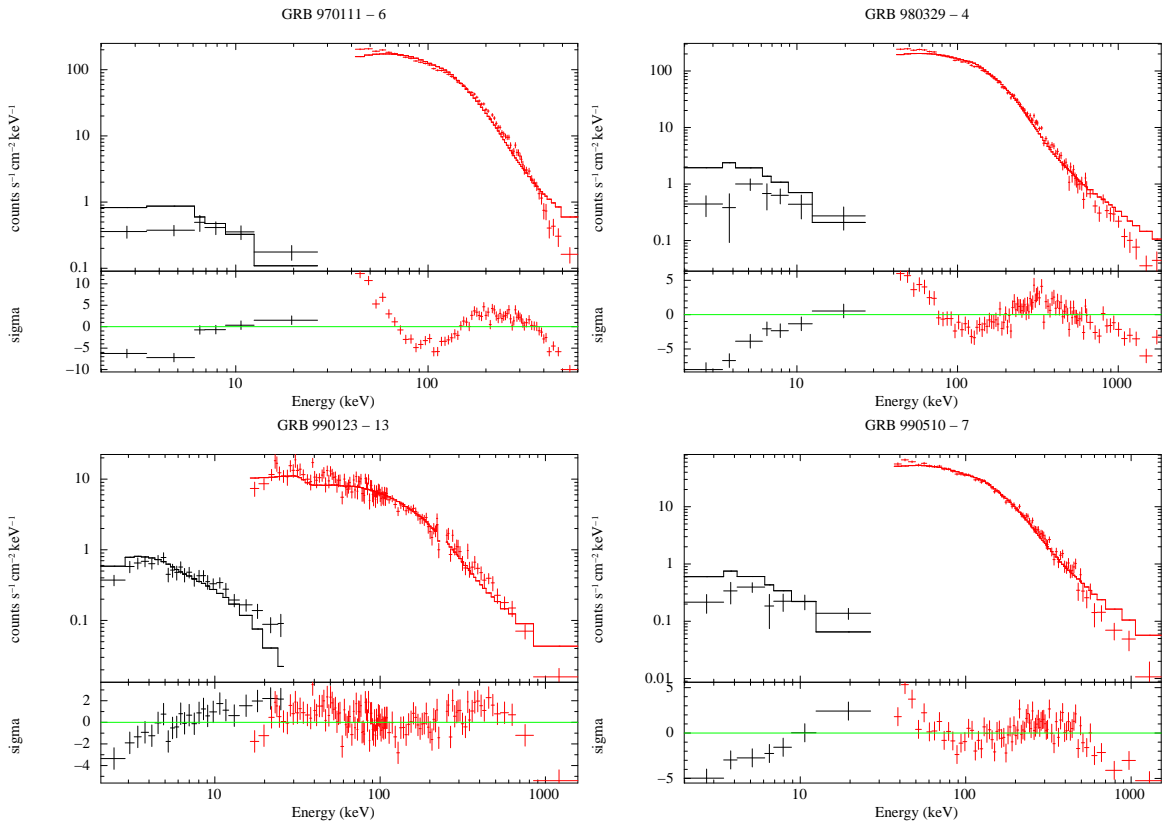

\begin{center}
\includegraphics[angle=-90.,width=.4\textwidth]{f5a.ps}
  \includegraphics[angle= -90.,width=.4\textwidth]{f5b.ps}
  \includegraphics[angle=-90.,width=.4\textwidth]{f5c.ps}
  \includegraphics[angle=-90.,width=.4\textwidth]{f5d.ps}
\end{center}
  \caption{Best--fit results of a few examples of 
time--resolved spectra fit with a BB$+$PL model
(see Table~\ref{t:results} for GRB interval identification number): 
GRB\,970111: interval 6; GRB\,980329: interval 4; 
GRB\,990123: interval 13; GRB\,990510: interval 7.} 
\label{f:bbpl_fit}
\end{figure}

\begin{figure}
\begin{center}
  \includegraphics[angle=-90.0,width=.4\textwidth]{f6a.ps}
  \includegraphics[angle=-90.0,width=.4\textwidth]{f6b.ps}

\vspace{0.5in}

  \includegraphics[width=.4\textwidth]{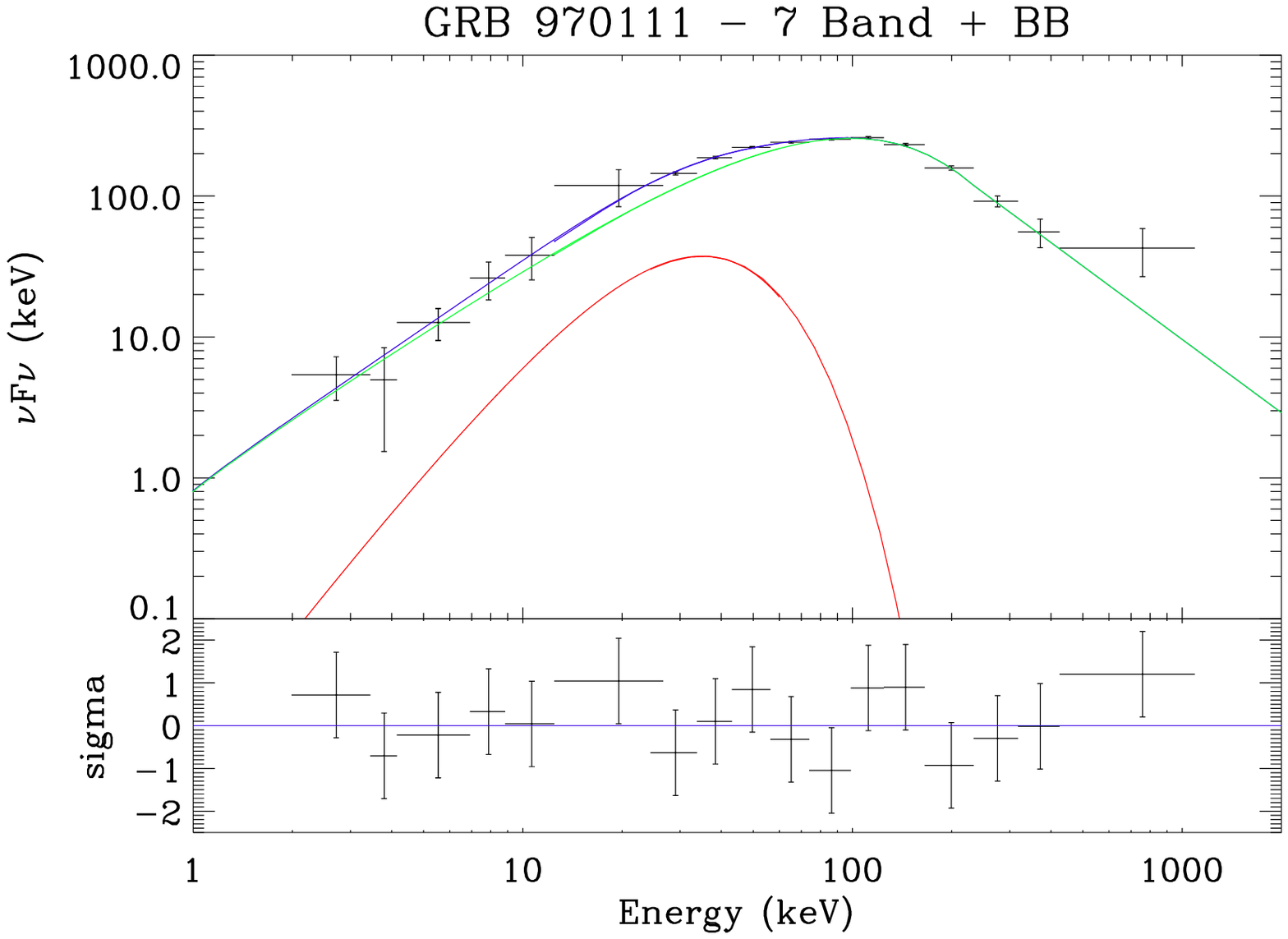}
  \includegraphics[width=.4\textwidth]{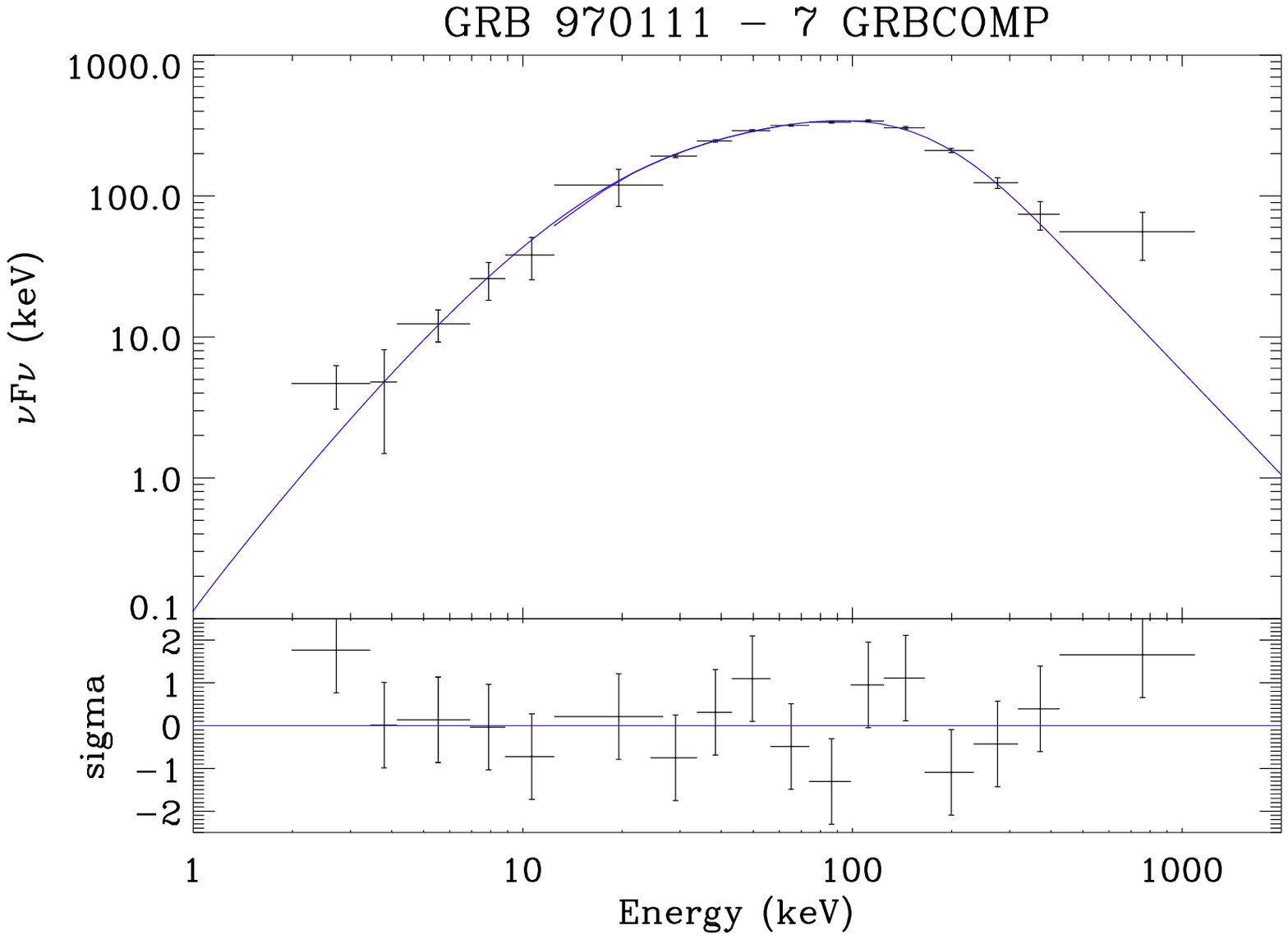}
\end{center}
  \caption{{\em Top panels}: count spectrum of the interval No. 7 of GRB\,970111  and its fit with {\sc bf} (left) and {\sc bf$+$bb} (center). {\em Bottom panels}: $E F(E)$ spectrum of the same interval, with, at the left side,  the {\sc bf$+$bb} best--fit  components and their sum,  at the right side, the {\sc grbcomp} best--fit curve.
} 
\label{f:970111-7}
\end{figure}

The fit results of the  {\sc grbcomp} model to the joint WFC$+$BATSE time--resolved spectra are also reported in Table~\ref{t:results}
(in square parenthesis, those parameters kept fixed in the fits). As it can be seen from this Table, and from the behavior of the Null Hypothesis Probability ($NHP$) reported in each of the four panels shown in  Fig.~\ref{f:time_evol},  and from the $NHP$ cumulative 
distribution shown in Fig.~\ref{f:NHP}, the {\sc grbcomp} model fits well, in some cases even better than {\sc bf} 
(see $NHP$ behavior in  Figs.~\ref{f:time_evol} and \ref{f:NHP}), almost all the available WFC$+$BATSE  
time--resolved spectra.
 
The goodness of the fit of the {\sc grbcomp} model to the data is exemplified in the bottom right panel of Fig.~\ref{f:970111-7}, where we show the fit of {\sc grbcomp} to the spectrum No. 7 of GRB\,970111. As it can be seen, this spectrum can be equally well fit with both {\sc bf$+$bb} and {\sc grbcomp}, with the advantage that {\sc grbcomp} can describe all the broad--band time spectra in our sample and is a physical model.

The fact that {\sc grbcomp} fits so well all the spectra,
stimulated us to investigate, in addition to the time evolution of the {\sc grbcomp} model parameters, also 
the search for possible correlations between {\sc grbcomp} parameters and between model parameters and measured flux (or luminosity when the GRB redshift is known).
  
The most relevant results of our investigation are summarized below.

\subsection{Time evolution of the {\sc grbcomp} parameters}

The time evolution of the {\sc grbcomp} best--fit parameters, shown (in red) in Fig.~\ref{f:time_evol},  is very suggestive. 
For all GRBs in our sample, the electron temperature in the observer frame $kT_{e,o}$ shows a time behavior almost similar to that of the corresponding peak energy $E_{p,o}$,  as obtained from the best--fit results  \citep{Frontera12a,Frontera12b} of the same data with {\sc bf}. 
It also immediately appears that $kT_{e,o}$ tracks the flux. 

We also find that the high energy index $\alpha_{boost}$ of the {\sc grbcomp} model shows  a time behavior similar to
that of the power--law photon index $\Gamma$ of the {\sc bb $+$ pl} model (see Table~\ref{t:results}). We warn, however, that this model does not fit most of the time--resolved spectra. However, the absolute values of $\Gamma$ are much lower than $\alpha_{boost}$, rising the energetics issue for the {\sc bb$+$pl} model that we will discuss in Section~\ref{s:discussion}.
 
From Fig.~\ref{f:time_evol} it also appears that the intrinsic seed photon temperature $kT_s$ and the photospheric radius $R_{ph}$ (see Section~\ref{s:discussion} for its definition) are anticorrelated: when $kT_s$  decreases with time, $R_{ph}$ increases. Instead the optical depth $\tau_{eff}$ is almost stable (990510) or slightly decreases with time.

Given that the {\sc grbcomp} model predicts the instantaneous spectral properties of the GRB prompt emission as a function of its main ingredients (outflow properties, soft photon temperature, etc; see Section~\ref{s:intro}), the consistency of our results with the model predictions can be only seen by investigating the correlations between time--resolved measured parameters. The results of this investigation are reported below.

\subsection{$kT_{e,o}$  vs. flux and other model parameters}

We find, as shown Fig.~\ref{f:ep-vs-kte} and Table~\ref{t:ep-vs-kte}, that, for three  GRBs (970111, 990123, 990510), $E_{p,i}$ and $kT_{e,i}$  are positively correlated with each other.  For two of them (990123 and 990510)  their correlation is well described by a power--law [$E_{p,i} = a (kT_{e,i})^m$], with a positive power--law index $m$. The values of $m$ are consistent with a weigthed average of $1.39\pm 0.14$. The correlation is almost absent for the fourth GRB (980329). 
 
In addition we find a positive power--law correlation between $kT_{e,o}$ (in the observer frame) and the 2--2000 keV flux [$kT_{e,o} = a (flux)^m$] for all GRBs, except the case of GRB\,980329, for which we find an almost absent correlation or, at most, with a value
of $kT_{e,o}$ that very slightly increases with flux. 
This is shown in Fig.~\ref{f:kte-vs-flux} and in Table~\ref{t:kte-vs-flux}. This correlation is reminiscent of the power--law correlation between $E_{p,o}$ and flux, that was already found and reported  for the same events \citep{Frontera12a,Frontera12b}, with the index $m$ of the $kT_{e,o}$--flux correlation basically lower  than that of the $E_{p,o}$--flux correlation: $0.65\pm0.09$ against $0.68\pm0.06$ for GRB\,970111, $0.04\pm0.07$ against $0.16\pm0.04$ for GRB\,980329, $0.37\pm0.04$ against $0.53\pm0.05$ for GRB\,990123, and $0.36\pm0.08$ against $0.81\pm 0.15$ for GRB\,990510. If we exclude GRB\,980329, the power-law index $m$ of the other three GRBs is statistically consistent with their weighted mean value $<m> = 0.41 \pm0.06$.         
The  fact that the $kT_{e,o}$ dependence on flux almost 
disappears in the case of GRB\,980329 
%
is discussed in Section~\ref{s:980329}. 


We do not find any correlation between $kT_{e,i}$ and 
the apparent photospheric radius $R_{ph}$ (970111: $\rho = -0.09$, $NHP = 0.82$; $NHP = 0.54$; 990123: $\rho = -0.02$, $NHP = 0.93$; 990510: $\rho = -0.12$, $NHP = 0.78$).

\subsection{Intrinsic peak energy $E_{p,i}$ and electron temperature $kT_{e,i}$ versus other model parameters}

In addition to the above discussed correlation  between $E_{p,i}$ and $kT_{e,i}$, we have performed the search for a possible correlation between $E_{p,i}$ and the optical depth $\tau_{eff}$, with negative results for 970111 ($\rho = -0.09$, $NHP = 0.82$), 990123 ($\rho =-0.33$, $NHP = 0.16$), and 990510 ($\rho = 0.57$, $NHP = 0.14$). Instead, a possible hint of correlation  (at a few percent significance) is found in the case of 980329 ($\rho = 0.82$, $NHP = 0.041$). 

We do not find any correlation between $kT_{e,i}$ and the seed photon temperature $kT_{s,i}$ for three GRBs: 980329
($\rho = -0.20$, $NHP = 0.70$), 990123 ($\rho =0.35$, $NHP = 0.20$), 990510 ($\rho = 0.02$, $NHP = 0.95$), while, in the case of 970111, we find a low significance correlation ($\rho = 0.78$, $NHP = 0.036$),  

Instead, we find a strong negative correlation between peak energy $E_{p,i}$ and the bulk parameter $\delta$, except in the case of the peculiar GRB\,980329, where this correlation is not statistically significant. This result is shown in Fig.~\ref{f:ep-vs-delta} and in Table~\ref{t:ep-vs-delta}.
In the light of the {\sc grbcomp} model, it is a consequence of the correlation between $E_{p,i}$ and $kT_{e,i}$, given 
that $\delta$ is inversely proportional to $kT_{e,i}$ (see 
eq~\ref{e:delta}).
But, independently of that, a negative correlation is also expected because a higher $\delta$ leads to softer spectra and thus to lower peak energies $E_{p,i}$.  In fact, the Comptonization spectra become softer  if they take into account  the bulk outflow effect, while they are getting harder in the case of a bulk inflow \citep{Laming04}. 

\subsection{Seed photon temperature $kT_s$ versus flux and photospheric radius}

The behavior of the seed photon temperature $kT_{s,o}$ with the 2--2000 keV flux is shown Fig.~\ref{f:kts-vs-flux}. As it can be seen from this figure and from the fit results reported in Table~\ref{t:kts-vs-flux}, $kT_s$ increases with flux not according to a power--law, except in the case of GRB\,980329, in which this law gives the best description of the data, with an index $m = 0.37\pm 0.12$.
  		
%
An outstanding anti-correlation between seed photon temperature $kT_s$ and the photospheric radius $R_{ph}$ is found (see Fig.~\ref{f:kts-vs-rph}), which is well described by a power--law, with  best--fit parameters reported in  Table~\ref{t:kts-vs-rph}. 
By summing together the data points of GRBs with known redshift (980329, 990123 and 990510), we find the very robust result ($\rho = -0.88$, $NHP = 1.7\times 10^{-10}$)  also shown in Table~\ref{t:kts-vs-rph}
and  Fig.~\ref{f:ave-kts-vs-rph}, with a weighted mean power--law index of $-0.92\pm 0.17$.

\section{Discussion}
\label{s:discussion}

By merging together the WFC and BATSE prompt emission spectra  of 20 GRBs  simultaneously observed  with 
both instruments, it has been possible to perform a 
time--resolved analysis in the broad energy band
from 2 keV to 2 MeV, still scarcely explored.
For each GRB time profile, we have obtained a number of 
time--resolved spectra. The highest number 
of time--resolved spectra has been obtained for the brightest GRBs 
in our sample: 970111 (8 spectra), 980329 (6), 990123 (19), and 990510 (8), with a total number of 41 spectra. 

With the brightest GRBs in our sample,
we tested three physical models: the {\sc bb}$+${\sc pl} model proposed by \citet{Ryde09}, 
in which {\sc bb} is the blackbody radiation emitted at the photospheric radius and {\sc pl}  could describe
a non--thermal component (like synchrotron or Comptonization of blackbody photons), the sum of a {\sc bb} plus the {\sc bf} function, and the
{\sc grbcomp} model proposed by T12. 

As discussed in Section~\ref{s:intro}, \citet{Ryde09} found that the {\sc bb$+$pl} model is consistent with the 
time--resolved spectra of about
50 GRBs detected with BATSE in the energy band from $\sim$25 to 1900 keV. This result was very appealing, even if the derived
values  of the power--law photon  index $\Gamma$, generally $<2$ (see Table~\ref{t:results}), would require an energy break beyond 2 MeV to avoid a divergence of the total emitted power. Such breaks generally are not observed in the GRBs detected with the Large Area Telescope (LAT)  aboard the \fermi\ satellite \citep[see, e.g.,][]{Abdo09a,Abdo09b,Abdo09c}, but they could be present in a significant fraction of GRBs not detected with LAT \citep{Ackermann12}. 

We find however that, when extending the energy band of the  time--resolved spectra down to 2 keV, the {\sc bb $+$ pl} model does not fit our spectra (see examples in Fig.~\ref{f:bbpl_fit}), except for a very few cases, 
near the GRB onset (see Table~\ref{t:results}).  

As shown in a previous paper \citep{Frontera12a,Frontera12b}, the {\sc bf} function gives a good description of most 
time--resolved spectra of the GRBs in our sample. In the few cases (6 of 41 spectra) in which this is not the case, we have added to {\sc bf} a {\sc bb} that resulted to describe well the prompt emission spectra of a few \fermi\ GRBs \citep{Guiriec11,Tierney13,Ghirlanda13}. In our case, we find that this addition  gives a significant improvement of the fit only in one case (interval No. 7 of GRB970111, see Table~\ref{t:bb-plus-bf} and Fig.~\ref{f:970111-7}). 

Instead, we find that the {\sc grbcomp} model fits very well almost all the time--resolved spectra, in some cases,
even better than the empirical {\sc bf} (see bottom panels of Fig.~\ref{f:time_evol} and Fig.~\ref{f:NHP}).
Given this result, we have further investigated this model also to understand if the best--fit parameter values derived, 
their time behavior and their correlation with other parameters of the same model or {\sc bf}, are physically acceptable.

According to {\sc grbcomp}, the intrinsic peak energy of the 
$E F(E)$ spectrum has a power--law dependence on the intrinsic outflow electron temperature according to the following equation
\begin{equation}
E_{p,i} = a (kT_{e,i})^b
\label{e:epi-vs-ktei} 
\end{equation}
where the parameters $a= a(t, \tau$) and $b = b(t, \tau)$ are both dependent on time and optical thickness $\tau$ at which the observed spectrum is coming out. 
In Figure~\ref{f:pred-ep-vs-kte} we show the model prediction and the dependence of the power--law parameters on the optical depth of the electron cloud. The best fit to the parameters data points derived from the numerical code is obtained by the two following empirical functions:

\begin{equation}
 a(\tau)=k_0-k_1~k_2~{\rm log}\Bigg\{{\rm exp}\Big[\Big(1-\Big(\frac{\tau}{k_3}\Big)^{k_4}\Big)/k_2\Big] +1\Bigg\}.
\label{e:a(tau)}
\end{equation}
where $k_0=4.15 \pm 0.05$, $k_1=1$ (frozen), $k_2 = 6.39 \pm 0.23$, $k_3=2.26\pm 0.25 $ and $k_4=1.65 \pm 0.10$,

and

\begin{equation}
b(\tau)=k_5-k_6~k_7~{\rm log}\Bigg\{{\rm exp}\Big[\Big(1-\frac{1}{(k_8~\tau)^{k_9}}\Big)/k_7\Big] +1\Bigg\}
\label{e:b(tau)}
\end{equation}
where $k_5=10$ (frozen), $k_6= 1$ (frozen), $k_7=12.289 \pm 0.003$, $k_8=0.266 \pm 0.003$, and $k_9=2.00 \pm 0.05$.

It is worth noting the asymptotic forms of these two functions for $\tau >> 1$, namely $a \sim 4$ and
$b \sim 1$, which lead to the well known relation,  in $E F(E)$ units, $E_p \sim 4 kT_e$, typical
of fully saturated Comptonization.

We find, as shown in Fig.~\ref{f:ep-vs-kte} and in Table~\ref{t:ep-vs-kte}, that  $E_{p,i}$ and $kT_{e,i}$  are indeed positively correlated with each other according to a power--law for three of the 4 GRBs (970111, 990123, and 990510), while this correlation is absent for GRB\,980329. This seemingly strange behavior of 980329 is discussed below (see Section~\ref{s:980329}). The measured power-law index $m$ ($\equiv b$) for 970111, 990123, 990510  (weighted mean value $<m> = 1.39\pm 0.14$) is  greater than 1, and, from the best--fit values found for $\tau_{eff}$ (see Table~\ref{t:results}), it is consistent, within its statistical uncertainties, with that expected.

In agreement with the {\sc grbcomp} model expectations, we find that the electron temperature $kT_{e,i}$ of the outflowing plasma is 
positively correlated with flux, except in the case of the peculiar event 980329. In this case, we find an almost flat dependence of $kT_{e,o}$ on flux, likely as a consequence of the behavior of $E_{p,i}$ versus $kT_{e,i}$ for this event (see Section~\ref{s:980329}). 

Consistently with the {\sc grbcomp} model predictions, we also find that the electron temperature of the outflowing plasma 
is not correlated with the seed temperature of the {\sc bb} photon bath.

It is noteworthy that the {\sc grbcomp} model gives a physical interpretation of the low-energy power--law photon index with its sign $\alpha_{bf}$  of  {\sc bf} \citep{Band93}. 
According to  {\sc grbcomp}, the corresponding energy index $-\alpha_{\rm bf} - 1$ is related to the index (slope) of the Comptonization Green function and thus to the bulk parameter $\delta$ (see T12):
\begin{equation}
- \alpha_{\rm bf} - 1 \approx -\frac{3-\delta}{2} + \left[\left(\frac{\delta + 3}{2}\right)^2+\gamma\right]^{1/2}
\label{index}
\end{equation}
where  $\gamma$ depends on the Comptonization level. In the case of saturated Comptonization, $\gamma \ll 1$, and $-\alpha_{{\rm bf}} - 1 \approx \delta$ . 
Thus an almost linear correlation between $\alpha_{{\rm bf}}$ and $\delta$ is predicted. 
Our results confirm the model prediction for  GRBs 970111, 980329, and 990123, while for GRB\,990510 the correlation is not significant (see Table~\ref{t:alpha-vs-delta} and Fig.~\ref{f:alpha-vs-delta}).  

In agreement with the model expectations, we find  a positive correlation between radiation flux and intrinsic seed photon temperature $T_s$ (see Fig.~\ref{f:kts-vs-flux}). This correlation is not well described by a power--law, except one case (980329) in which a power--law ($kT_s = a (flux)^m$) gives a good fit, with an index $m$ consistent with 0.25. This index is expected when all the parameters which determine the GRB luminosity [see eq.~(\ref{e:lum})], but the BB temperature $T_s$, are fixed during the event (see Sect.~\ref{s:980329}).  In general, the luminosity and thus the flux, for a given $\beta$ (see Fig.~\ref{f:spectrum-vs-beta}),  depend on $T_s$, on the BB emitting surface $A_s$ and on the Comptonization enhancement factor $\eta_{Comp}$, which is proportional to the electron plasma temperature (T12). Indeed, for 3 of the 4 events in our sample, assuming a power--law (but the fit is not good),  the index $m$ is found to be consistent with 0.5.
%
%
   
Another important result of our test  is shown in Fig. \ref{f:kts-vs-rph}: the seed photon temperature $kT_s$ anticorrelates, following a power--law, with the photospheric 
radius $R_{ph}$ ($kT_s \propto (R_{ph})^{-n}$), with an average value of the power--law index $<n> = 0.92\pm 0.17$ for the three GRBs with known redshift (see Table~\ref{t:kts-vs-rph}).  
%
%
In the scenario of the {\sc grbcomp} model, this
negative correlation, found to occur during the entire GRB prompt phase, is a consequence of the dependence of the BB luminosity $L_s$ on $T_s$ and on the BB emitting area $A_s$, which is proportional to $R_{ph}^2$. Thus, if $L_s$ would be constant, then $kT_s \propto R_{ph}^{-0.5}$. But, this a special case. The fact that average power--law index is about 0.9 is a strong hint that $L_{s}$ is not constant during the prompt emission. In any case, what is important is the fact that, when $A_s$ increases, the seed photon temperature drops.   The photospheric radius is found to increase 
from $10^{13}$ to $10^{15}$ cm for all GRBs of our sample, but GRB\,980329 (see Sect.~\ref{s:980329}). This range of values is consistent with other model predictions \citep[e.g.,][]{Zhang09}. 

The seed photon temperature we found could appear not consistent with an origin from the star photosphere, dominated by optical-UV photons. In fact, when the star explodes, its parts next to the outflow should be hotter than the effective temperature of the star atmosphere because the subrelativistic outflow heats them.
More precisely, as the outflow passes throughout the star atmosphere while moving outwards, 
together with  kinetic energy dragging, the development of
turbulent mixing dictated by  Kelvin-Helmholtz instabilities is expected.
Consequently, the outer parts of the hot outflow, while interacting with the much cooler star photosphere, unavoidably are expected to cool down, leading to the formation of a seed photon population with characteristic energy in the range between that of the optical-UV star photons and of the outflow  electrons
at hundreds keV. 
This photon field presumably illuminates the outflow either from backward or,
at most, at right angles (depending on the distance from the star where
spectral formation mostly occurs) and is different from the isotropic
optical-UV photon field assumed by \citet{Lazzati00} to be present since before the GRB event.

\subsection{GRB\,980329: does this event confirm the {\sc grbcomp} predictions?}
\label{s:980329}

\begin{figure}
\begin{center}
  \includegraphics[width=.4\textwidth]{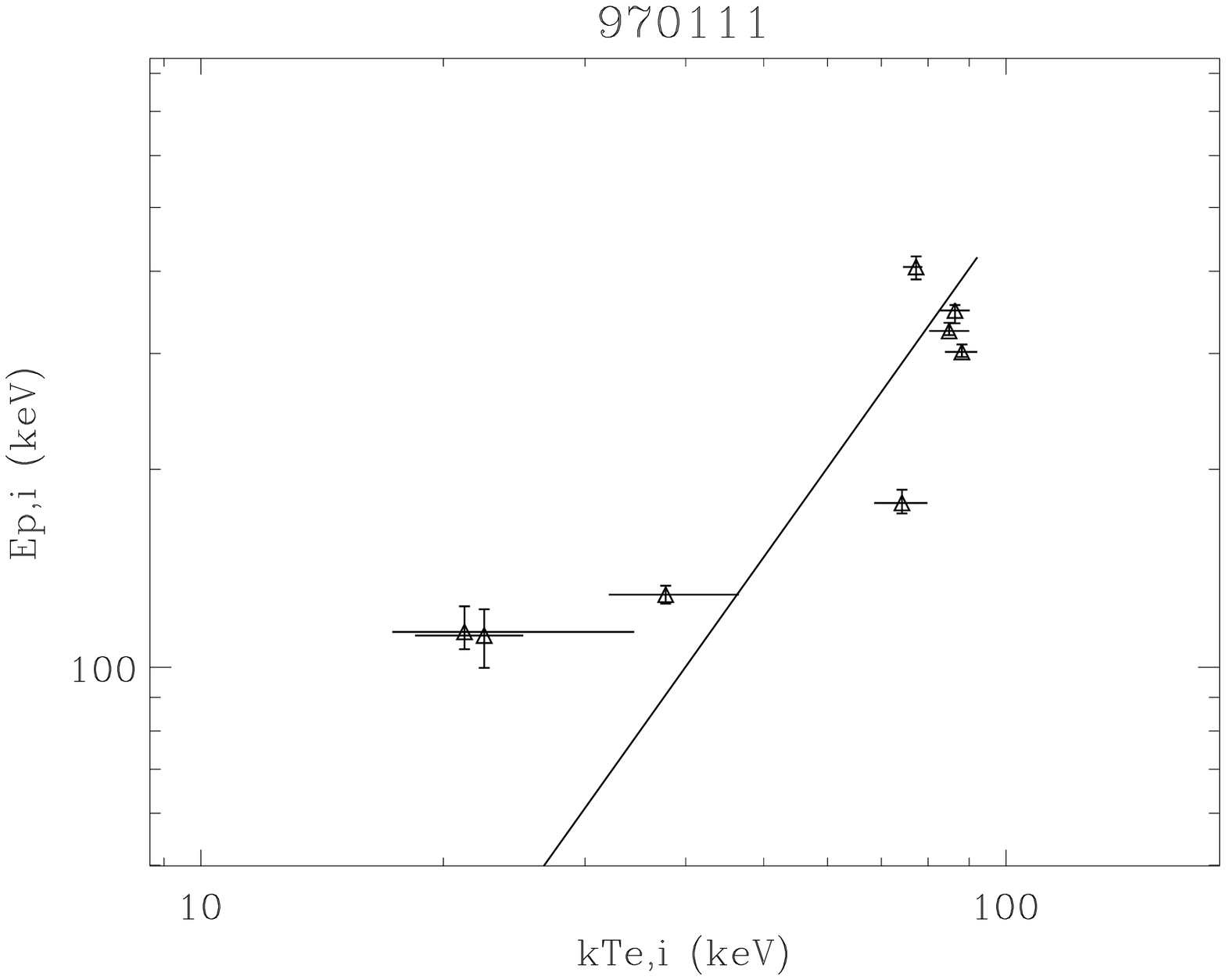}
  \includegraphics[width=.4\textwidth]{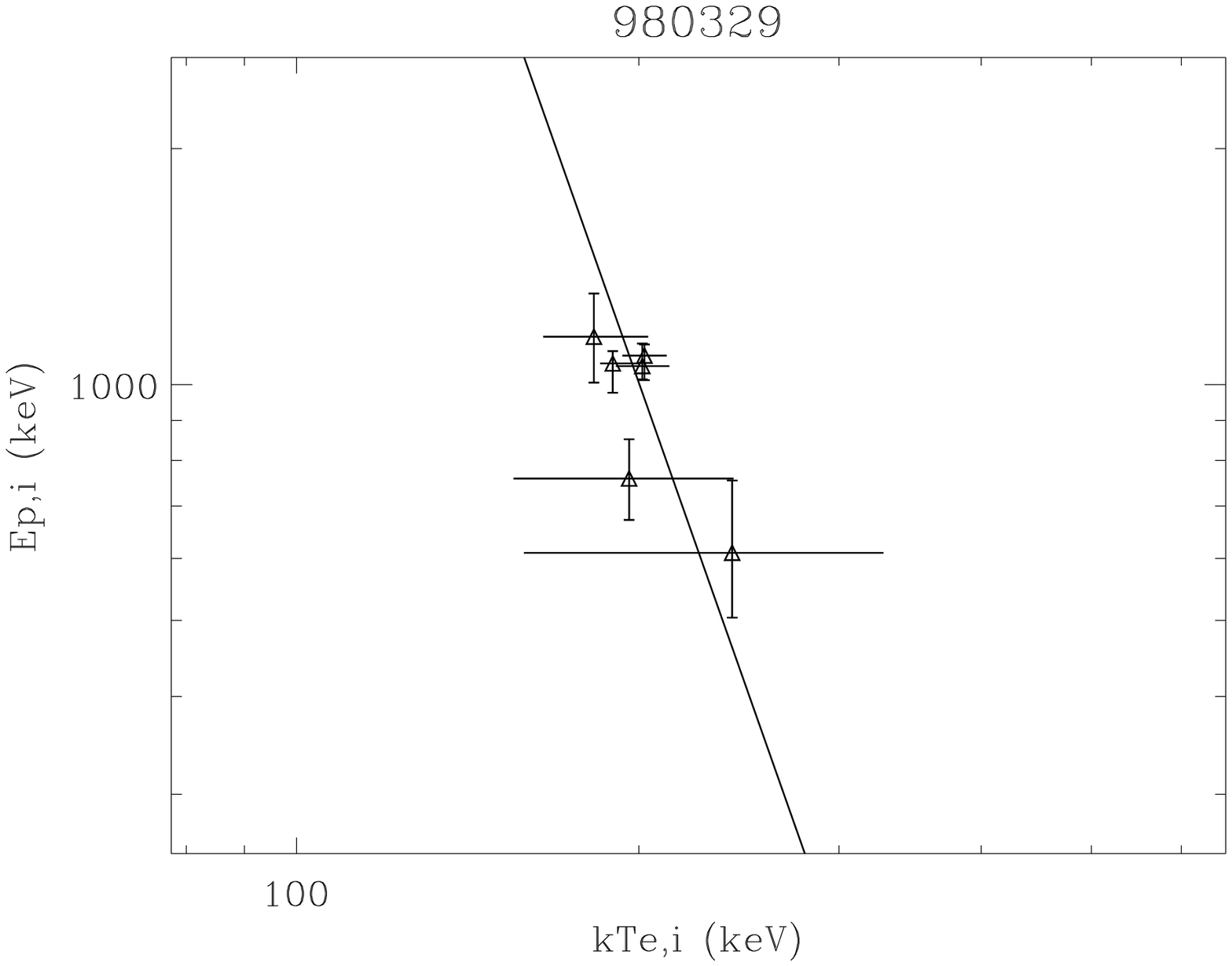}
  \includegraphics[width=.4\textwidth]{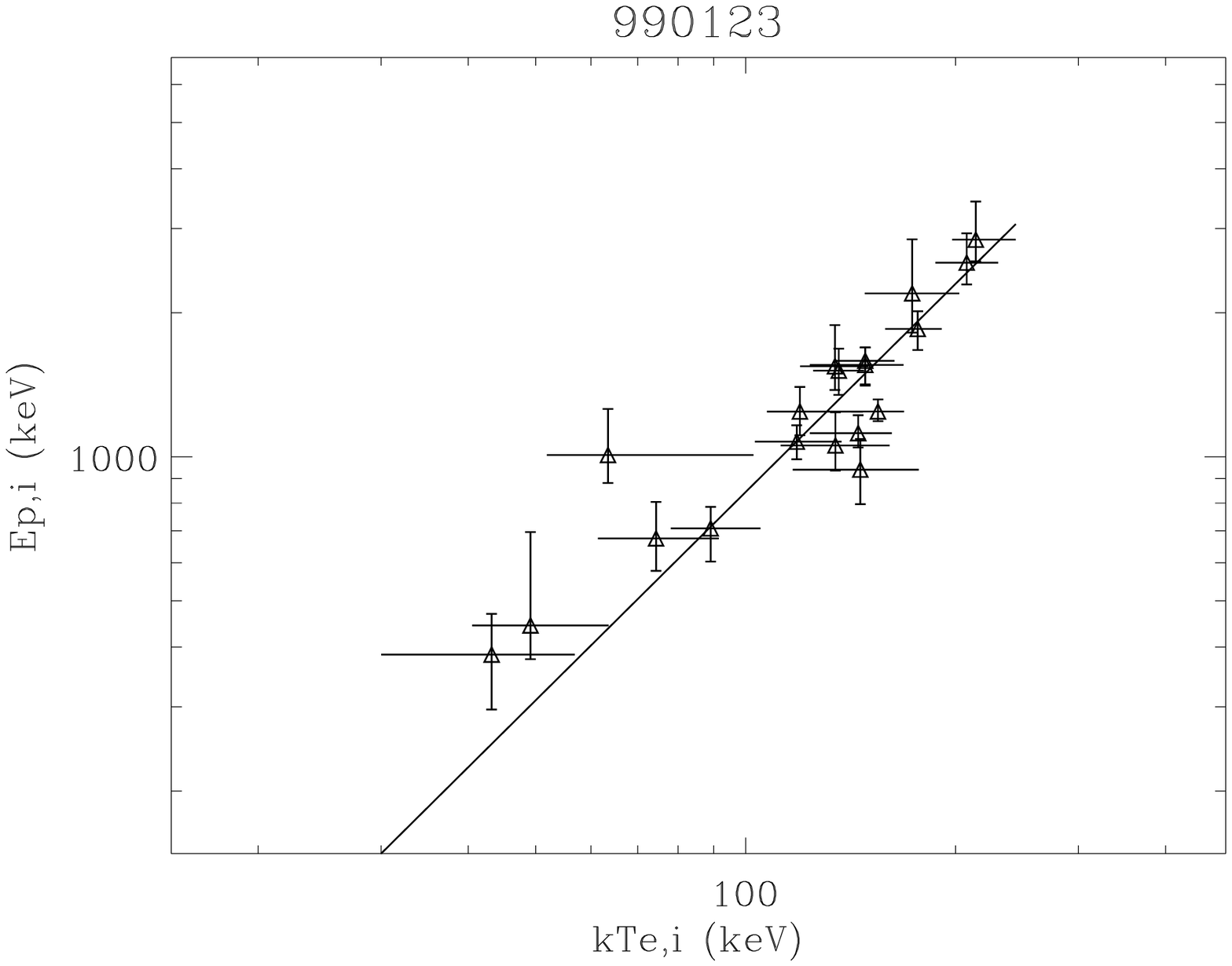}
  \includegraphics[width=.4\textwidth]{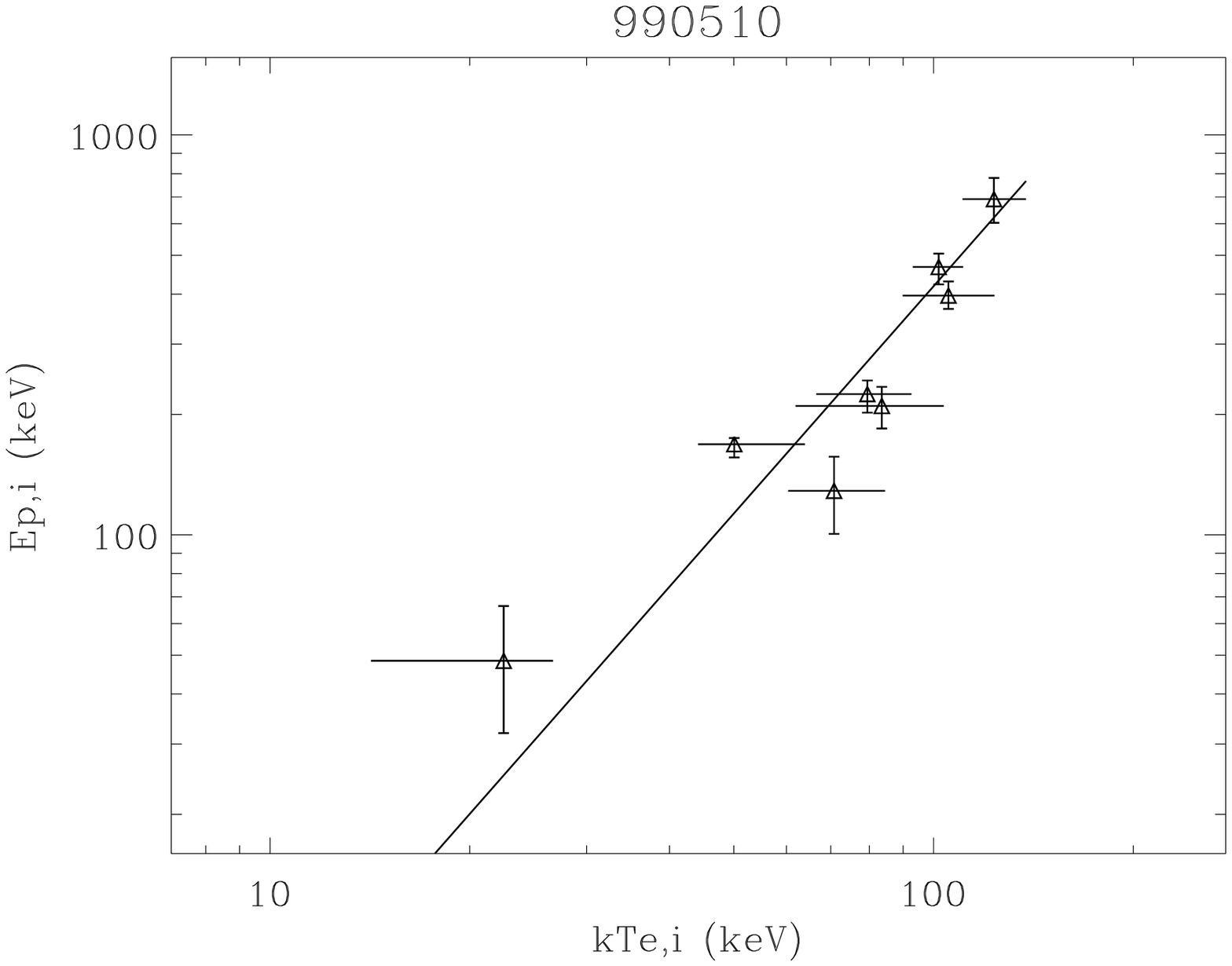}
\end{center} 
\caption{Correlation between $kT_{e,i}$ and $E_{p,i}$ for each of the GRBs 970111, 980329, 990123, and 990510.
The best--fit power--law curve is also shown.}
\label{f:ep-vs-kte}
\end{figure}

\begin{figure}
\begin{center}
  \includegraphics[width=.4\textwidth]{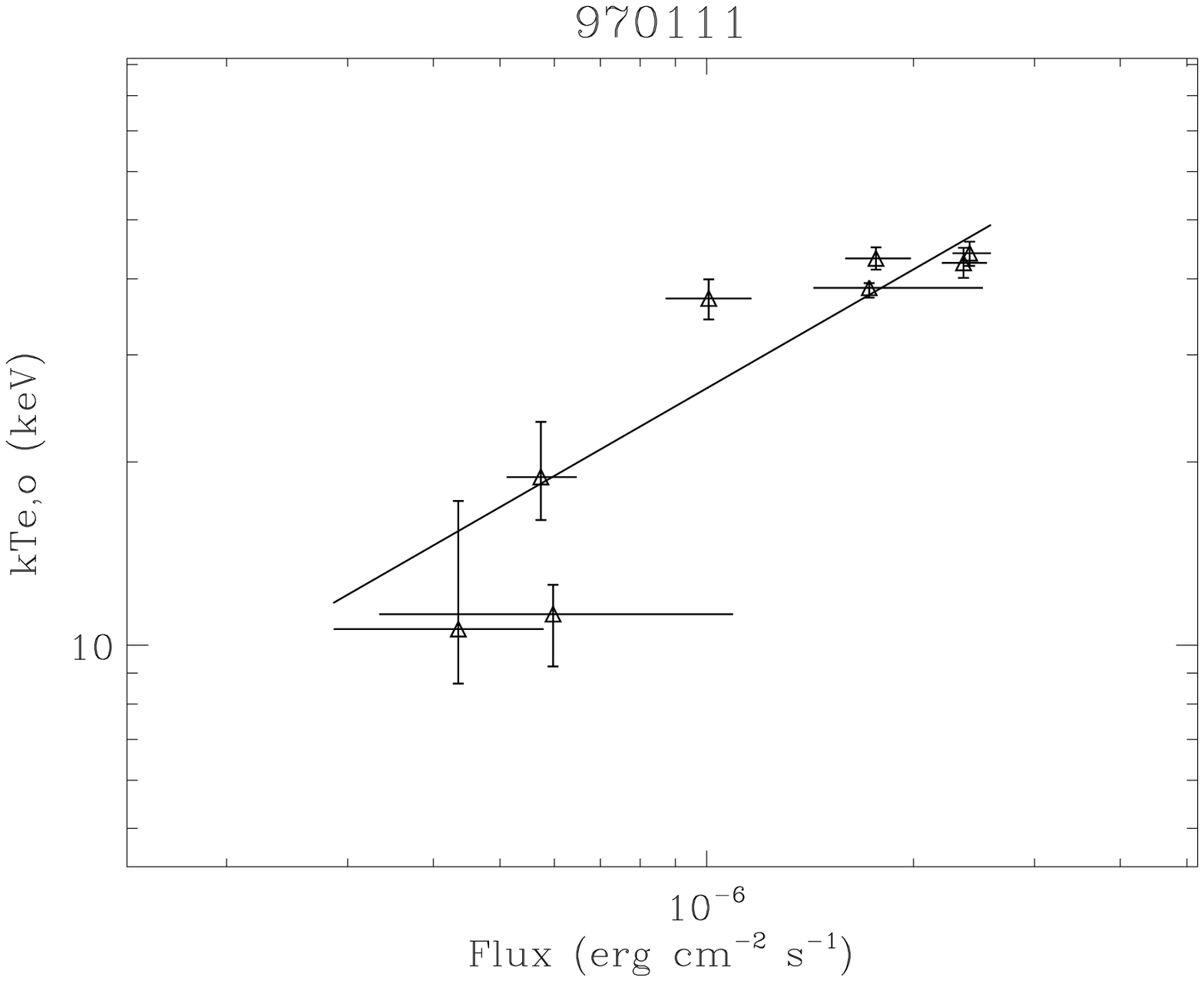}
  \includegraphics[width=.4\textwidth]{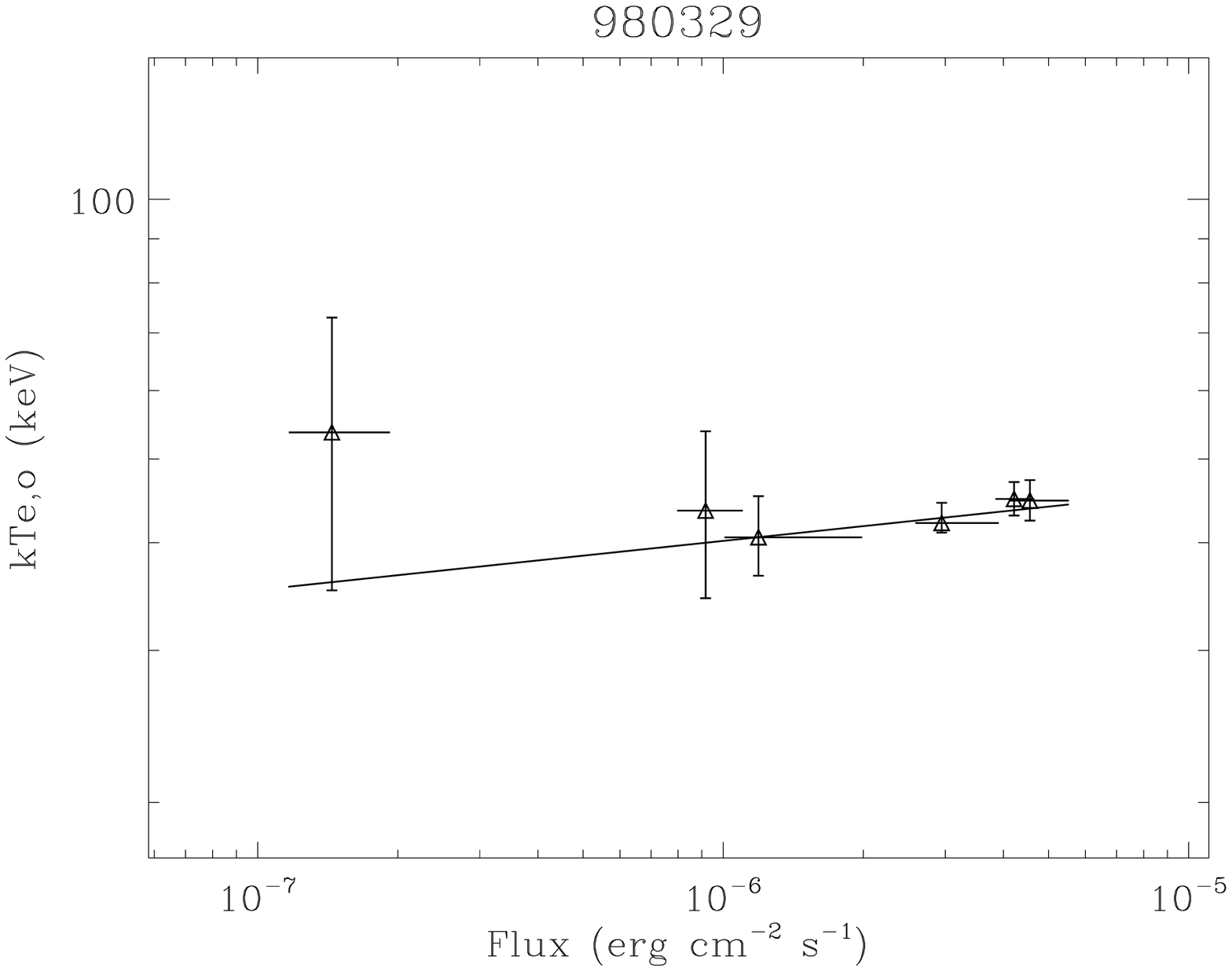}
  \includegraphics[width=.4\textwidth]{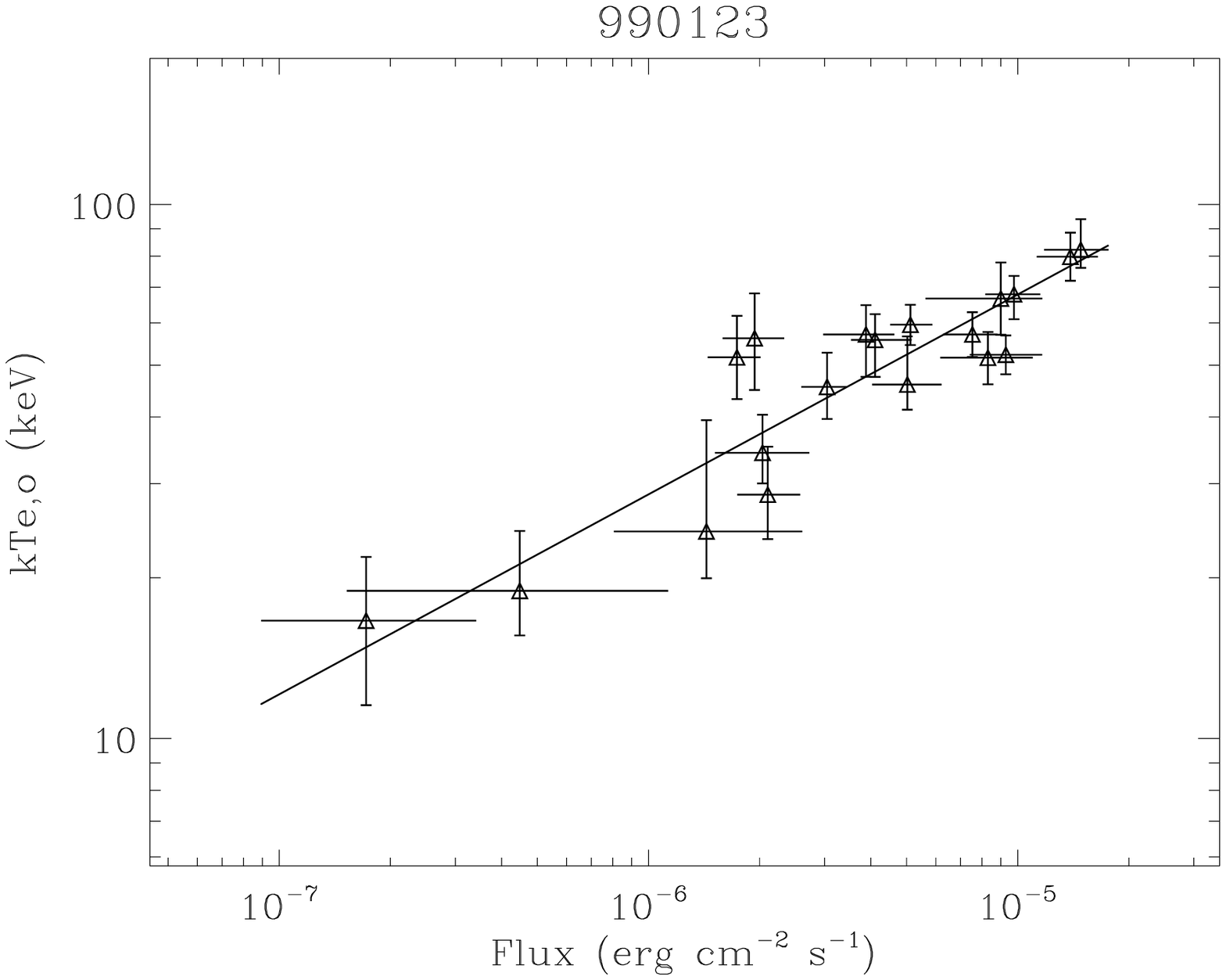}
  \includegraphics[width=.4\textwidth]{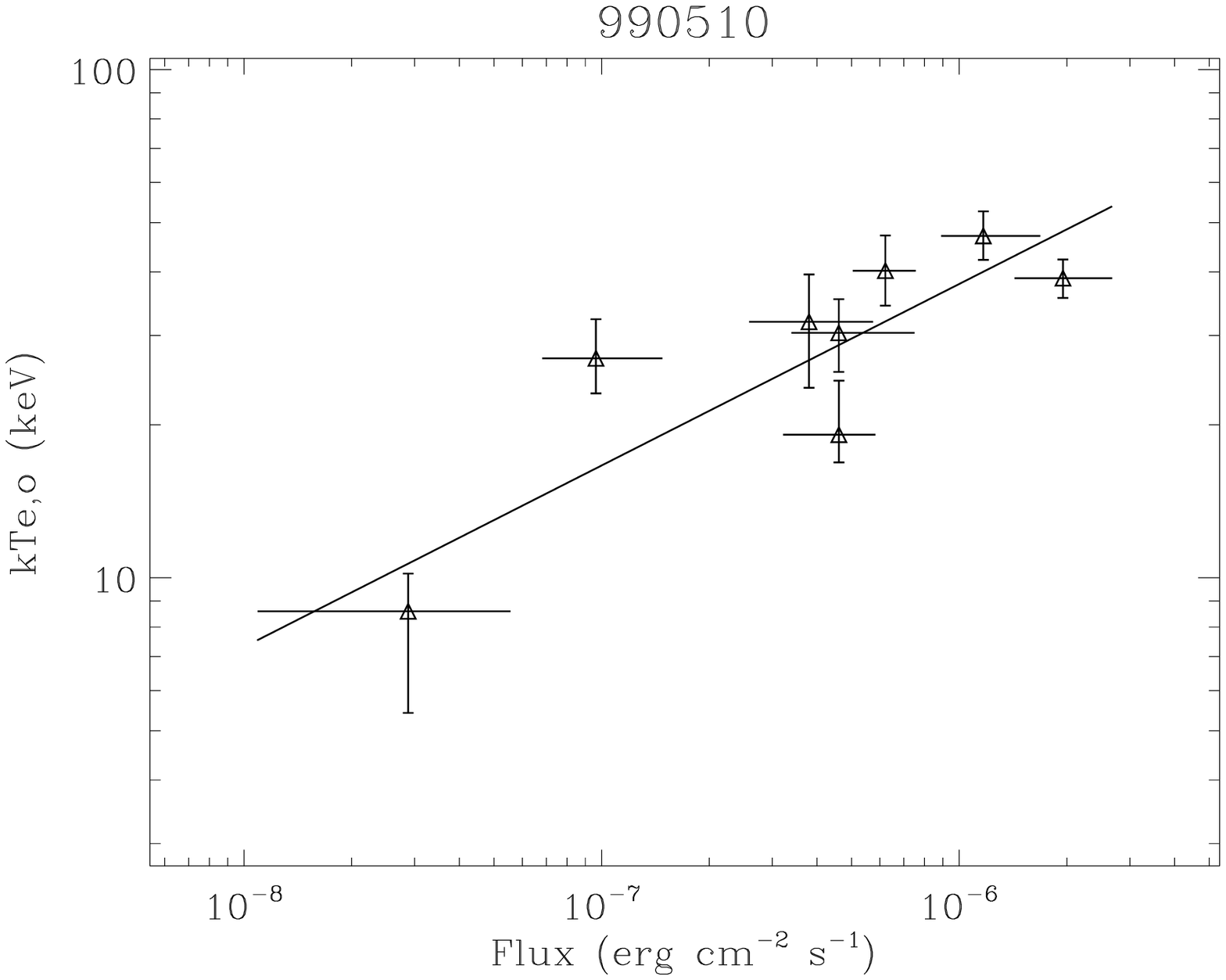}
\end{center}
  \caption{Dependence of the time--resolved electron temperature $kT_{e,o}$, obtained from the best fit 
of the {\sc grbcomp} to the joint WFC$+$BATSE spectra, on the 2--2000 keV flux measured in the corresponding interval.
The best--fit power--law curve is also shown.} 
\label{f:kte-vs-flux}
\end{figure}

\begin{figure}
\begin{center}
  \includegraphics[width=.4\textwidth]{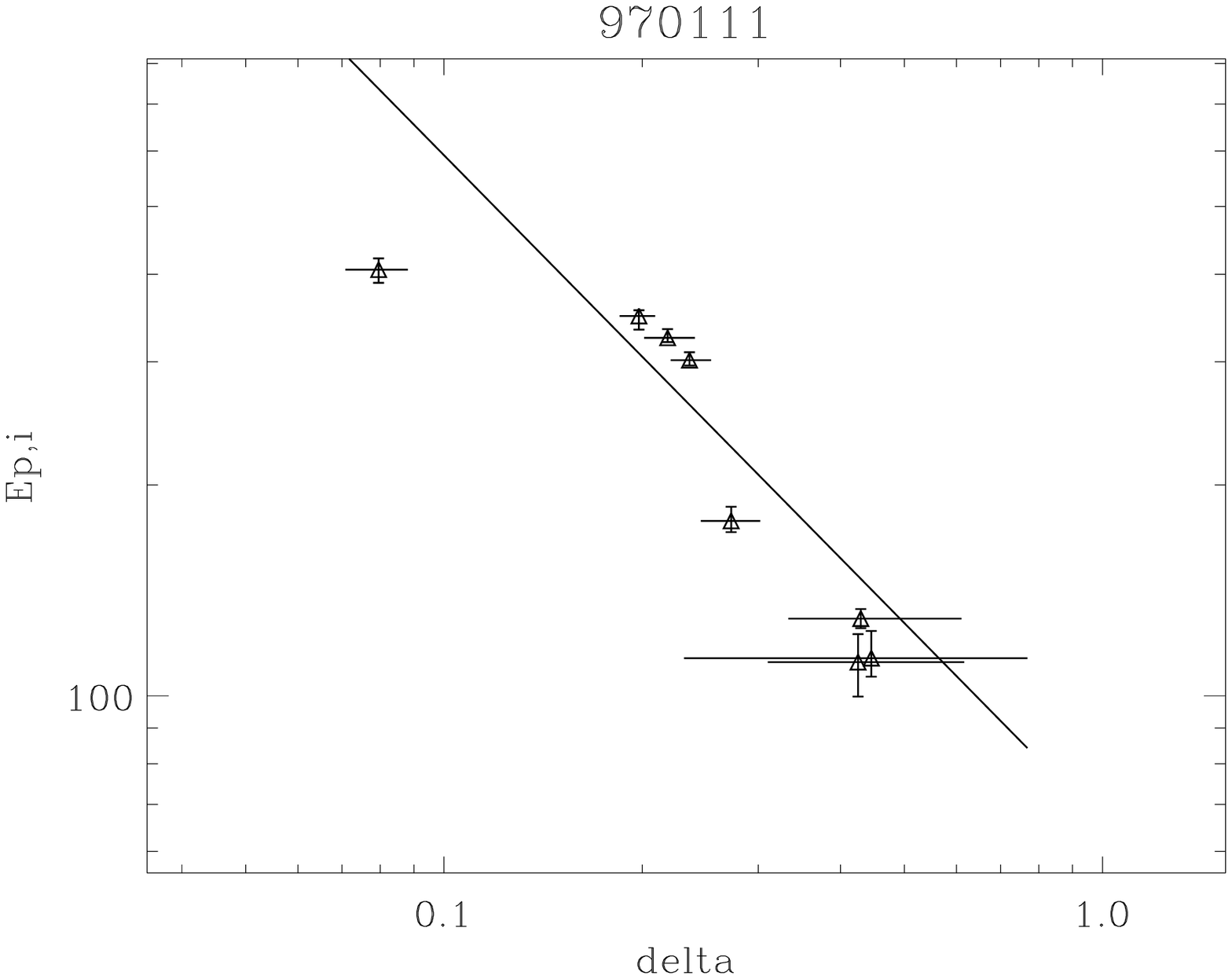}
  \includegraphics[width=.4\textwidth]{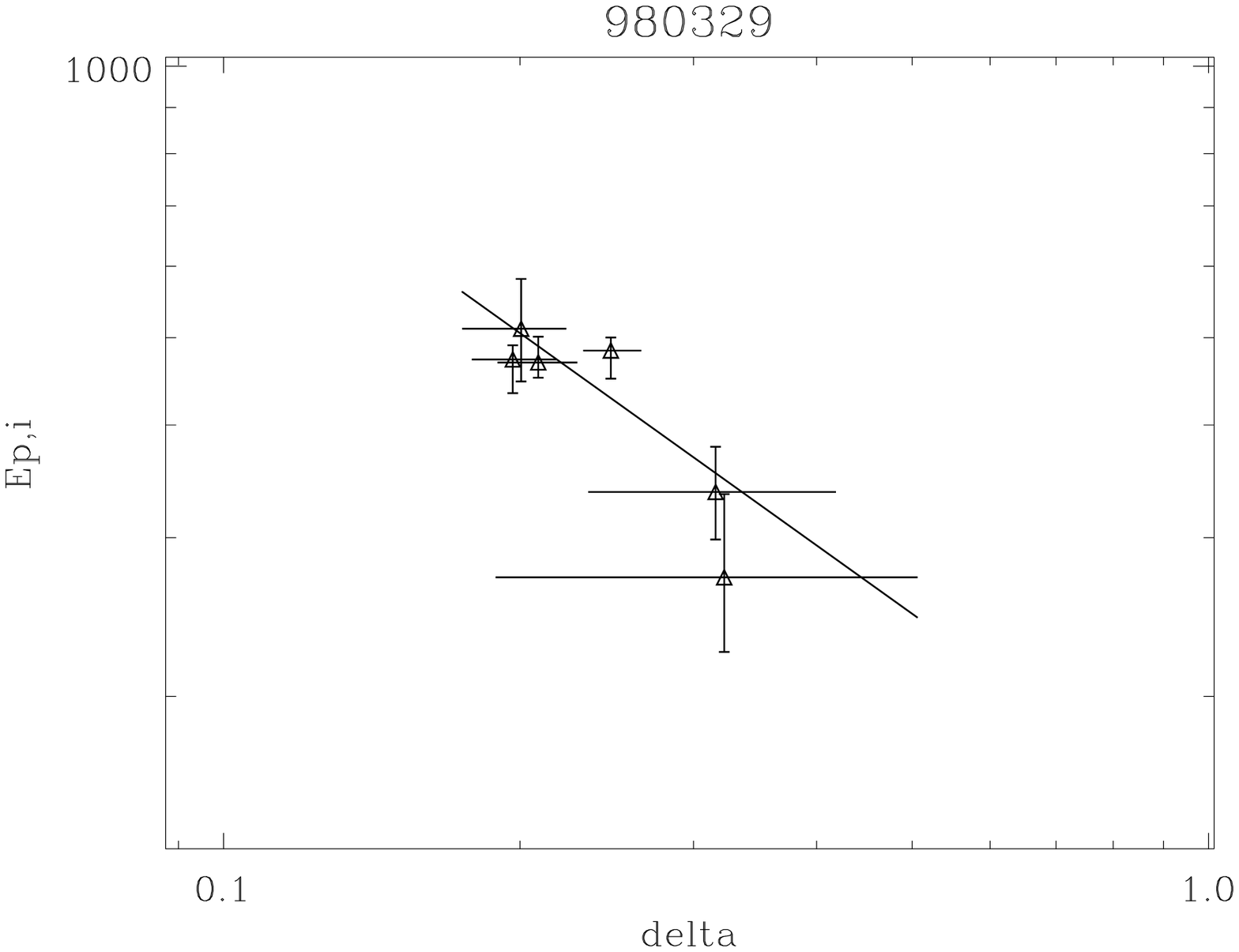}
  \includegraphics[width=.4\textwidth]{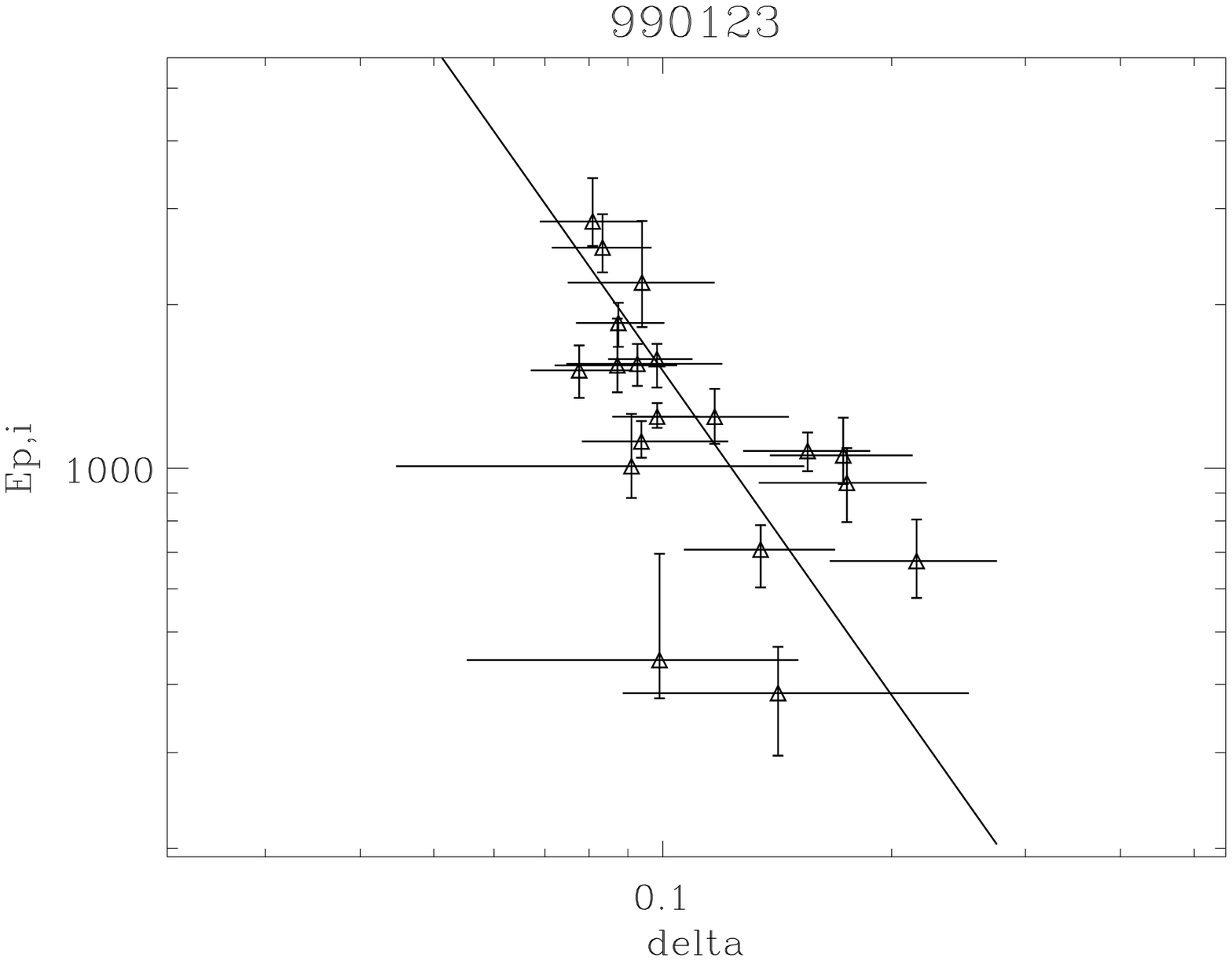}
  \includegraphics[width=.4\textwidth]{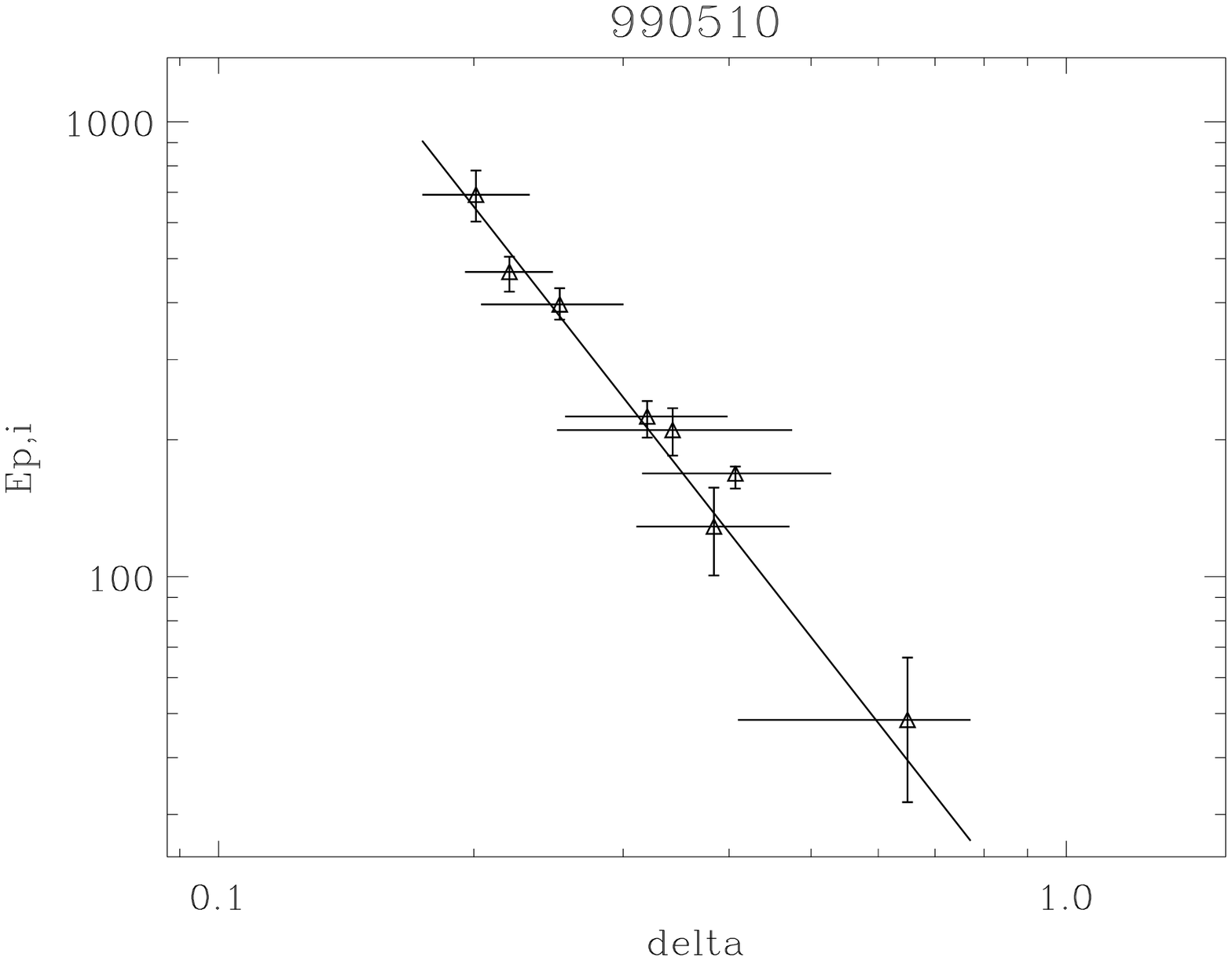}
\end{center}
 \caption{Correlation between the peak energy derived from the {\sc bf} fit to the time--resolved spectra
 and the bulk parameter $\delta$ of the {\sc grbcomp} model, for GRBs 970111, 980329, 990123, and 990510. The best--fit power--law curve is also shown.}
\label{f:ep-vs-delta}
\end{figure}

\begin{figure}
\begin{center}
  \includegraphics[width=.4\textwidth]{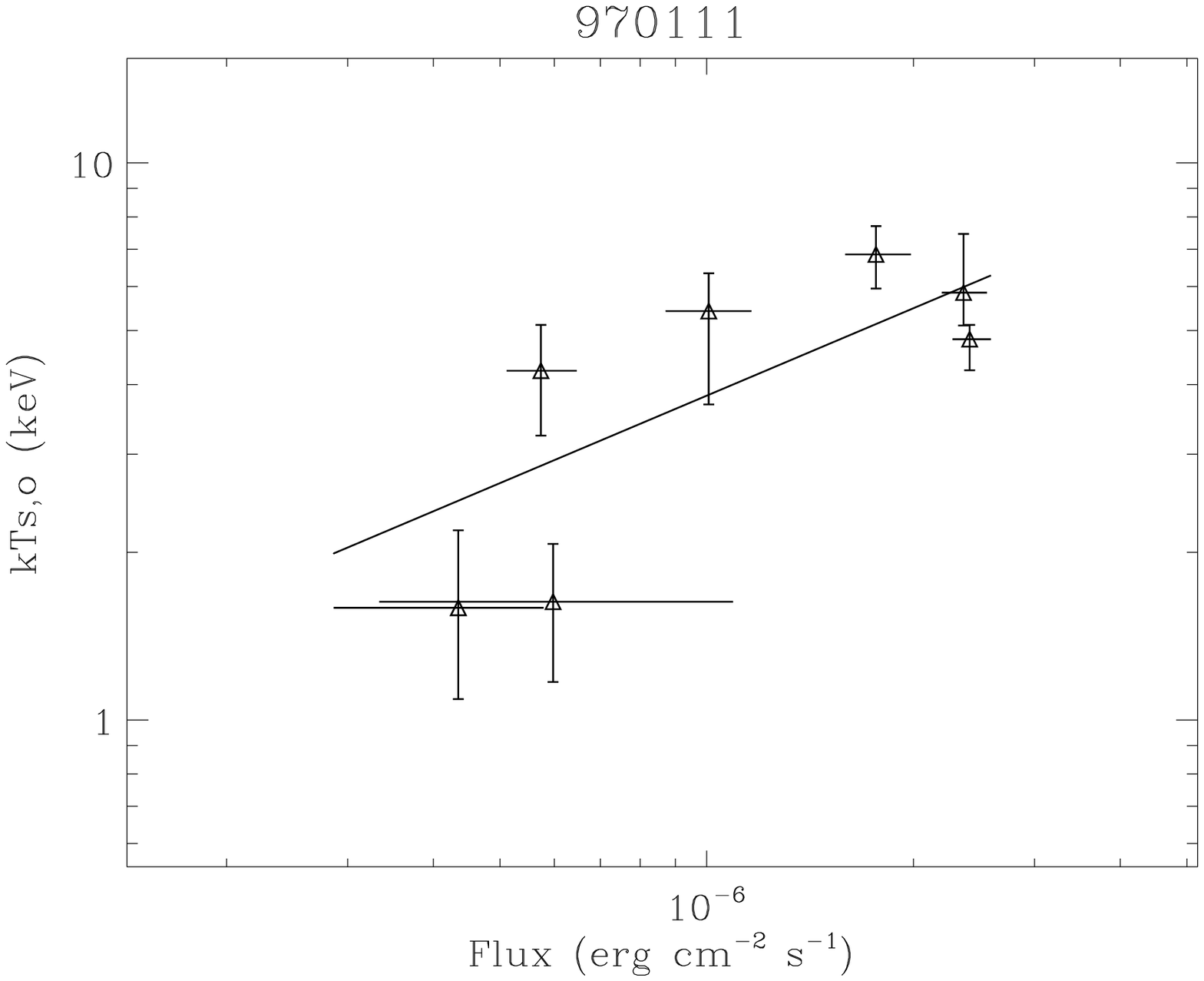}
  \includegraphics[width=.4\textwidth]{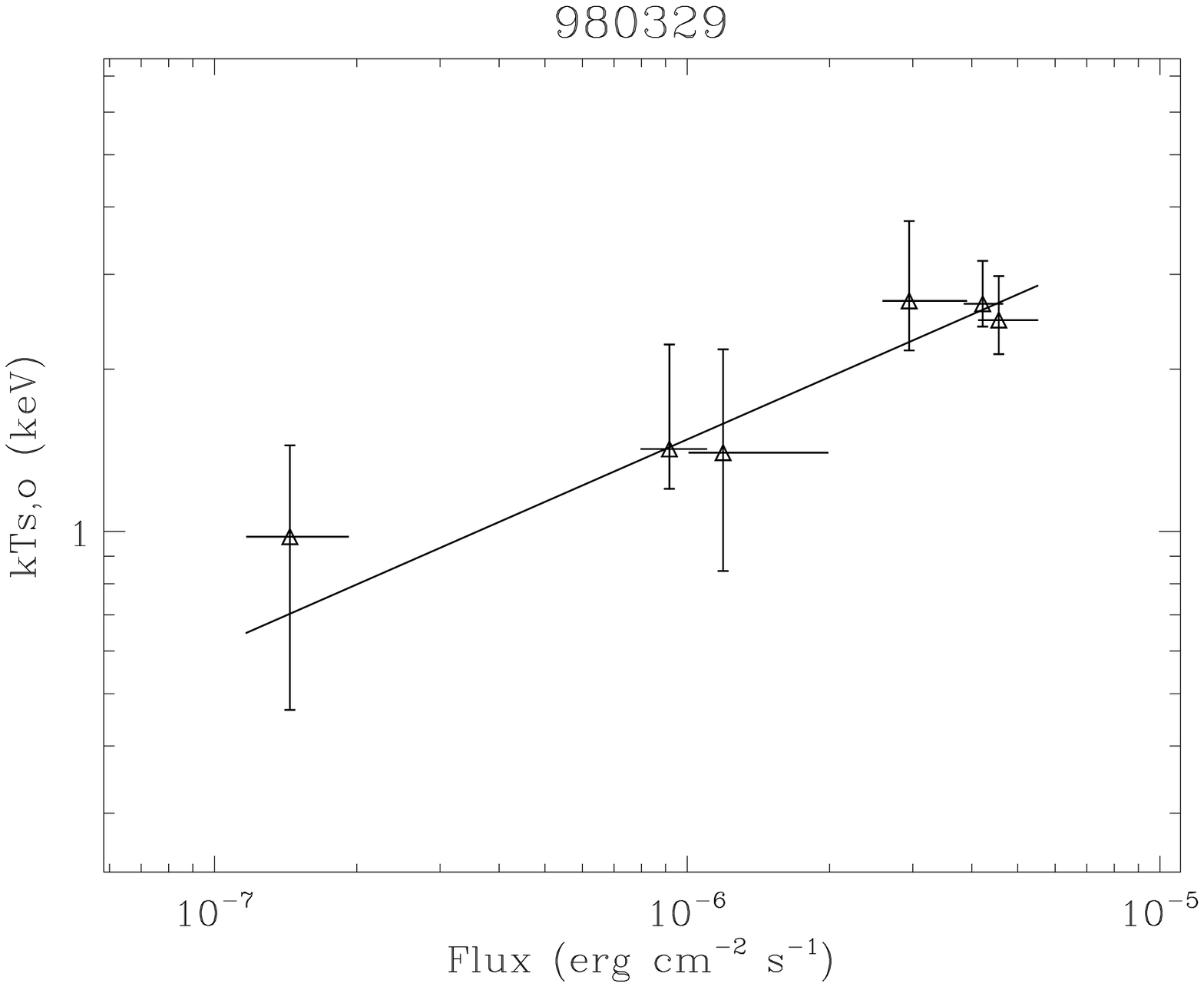}
  \includegraphics[width=.4\textwidth]{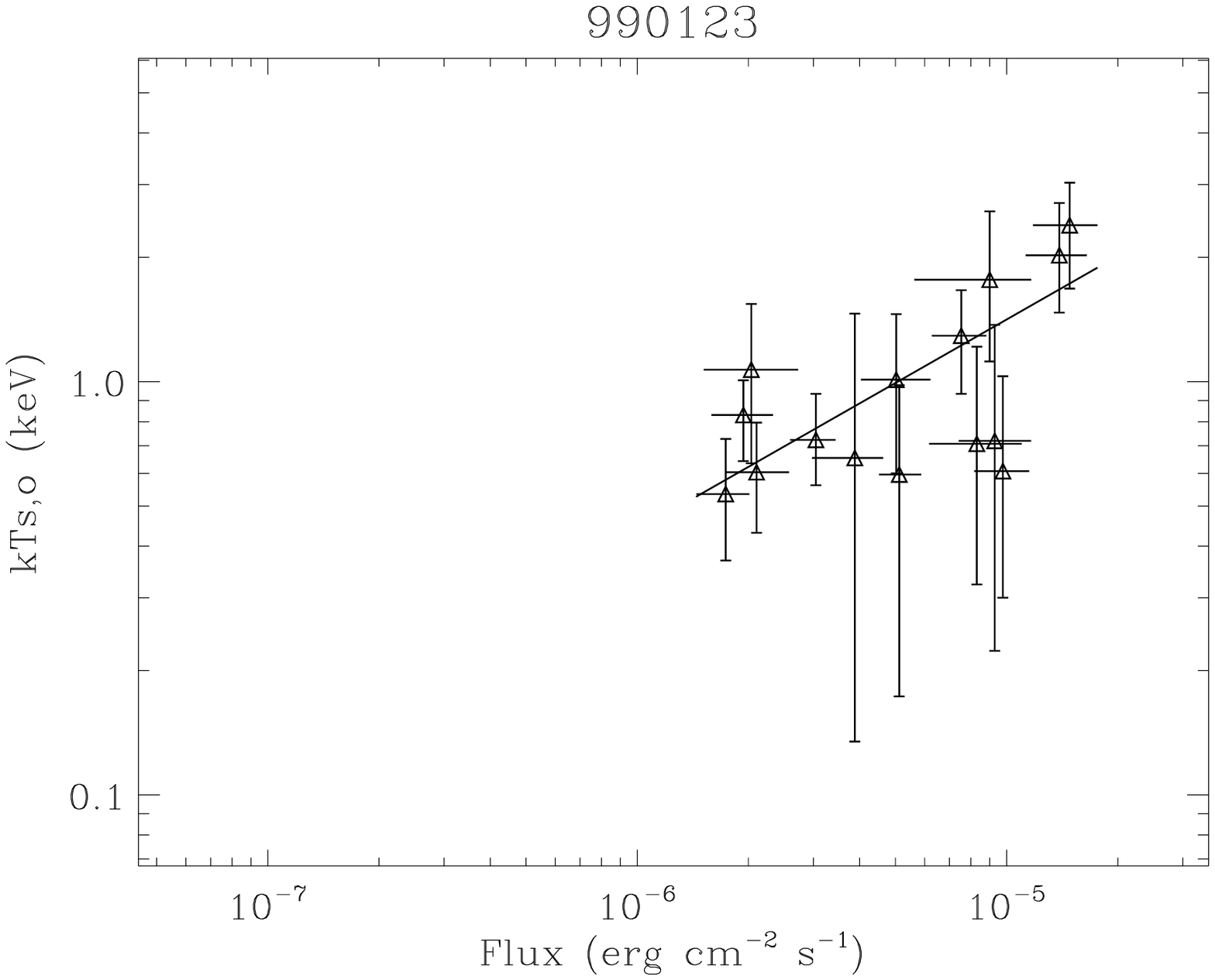}
  \includegraphics[width=.4\textwidth]{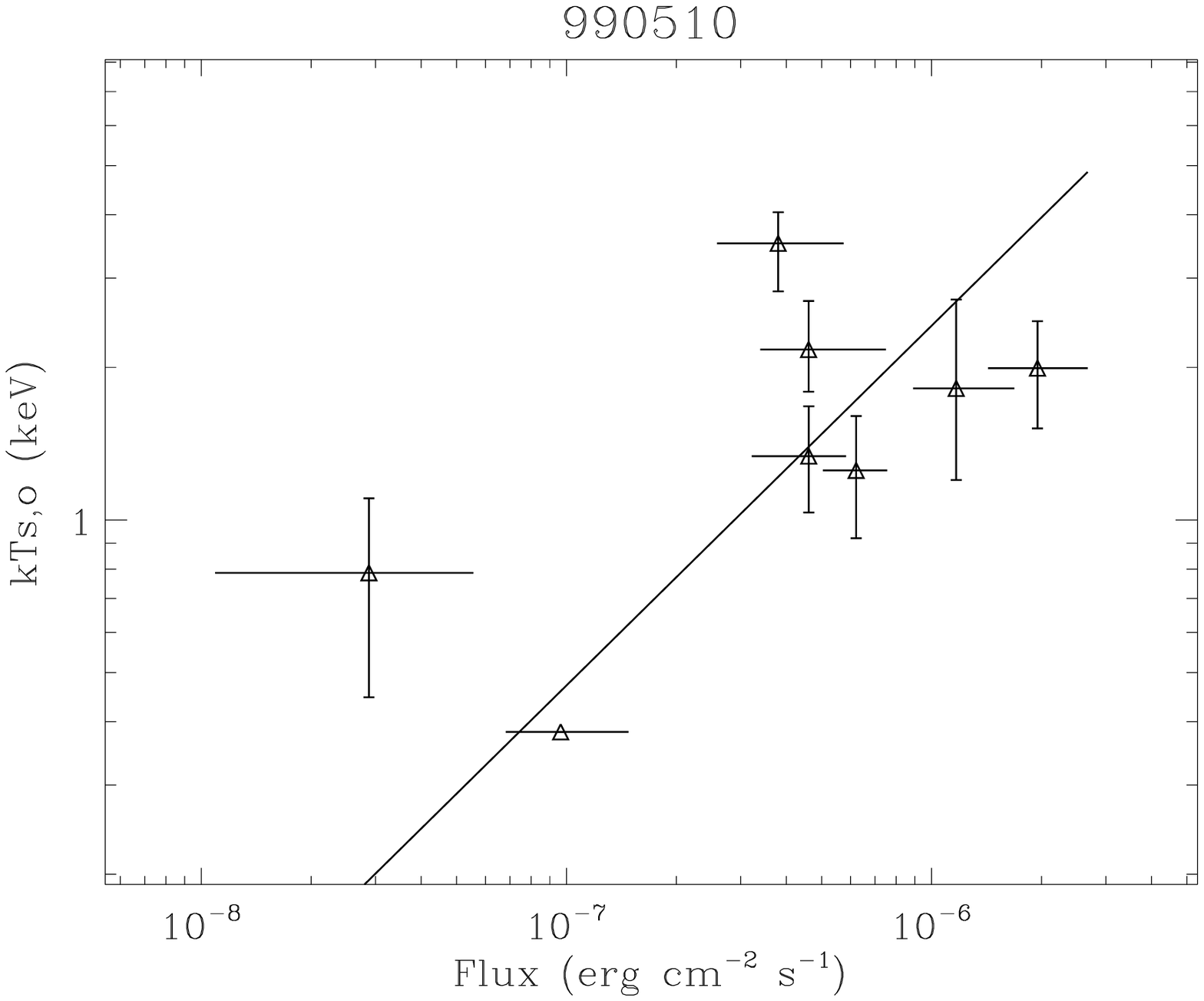}
\end{center}
 \caption{Seed photon temperature $kT_s$ as a function of the prompt emission time--resolved flux
 for GRBs 970111, 980329, 990123, and 990510. The best--fit power--law curve is also shown. }
\label{f:kts-vs-flux}
\end{figure}

\begin{figure}
\begin{center}
  \includegraphics[width=.4\textwidth]{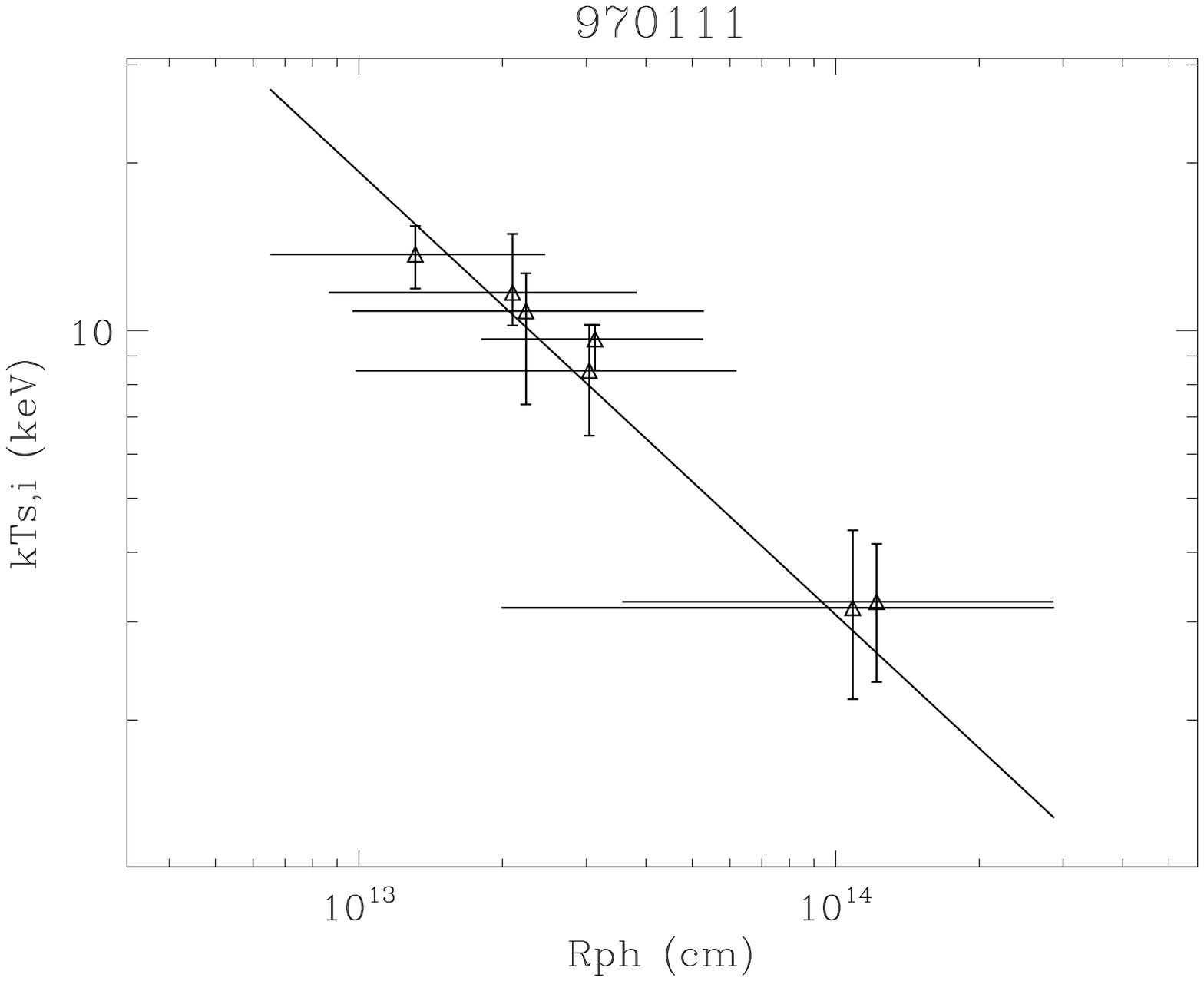}
  \includegraphics[width=.4\textwidth]{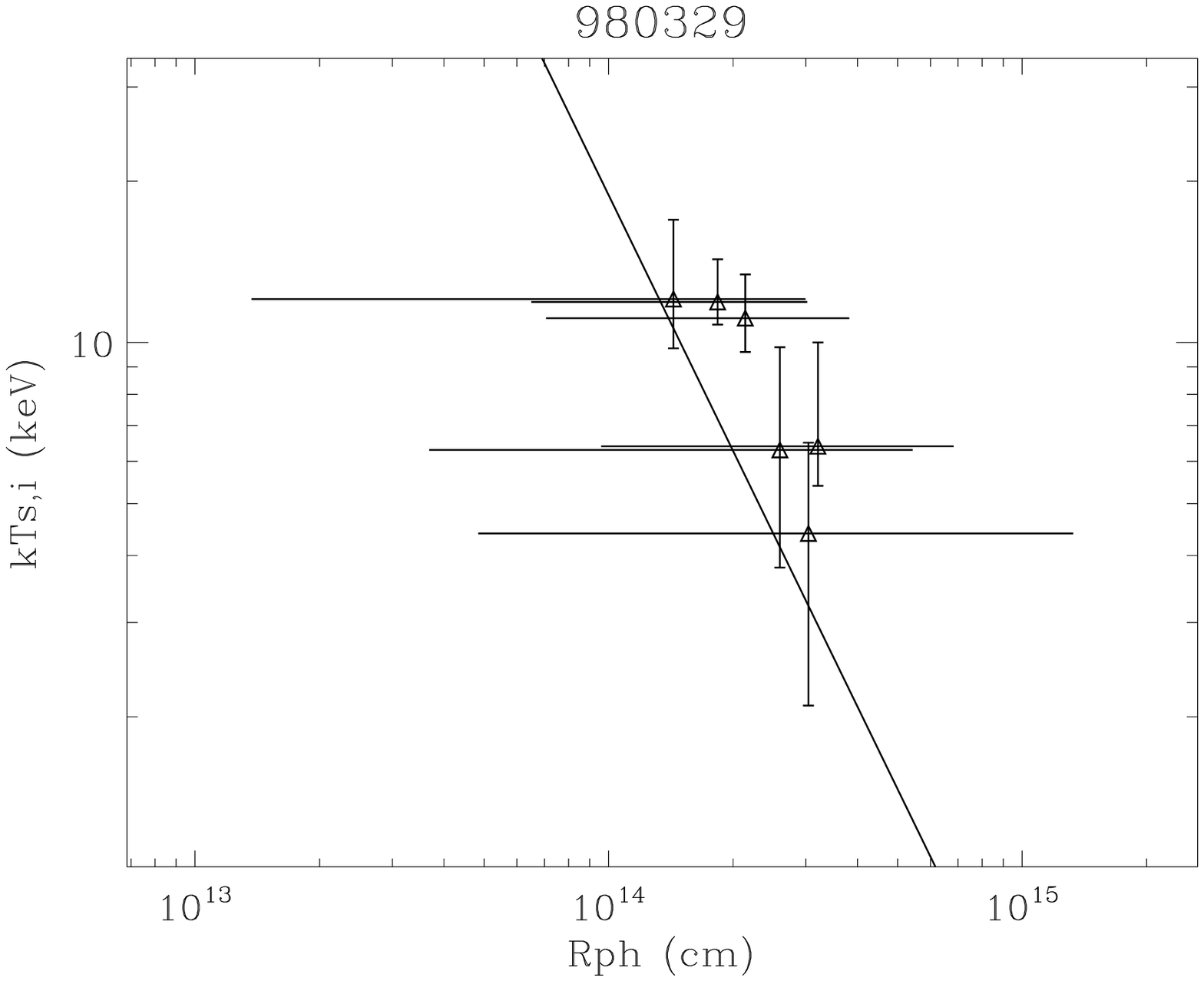}
  \includegraphics[width=.4\textwidth]{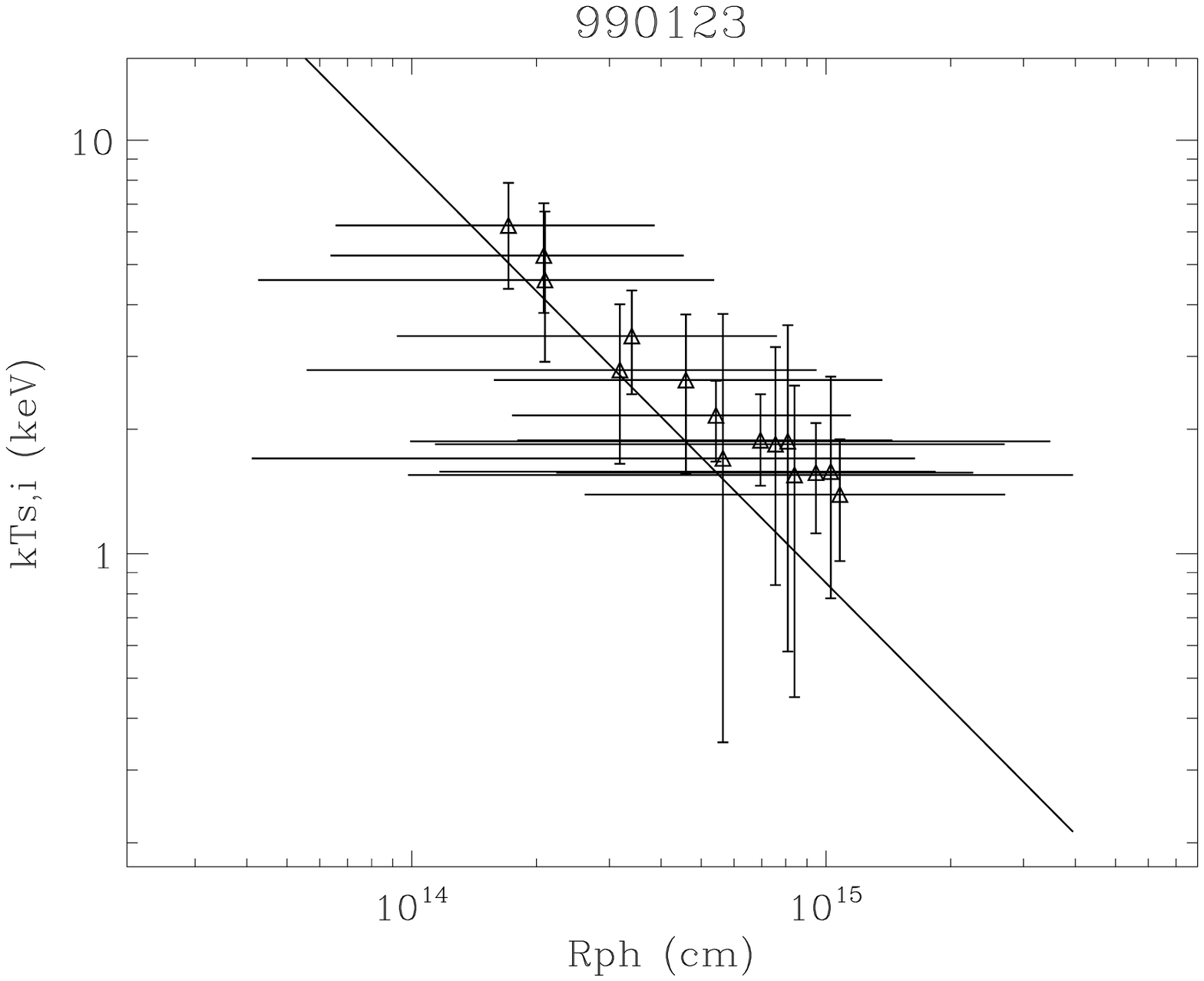}
  \includegraphics[width=.4\textwidth]{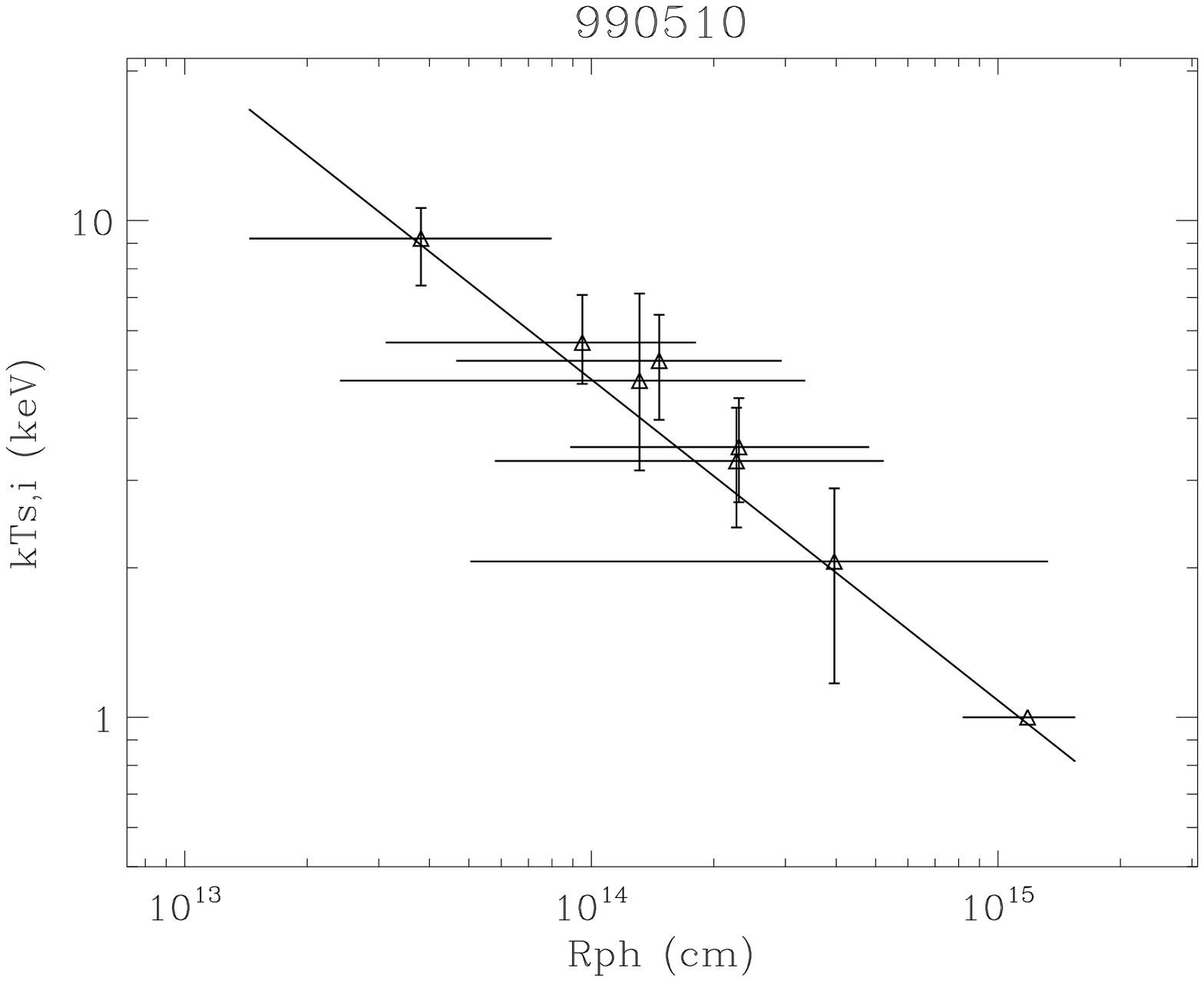}
\end{center}
 \caption{Intrinsic seed photon temperature $kT_{s,i}$ as a function of the photosperic radius at which seed photons 
are emitted, for GRBs 970111, 980329, 990123, and 990510. In the case of GRBs 970111, a redshift $z = 1$ was assumed. The best--fit power--law curve is also shown. }
\label{f:kts-vs-rph}
\end{figure}

\begin{figure}
\begin{center}
  \includegraphics[width=.8\textwidth]{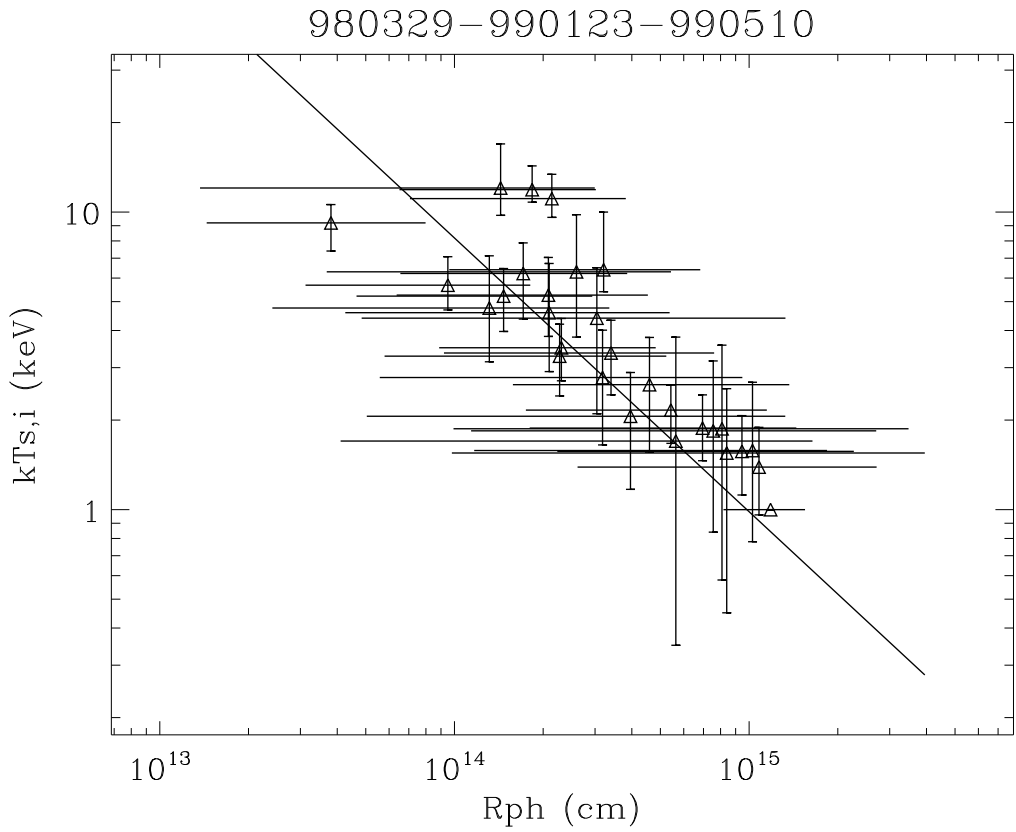} 
\end{center}
 \caption{Superposition of the intrinsic seed photon temperature $kT_{s,i}$ points for GRBs with known redshift (980329, 990123 and 990510) versus the corresponding photosperic radii $R_{ph}$ at which seed photons 
are emitted. The best--fit power--law curve is also shown. }
\label{f:ave-kts-vs-rph}
\end{figure}

\begin{figure}
\begin{center}

\includegraphics[angle=0.0,width=.6\textwidth]{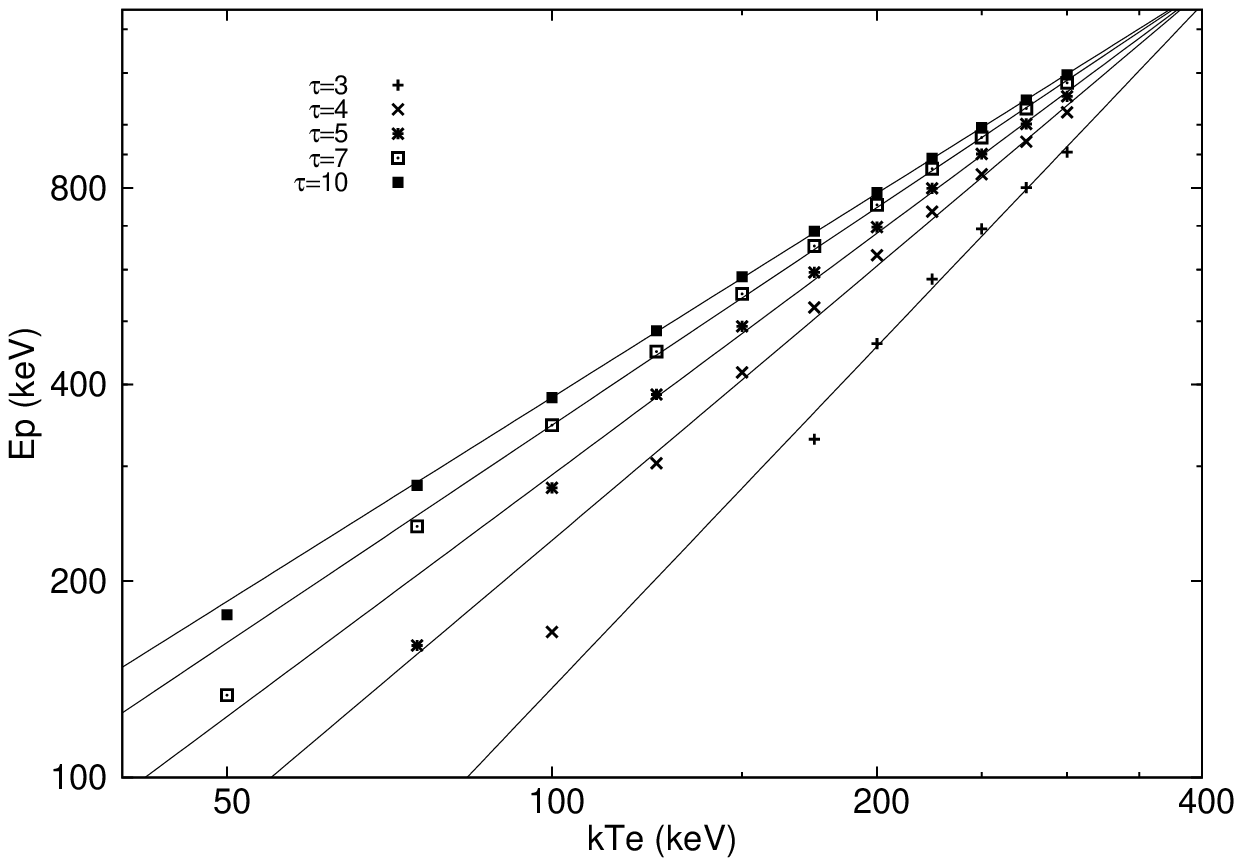}
\includegraphics[angle=-90.0,width=.6\textwidth]{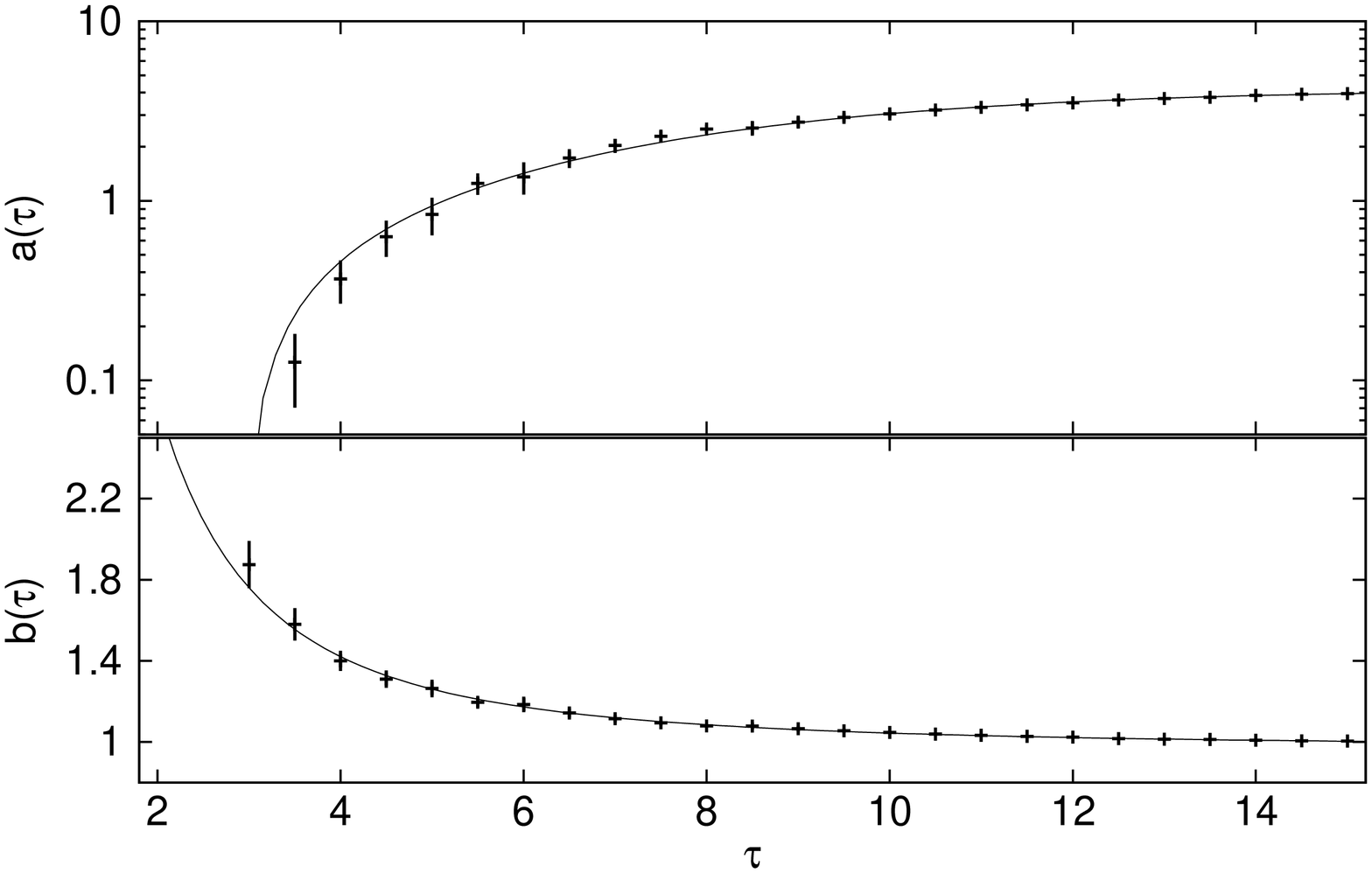}
\end{center}
\caption{{\em Top panel}: Theoretical dependence of the intrinsic peak energy $E_{p,i}$ of the $E F(E)$ spectrum on the electron temperature $kT_{e,i}$ as derived from numerical simulations of the {\sc grbcomp} model.  The data can be fitted by a power--law function  $E_{p,i}= a(\tau)(kT_{e,i})^{b(\tau)}$. An outflow velocity $\beta=0.2$ is assumed.
{\em Bottom panel}: dependence of the normalization factor $a(\tau)$ and index $b(\tau)$ as a function of 
the optical depth $\tau$.  The best--fit curves of $a(\tau)$ and $b(\tau)$ are also shown, while the expression of the fitting functions and best-fit parameters are reported in the text. Note that $b(\tau)$ asymptotically tends to 1, leading to a linear relation between
$E_{p,i}$ and $kT_{e,i}$, as expected for saturated Comptonization.}   
\label{f:pred-ep-vs-kte}
\end{figure}

\begin{figure}
\begin{center}
  \includegraphics[width=.4\textwidth]{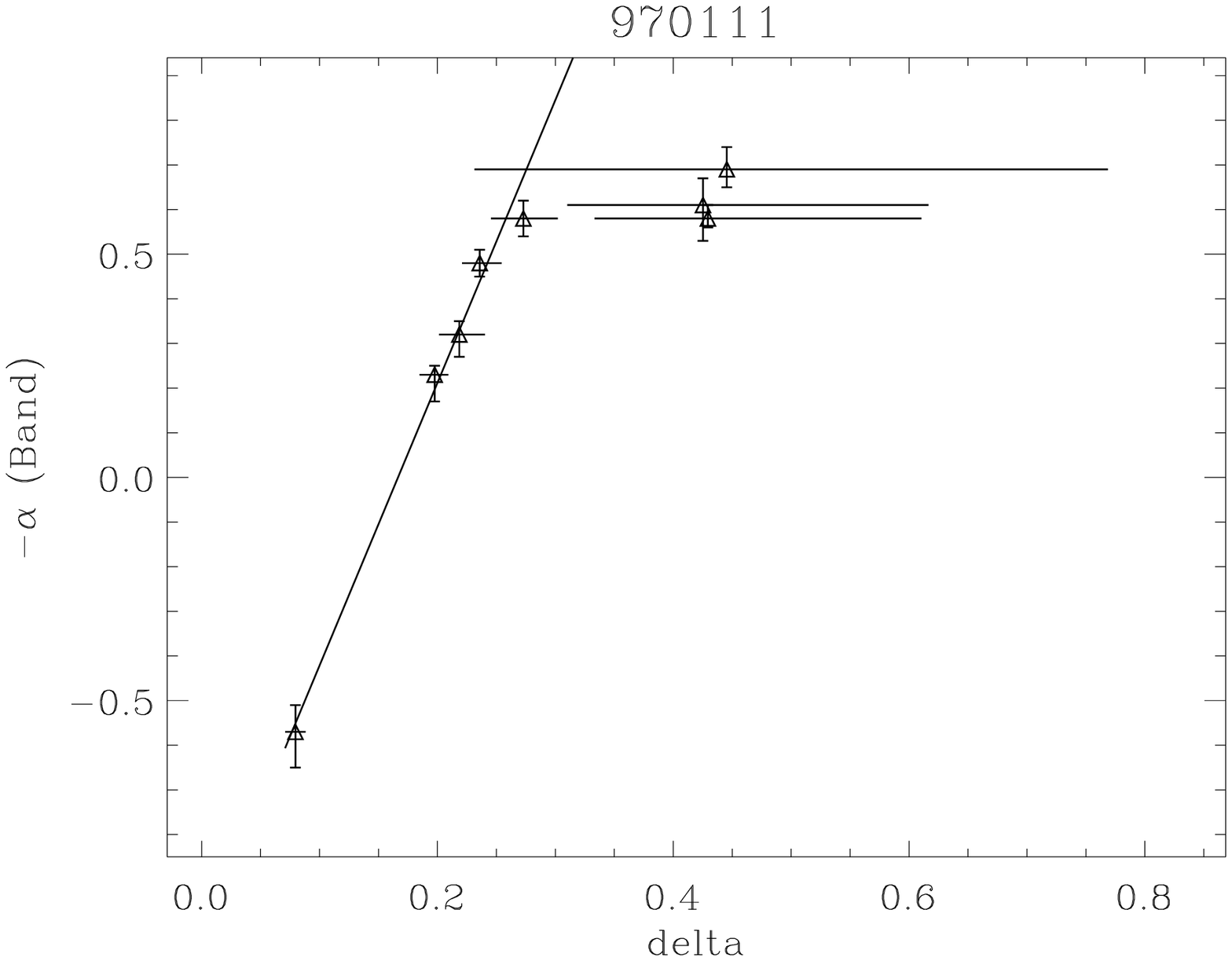}
  \includegraphics[width=.4\textwidth]{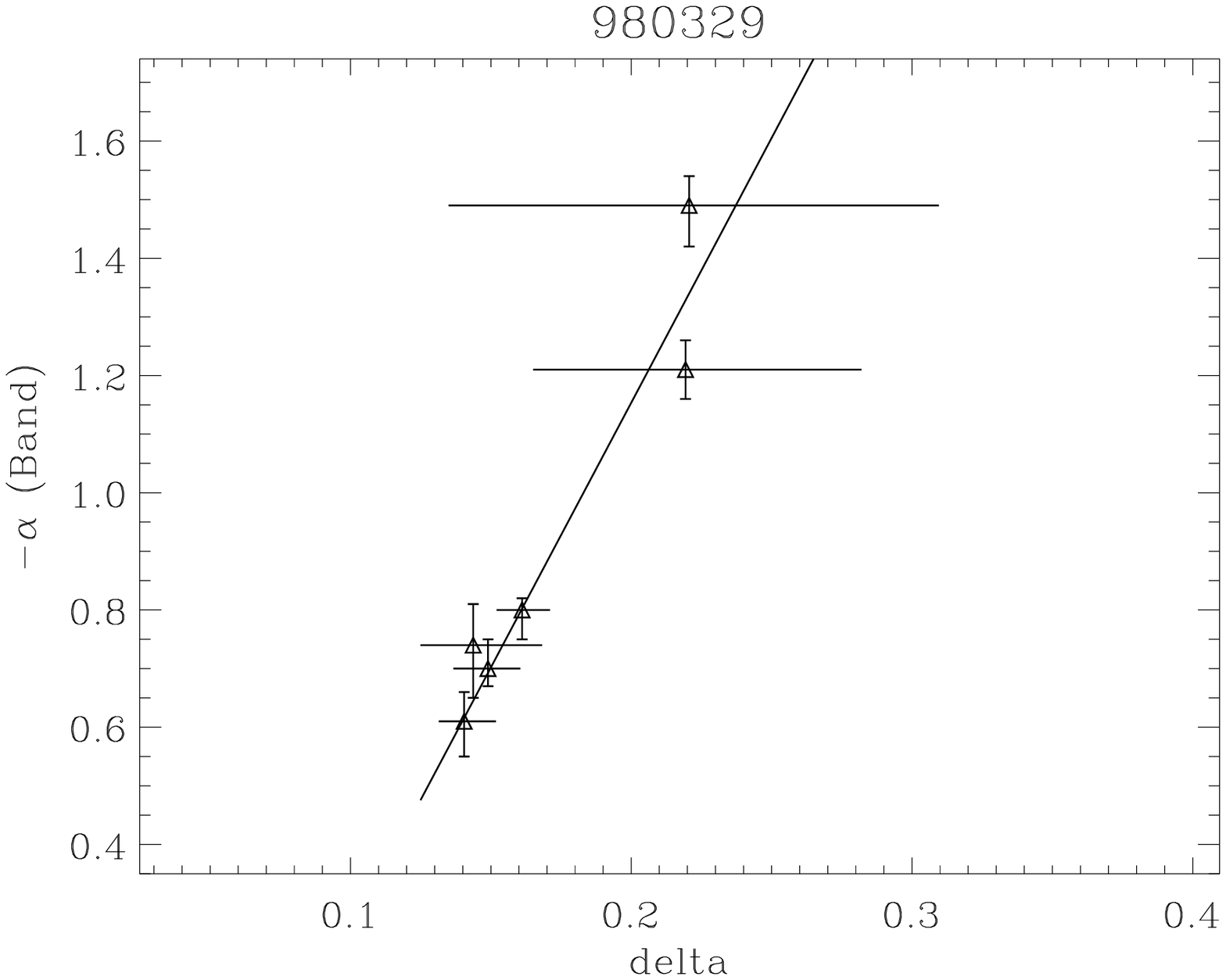}
  \includegraphics[width=.4\textwidth]{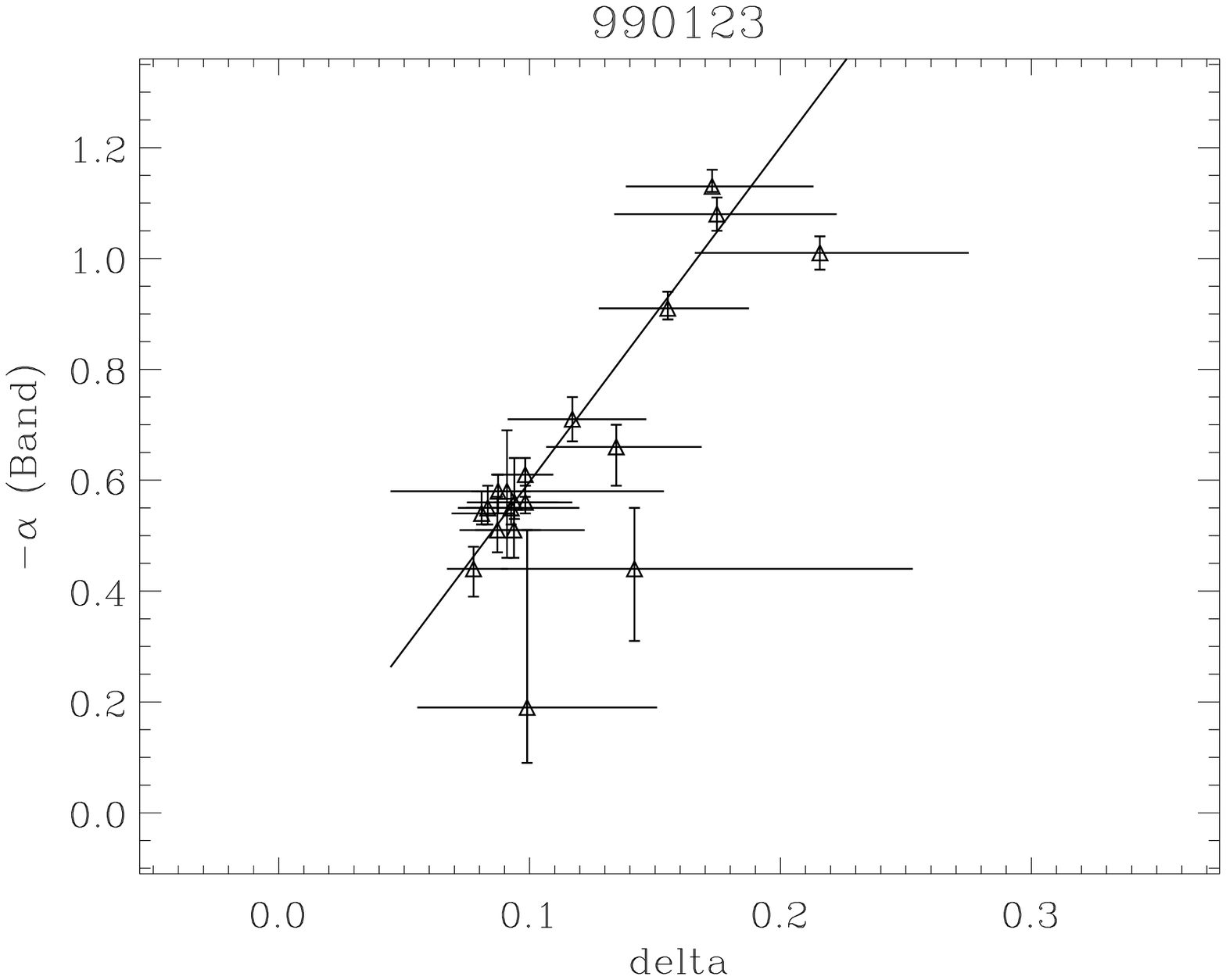}
  \includegraphics[width=.4\textwidth]{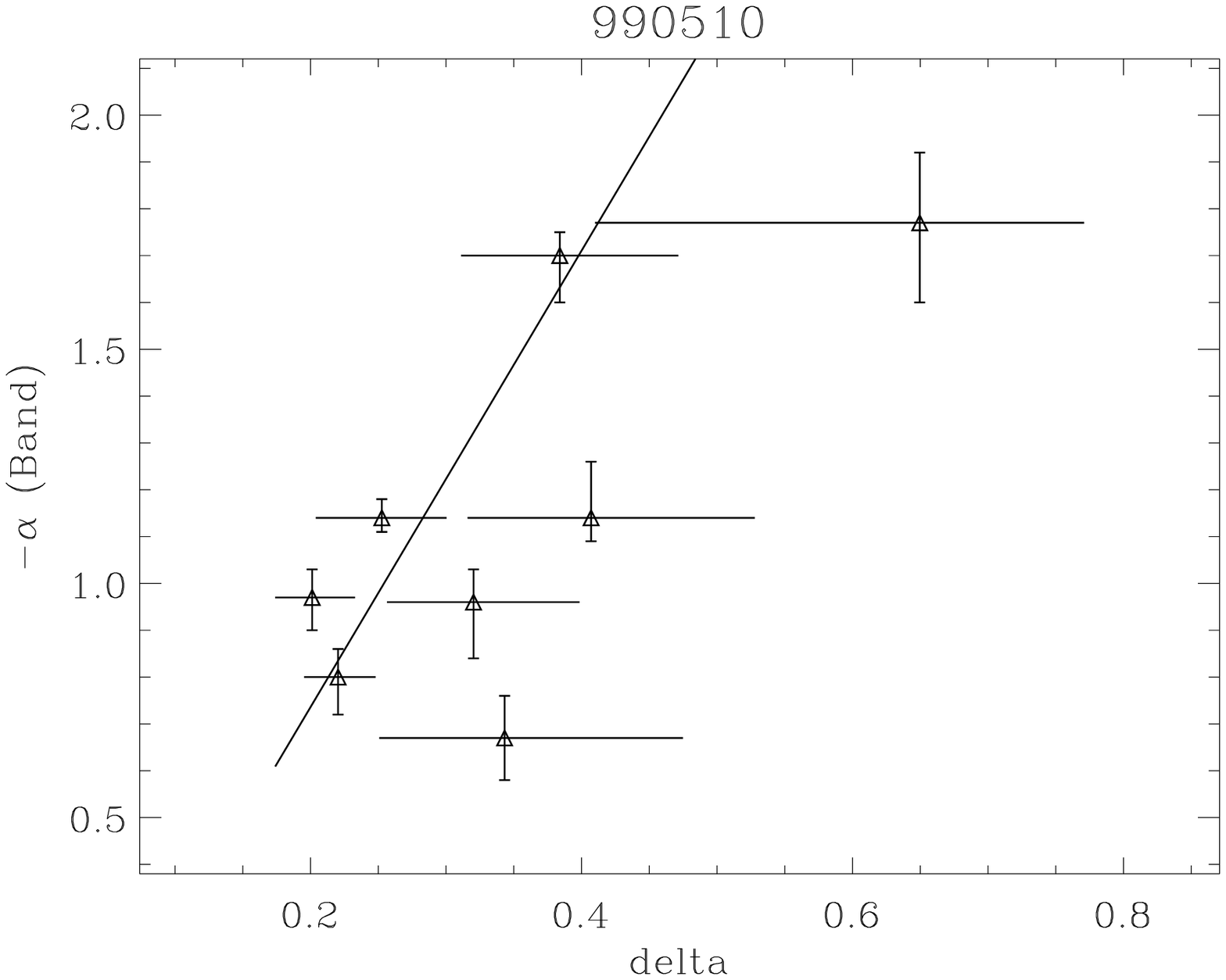}
\end{center}
 \caption{Correlation between the low-energy  index $\alpha_{bf}$ derived from the best fit of the {\sc bf}
to the time--resolved spectra and the bulk parameter $\delta$ of the {\sc grbcomp} model, for GRBs 970111, 980329, 990123, 990510. The best--fit curve is also shown.}
\label{f:alpha-vs-delta}
\end{figure}

As discussed above, GRB\,980329 show peculiar properties, seemingly in contrast with the {\sc grbcomp} predictions. Unlike the other three events in our sample, its intrinsic peak energy $E_{p,i}$ is almost independent on $kT_{e,i}$ (see Fig.~\ref{f:ep-vs-kte}); in addition it does show an almost insignificant anticorrelation between $E_{p,i}$ and the bulk parameter $\delta$ (see Fig.~\ref{f:ep-vs-delta}), and between the seed photon temperature $T_s$ and the photosperic radius
$R_{ph}$.
Other peculiarities, that we have already discussed, concern an almost absent dependence of $kT_{e,o}$ on flux, and a dependence of $kT_{s,o}$ on flux according to a power--law with an index of 0.25. 
 
All these peculiarities are found to be related to the fact that $kT_{e,i}$ is constant during the event (see Fig.~\ref{f:time_evol}) and that also the {\sc bb} emission region has to be almost constant. These features are a matter of fact, are not required by the model and  may depend on the properties of the electron plasma and phenomenon evolution.

\begin{figure}
\begin{center}
  \includegraphics[width=.8\textwidth]{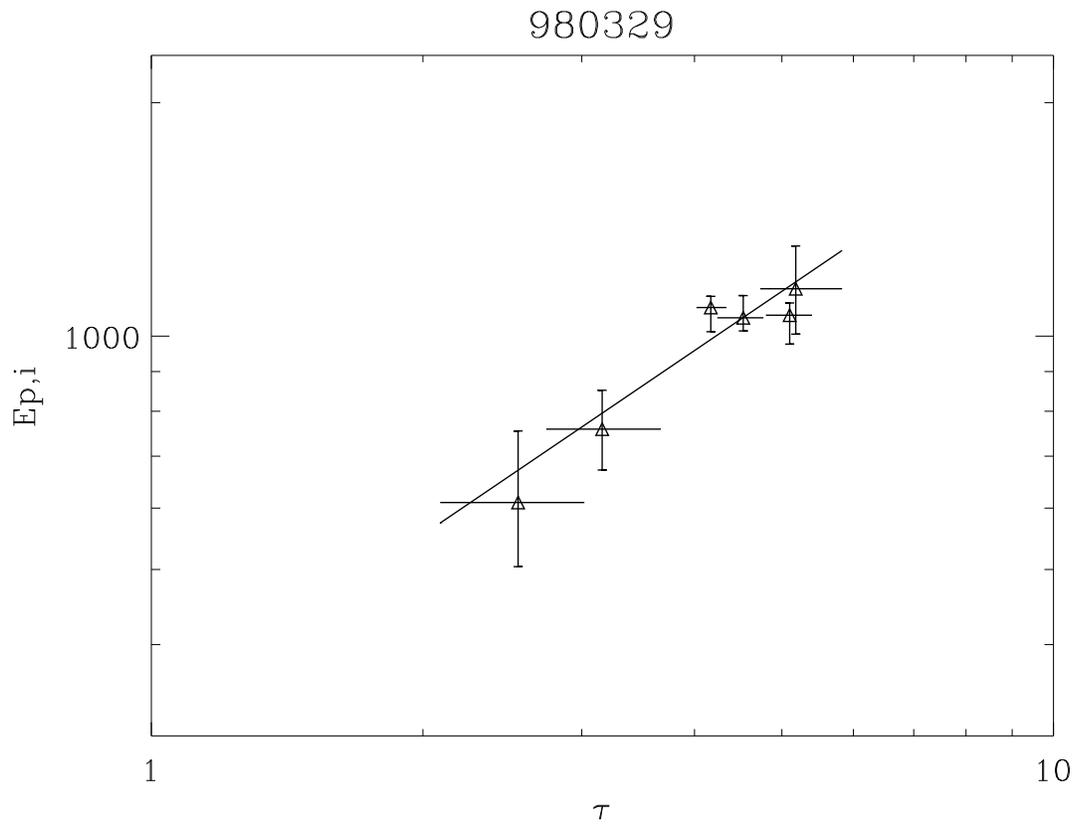} 
\end{center}
 \caption{GRB\,980329 $E_{p,i}$ dependence on optical thickness $\tau$.}
\label{f:ep-vs-tau}
\end{figure}

In the context of the {\sc grbcomp} model,
the consequences of a constant electron-temperature behaviour as
a function of time unavoidably are expected to impact, in
a well-predictable way,  on the other parameters.
We have established these consequences and found that all of them are satisfied by the data. 
If $kT_{e,i}$ is almost constant, we do not expect a correlation between $E_{p,i}$ and $kT_{e,i}$, and this result is found (see Fig.~\ref{f:ep-vs-kte}). In addition, we expect a weak or no dependence of the measured  2--2000 keV flux on $kT_e$, which is also found (see Fig.~\ref{f:kte-vs-flux}). We expect a weakening of the  correlation between $E_{p,i}$ and $\delta$ (see eq.~(1)), given that only the electron cloud optical thickness $\tau$ variations  influence $\delta$, which is also observed (see Fig.~\ref{f:ep-vs-delta}). Thus we expect a dependence of $E_{p,i}$  on $\tau$, which is found in the data, as it can be seen in Fig.~\ref{f:ep-vs-tau}.

\begin{figure}
\begin{center}
  \includegraphics[width=.8\textwidth]{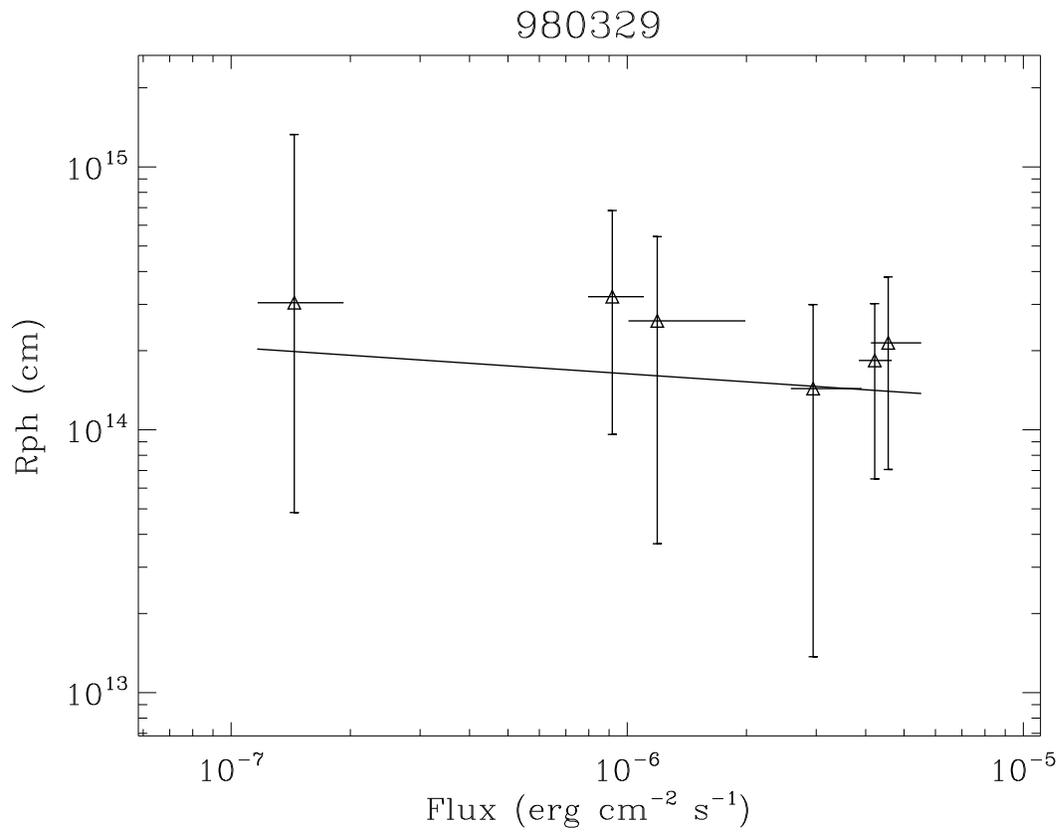} 
\end{center}
 \caption{GRB\,980329 $R_{ph}$ dependence on 2--2000 keV flux.}
\label{f:rph-vs-flux}
\end{figure}

The constancy of the {\sc bb} emission region during the prompt emission, derives from (and tests) the main assumption of the {\sc grbcomp} model, i.e., that the GRB luminosity (see eq.~(\ref{e:lum})), as discussed above,
depends, in addition to the Comptonization  enhancement factor, on the seed photon luminosity $L_s$ assumed to be a blackbody emission ($L_s \propto T_s^4 A_s$). In the case of GRB\,980329, given that the electron temperature does not significantly change with time, the received flux should depend  on the BB temperature $T_s$ and  on the blackbody emission region size $A_s$ ($\propto R_{ph}^2$). If also this surface does not change during the prompt emission, we expect no correlation between flux and $R_{ph}$, no correlation between $kT_s$ and $R_{ph}$, and a power--law dependence of $kT_{s,i}$ on flux with an index of 0.25. All these conditions are satisfied in the case of GRB\,980329: $R_{ph}$ almost independent of  flux (see Fig.~\ref{f:rph-vs-flux}), no correlation between $kT_s$ and $R_{ph}$ (see Table~\ref{t:kts-vs-rph} and Fig.~\ref{f:kts-vs-rph}) and a power--law dependence of $kT_s$ on flux with an index $m = 0.37 \pm 0.12$, which is consistent, within 1$\sigma$ with the expected value of 0.25.

 A legitimate question is what observationally distinguishes GRB\,980329 from the other GRBs in our sample. Our answer is that, unlike other events, 980329 shows a fast rise and exponential decay (FRED) light curve. It would be very interesting to test whether this feature is the observational signature of the behaviour we have observed in the correlations between {\sc grbcomp} parameters or between these parameters and measured quantities (e.g., flux). Unfortunately,  broad band observations extending down to 2 keV are not available in the current data sets. Future GRB missions with a detection threshold of 2 keV or less  are desirable \citep[see, e.g.,][]{Amati13}.

\subsection{Interpretation of the $E_{p,i}$--$L_{iso}$ relation}

The {\sc grbcomp} model gives a physical interpretation of the  $E_{p,i}$--$E_{iso}$ relation (T12). 
The model interprets the peak energy $E_{p,i}$ of the $E F(E)$ spectrum in terms of the temperature $kT_{e,i}$ of the electron plasma modified by the plasma outflow bulk velocity and other properties, all expressed through the bulk motion Comptonization parameter $\delta$. In this view, the 
$E_{p,i}$--$E_{iso}$ relation  is interpreted as a result of the Stephan--Boltzmann law modified by the Comptonization enhancement factor $\eta_{comp} = L_{grb}/L_{s}$, which, as discussed by T12, in the case of saturated Comptonization, is proportional to $kT_e$. The final result is that, on average,  $L_{grb}\propto kT_{e,i}^2$, or that $kT_{e,i} \propto L_{grb}^{1/2}$ and thus $kT_{e,i} \propto E_{grb}^{1/2}$ . 

This interpretation can be extended to the time--resolved
$E_{p,i}$--$L_{iso}$ relation found within each GRB by us 
\citep{Frontera12a,Frontera12b} and other authors 
\citep{Ghirlanda10,Lu12}. 
Indeed, according to {\sc grbcomp}, the GRB luminosity at a given time $t$ can be written
as
\begin{equation}
L(t) = A_s(t) ~(kT_s)^4(t) ~\eta_{comp}(t,\tau)
\label{e:lum}
\end{equation}
where $A_s(t)$ is the emitting area, and the other (intrinsic) quantities have already been defined above. 
However we have seen that $E_{p,i}$  is related to the outflow temperature $kT_{e,i}$ through  eq.~(\ref{e:epi-vs-ktei}), with the parameters $a(t,\tau)$ and $b(t,\tau)$ that may depend on time, because they are functions of $\tau$ which in general is not constant through the prompt emission phase.

For a given observing time interval ($t_1$, $t_2$), which is short enough that all intrinsic (we omit the index $i$) quantities can be considered almost constant, the average luminosity and intrinsic peak energy are given by
\begin{equation}
<L> = (kT)^4_s ~\eta_{comp}(\kte, \tau) ~\frac{\int^{t_2}_{t_1}A_s(t) \mathrm{d}t}{t_2-t_1}
\label{e:eq_lumin_aver}
\end{equation}

\begin{equation}
<E_{p}> = a(\tau)~ kT_{e}^{b(\tau)}
\label{e:eq_ep_aver}
\end{equation}

If now the emission area follows the law $A(t) \propto \kte~t^2$ (see equation [24] of T12), 
equation (\ref{e:eq_lumin_aver}) becomes

\begin{equation}
 <L> \propto ~(kT)^4_s ~\eta(\kte, \tau) ~\kte ~f(t)
\label{e:aver_lumin_a}
\end{equation}
where $f(t)$ is the factor obtained by simply integrating
the function $t^2$ over the interval ($t_1$, $t_2$).

\begin{figure}
\begin{center}

\includegraphics[angle=0.0,width=.6\textwidth]{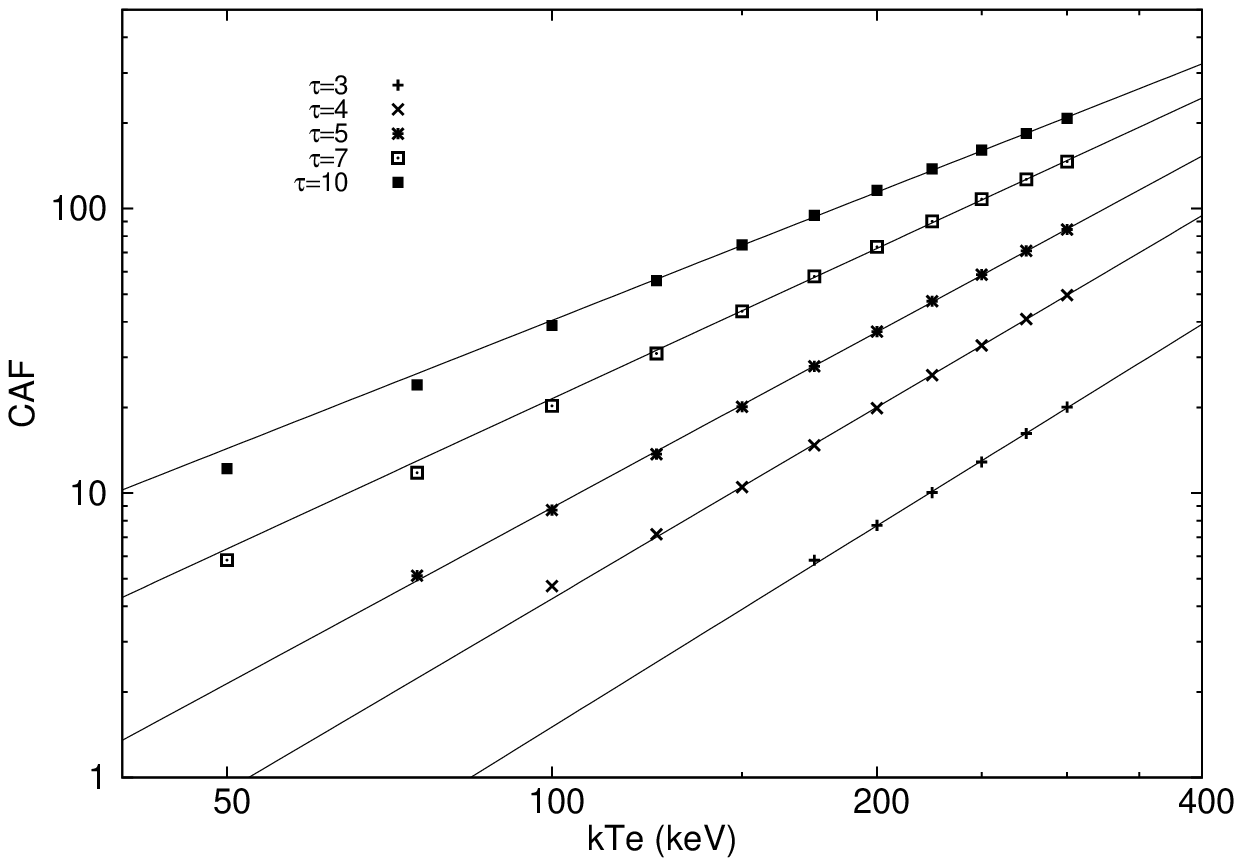}
\includegraphics[angle=-90.0,width=.6\textwidth]{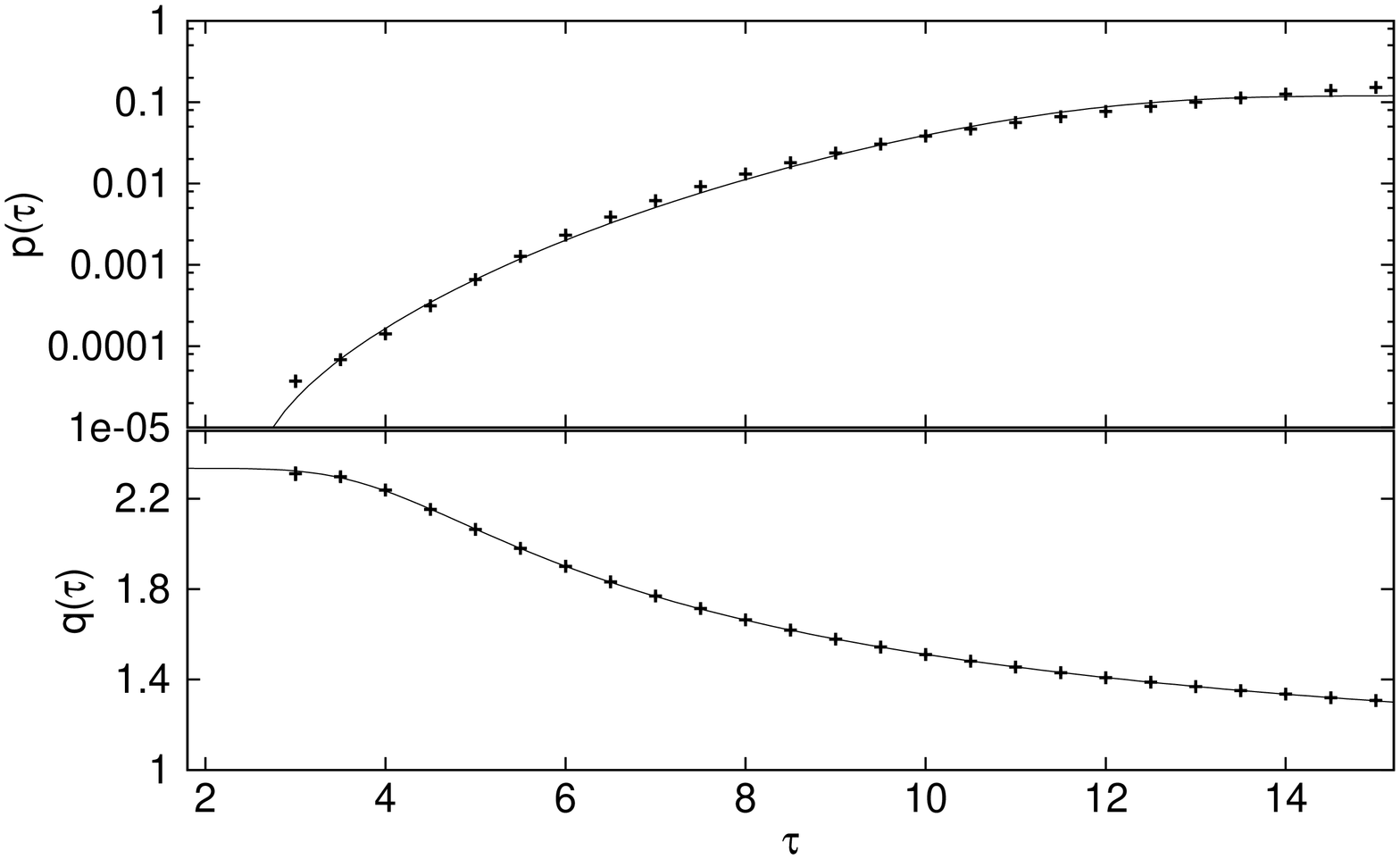}
\end{center}
\caption{{\em Top panel}: Theoretical dependence of the Comptonization amplification factor (CAF) as a function of the electron temperature $kT_{e,i}$ obtained with the {\sc grbcomp} model. The obtained values are described by a power--law function$\eta_{comp}= p(\tau)(kT_{e,i})^{q(\tau)}$.  An outflow velocity $\beta=0.2$ is assumed.
{\em Bottom panel}: dependence of the power-law parameters $p(\tau)$ and $q(\tau)$ 
on the optical depth $\tau$. The best--fit curves are also shown and the empirical fitting functions of the parameters are given in the text.}   
\label{f:caf-vs-ktei}
\end{figure}

The Compton amplification factor (CAF) has a general dependence on the electron temperature and 
the optical depth, which can be approximated, as shown in Fig.~\ref{f:caf-vs-ktei} obtained with a numerical code, by a power--law:

\begin{equation}
\eta_{comp}(\kte, \tau) = p(\tau)~ \kte^{q(\tau)}.
\label{e:eq_eta}
\end{equation}
where the dependence on $\tau$ of the parameters $p(\tau)$ and $q(\tau)$ can be best fit by two following empirical functions, similar to those describing $a(\tau)$ and $b(\tau)$ (see eqs.~\ref{e:a(tau)} and \ref{e:b(tau)}):

\begin{equation}
p(\tau)=k_0+k_1~k_2~{\rm log}\Bigg\{{\rm exp}\Big[\Big(1-\Big(\frac{\tau}{k_3}\Big)^{k_4}\Big)/k_2\Big] +1\Bigg\}
\end{equation}
where $k_0= (3.2\pm 1.43) \times 10^{-5}$, $k_1=1.060 \pm 0.035$, $k_2 = 0.14 \pm 0.01$, $k_3= 0.076\pm 0.002$, $k_4 =0.59\pm 0.03$,

and

\begin{equation}
 q(\tau)=k_5-k_6~k_7~{\rm log}\Bigg\{{\rm exp}\Bigg[\Bigg(1-\Big(\frac{1}{k_8~\tau}\Big)^{k_9}\Bigg)/k_7\Bigg] +1\Bigg\}
\end{equation}
where
$k_5 = 2.334+/- 0.004$, $k_6=1.444 \pm 0.006$, $k_7=0.152 \pm 0.004$, $k_8=0.230 \pm 0.001$.

Substituting eq.~(\ref{e:eq_eta}) into eq.~(\ref{e:aver_lumin_a}) we obtain

\begin{equation}
 <L> \propto ~(kT)^4_s  ~\kte^{q(\tau)+1} ~f(t),
\label{e:aver_lumin_b}
\end{equation}
which can be reverted to
\begin{equation}
 \kte \propto <L>^{\frac{1}{1+q(\tau)}} ~ (kT_s)^{-\frac{4}{1+q(\tau)}} ~ f(t)^{\frac{1}{1+q(\tau)}}
\label{e:kte_vs_lumin}
\end{equation}

However, combining eq.~(\ref{e:eq_ep_aver}) and eq.~(\ref{e:kte_vs_lumin}) we finally obtain
\begin{equation}
\ep \propto  <L>^{\frac{b(\tau)}{1+q(\tau)}} ~ (kT)_s^{-\frac{4b(\tau)}
{1+q(\tau)}} ~f(t)^{-\frac{b(\tau)}{1+q(\tau)}}
\label{e:ep_vs_liso}
\end{equation}

The intriguing implication of the last result is that the behavior of the parameters $b(\tau)$ and  $q(\tau)$, shown in 
Figs.~\ref{f:pred-ep-vs-kte} and \ref{f:caf-vs-ktei},
is such that the exponent
of the $<L>$ term is always $\sim 0.5$ for any value of the optical depth $\tau$, and not only in the asymptotic limit $\tau \gg 1$
where $b(\tau) \rightarrow 1$ and $q(\tau) \rightarrow 1$.
%
 
However, from eq.~(\ref{e:kte_vs_lumin}), only in the asymptotic case of very large optical depth when $q(\tau) \rightarrow 1$, we get

\begin{equation}
 \kte \propto <L>^{\frac{1}{2}} ~ (kT_s)^{-2} ~ f(t)^{\frac{1}{2}}.
\label{e:kte_vs_lumin_taugg1}
\end{equation}

We actually find that, within each GRB, the measured flux is correlated with the electron temperature, while a correlation between flux and peak energy $E_{p,o}$ within each GRB is well established 
\citep[see, e.g.,][] {Frontera12a,Frontera12b}. 
As also expected, the correlation between flux and $kT_{e,o}$ is  almost absent when the electron temperature is almost constant during the event, as found for GRB\,980329.

\begin{figure}
\begin{center}
  \includegraphics[width=.4\textwidth]{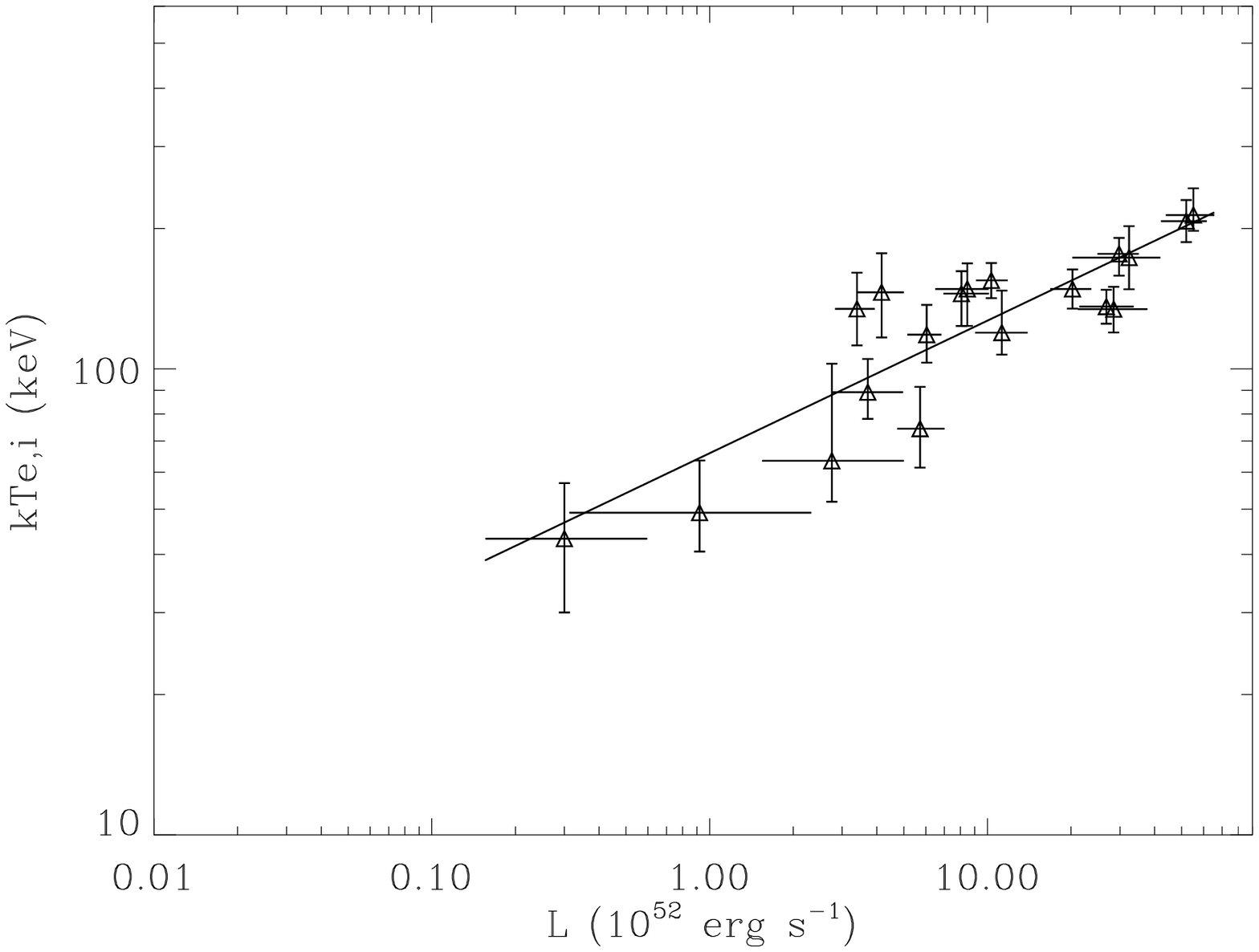}
  \includegraphics[width=.4\textwidth]{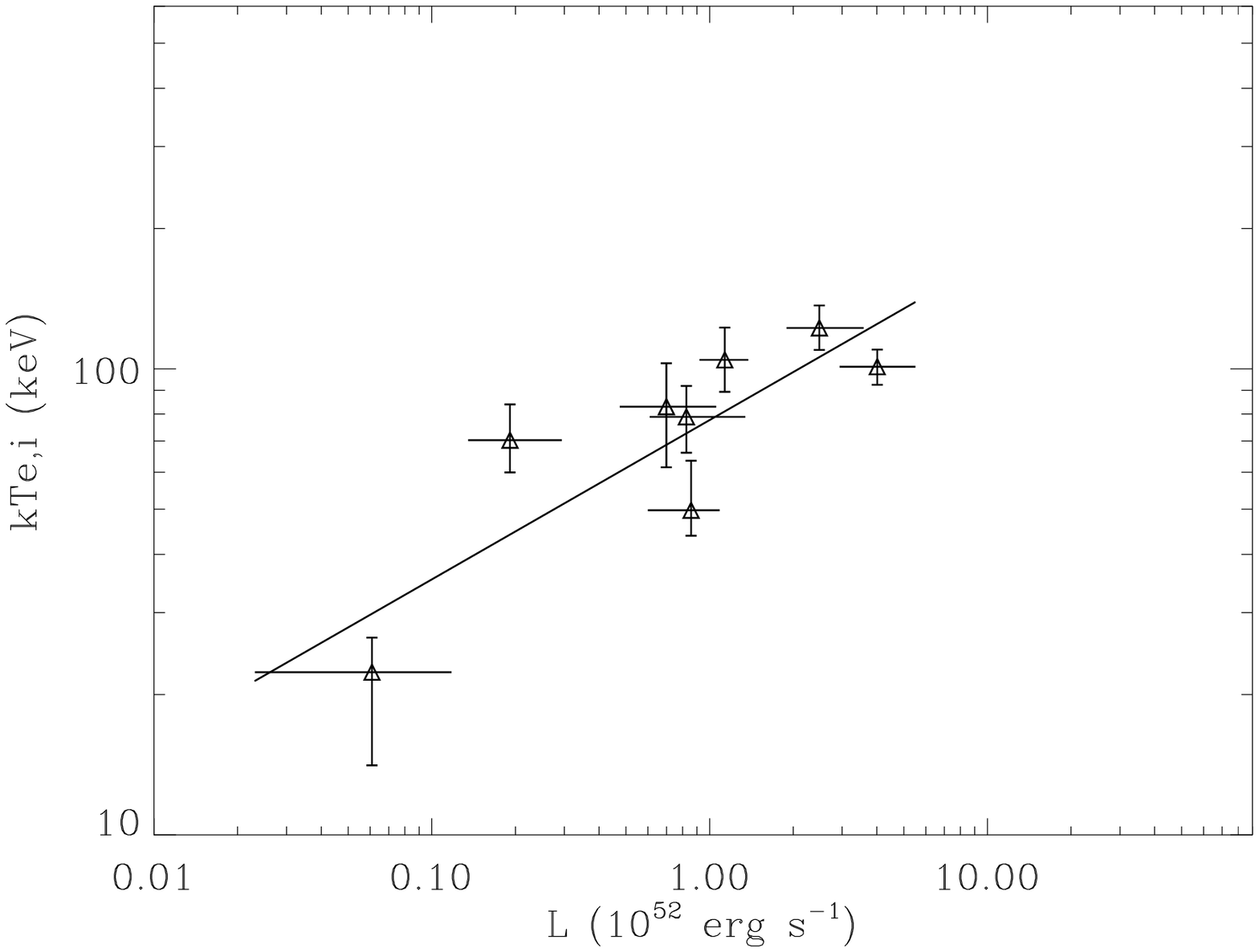}
  \includegraphics[width=.4\textwidth]{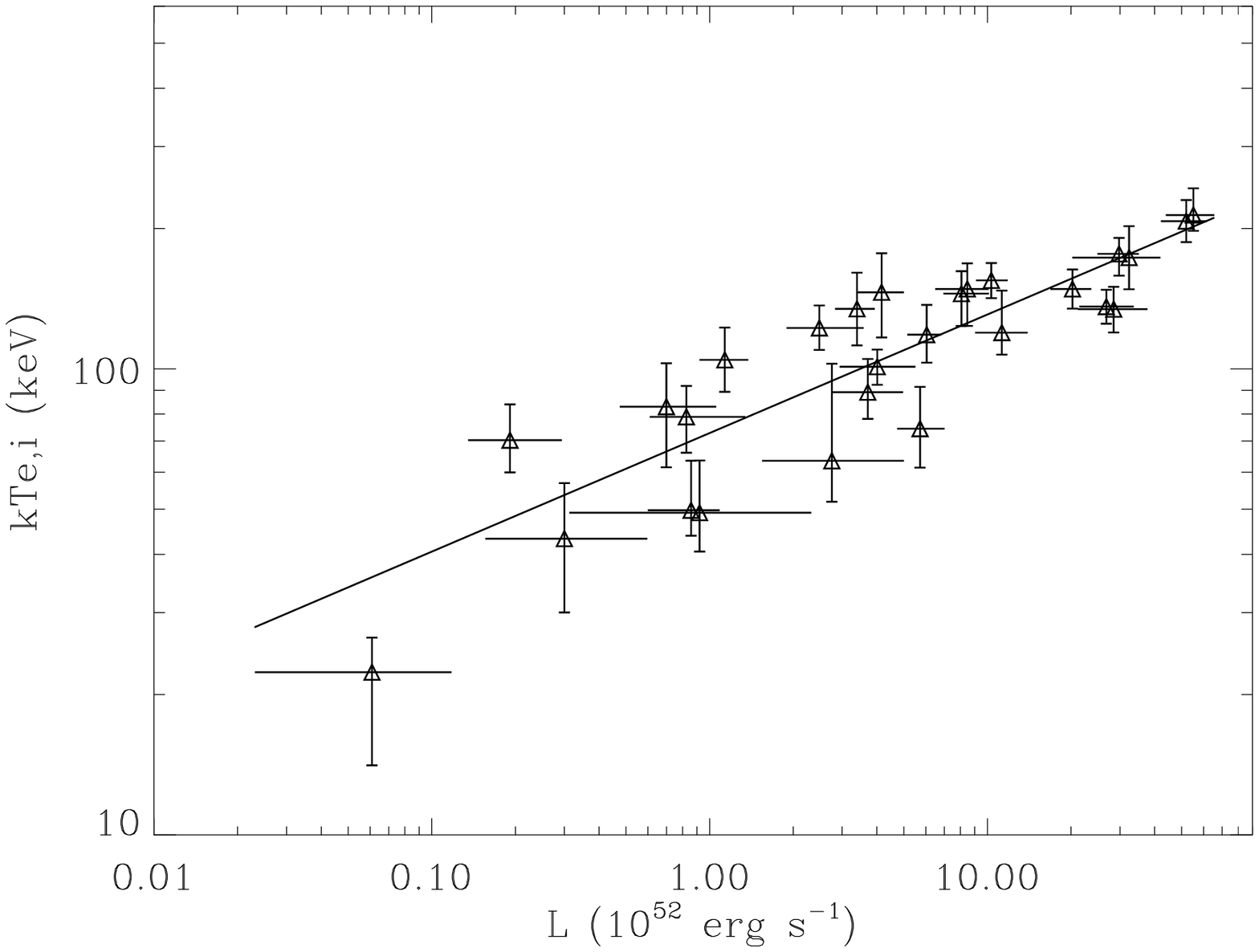}
\end{center}
 \caption{{\em Top panels}: Intrinsic $kT_{e,i}$ 
 as a function of the 2--2000 keV luminosity for GRBs 990123 ({\em left panel}) and 990510 ({\em right panel}).
{\em Bottom panel}: Average intrinsic electron temperature $kT_{e,i}$ of the {\sc grbcomp} model as a function of the 2--2000 keV isotropic luminosity $L_{iso}$, 
obtained by merging together the data points of GRBs with known redshift (990123, 990510).
The continuous line gives the best--fit power--law curve (see Table~\ref{t:kTei-vs-L}).} 
\label{f:kTei-vs-L}
\end{figure}

Similarly, for GRBs with time variable $kT_{e,i}$ and 
known redshift (990123 and 990510), the intrinsic temperature $kT_{e,i}$ of each GRB is correlated with the corresponding isotropic luminosity $L_{iso}$ (see top panels of Fig.~\ref{f:kTei-vs-L} and Table~\ref{t:kTei-vs-L}) according to a power--law ($kT_{e,i} \propto (L_{iso})^m$). When we merge these data  (see bottom panel of Fig.~\ref{f:kTei-vs-L} and Table~\ref{t:kTei-vs-L}),  the power--law dependence of $kT_{e,i}$ on $L_{iso}$, becomes more robust (see Table~\ref{t:kTei-vs-L}) with an average index $m = 0.31 \pm0.02$, as expected in the scenario underlying the {\sc grbcomp} model in the case the optical depth is not so large. Indeed, looking at the theoretical behaviour of $q(\tau)$ (see 
Fig.~\ref{f:caf-vs-ktei}) and at the best fit 
values of $\tau$ for 980329, 990123 and 990510 
(Table~\ref{t:results}), it becomes evident from 
eq.~(\ref{e:kte_vs_lumin}) that the relation 
$kT_{e,i} \propto (L_{iso})^m$ should 
have $m < 1/2$, which is in full agreement with what we found.

Given the importance of the $kT_{e,i}$ time behavior during the prompt emission and the change of $kT_s$ from one GRB to other, the dispersion of these two parameters, in addition to $\tau$, could be at the origin of the dispersion of the average $kT_{e,i}$--$L_{iso}$ relation and eventually of the $E_{p,i}$--$L_{iso}$ relation.

\section{Conclusions}

With a sample of 4 strong GRBs (970111, 980329, 990123, and 990510) simultaneously detected with both \sax\ WFCs and BATSE we have performed a time--resolved spectral analysis in the energy band from 2 keV to 2000 keV. In total we derived 41 spectra, all used to test three different models: {\sc bb$+$pl}, {\sc bf$+$bb}, and the recently proposed model {\sc grbcomp}.

We do not expect significant systematic errors from this joint analysis. 
The response functions of both instruments are well known. Indeed, in the fits 
the cross-calibration factor was found to be consistent with 1, in spite of being left free to 
vary between 0.8 and 1.2 (see Section~\ref{s:models}). This is not the first time that
a joint WFC/\batse\ spectral analysis has been performed. Results of similar analyses have been 
reported in the past by the \batse\ team \citep{Briggs00,Kippen03,Kippen04}. In addition, the 
\batse--deconvolved spectra of bright GRBs were cross-checked with those obtained with
\sax\ GRBM \citep{Frontera09}; these in turn were cross-calibrated with WFC, 
with many published results 
\citep[e.g.][]{Frontera98a,Frontera00a}.

The result is that the {\sc bf} function fits almost all spectra \citep{Frontera12a,Frontera12b}. In the few cases (14\%) in which this function does not work, following other authors \citep{Guiriec11,Axelsson12}, we have added a {\sc bb}. In  one case we obtained a significantly better fit.

The fit of the time--resolved spectra with the {\sc bb$+$pl} model has provided negative results. Only in few cases this model is suitable to describe the time--resolved 2--2000 keV spectra, thus this model does not appears to interpret the 2--2000 keV GRB prompt emission spectra.
 
Instead, we find that the {\sc grbcomp} model,
even better than {\sc bf}, describes almost all the 41 
time--resolved spectra we have analyzed. 
Most of the predictions of this model have been confirmed by the discussed correlation results. Very constraining expectations of the {\sc grbcomp} model have been tested and verified. Even the behavior of one of the four GRBs in our sample (980329), which is different from that of the other three, fits very well the expectations of the model. Moreover, this GRB gives the best opportunity to verify the validity of the model.
 In addition, the physical interpretation of the 
time--resolved $E_{p,i}$--$L_{iso}$ relation given by {\sc grbcomp} is confirmed by our results.

The very good description of our time--resolved spectra with
 {\sc grbcomp} model strengthens the physical  scenario assumed by the model: an early subrelativistic expanding outflow that interacts with a bath of black--body seed photons whose temperature is a free parameter. This scenario is fully compatible with a supernova explosion.  From the fit of the model to our data, we find that this temperature is in the keV range, which is not usual in a common supernova explosion \citep[see, e.g.,][]{Lazzati00}. A likely interpretation is that the seed photons  are energized by the jet. The best fit of our spectra with the {\sc grbcomp} model is also in favor of an early  non--relativistic phase of the outflow velocity. This is not the standard assumption in the collapsar model. However, as discussed by T12, whether the early expansion phase is soon relativistic may depend on several parameters, such as
the initial jet energy, the chemical composition of the star envelope (in particular the presence or 
not of a H-envelope), the core angular momentum, and the influence of magnetic torques \citep[]{Gehrels09, Woosley11}.
Independently of all that, the important result of our test is that a Comptonization process of an initially 
non--relativistic expanding outflow with seed photons in a region close to the photospheric radius well describes the low energy part ($<E_{p,i}$) of more than 40 time--resolved spectra. A further convolution of the Comptonized part of the seed spectrum with a Green function, physically representing a relativistic Inverse Compton (T12), well describes the 
high--energy part of our time--resolved spectra. T12 do not analyze in detail the possible mechanisms to get in the late phase a relativistic outflow. However the peak energy of the $E F(E)$ spectra is mainly determined by the temperature $kT_e$ of the electron plasma non--relativistically outflowing. Possible scenarios for the formation of a relativistic phase are discussed in Section~\ref{s:intro}. The positive results obtained from the test of this model motivate us to study more in detail the formation of the relativistic stage of the model.

Unfortunately, our data are inadequate to perform an extended test of the {\sc grbcomp} model. This test
should be done with a large sample of GRBs observed in a very broad--energy band extending down to 2 keV like the 
time--resolved spectra obtained for the GRBs in our sample, in order to discriminate between the {\sc grbcomp} and other models. Indeed, an important feature of the predicted {\sc grbcomp} spectra is their break at low energies (see Fig. 7 of T12), where the blackbody spectrum of the seed photons takes over. Even the GBM experiment aboard \fermi\ is not fully adequate to test this model, having a passband with a lower threshold at about 10 keV. New satellite missions devoted to GRBs should include detectors with a broad passband extending down to at least 1 keV or even better below.

\begin{acknowledgments}
We are grateful to Pawan Kumar and Tsvi Piran for valuable discussions and suggestions, and Jean in't Zand for providing us the WFC data. The \sax\ satellite
was a joint effort of ASI and Netherland Space Agency. This research made use
of data obtained through the HEASARC Online service prodived by the NASA Goddard Space Flight Center.
This work was supported by PRIN MIUR 2009 project on
"{\em Gamma Ray Bursts: from progenitors to the physics of the
prompt emission process}" (Prot. 2009 ERC3HT). RL acknowledges financial support from Italian Space Agency (ASI) under contract ASI I/033/10/0.  

\end{acknowledgments}

\bibliographystyle{apj}
\bibliography{apj-jour,grb_ref}

\clearpage

%
%
\begin{deluxetable}{lcccc}
\tabletypesize{\small}
\tablewidth{0pt}
\tablecaption{GRB Sample Chosen for the Time--Resolved Spectral Analysis.}
\tablehead{
\colhead{GRB} & \colhead{Redshift} & \colhead{Fluence} &
\colhead{No. of Intervals} & \colhead{No. of Useful Intervals} \\
 & &  \colhead{($\times 10^{-6}$~cgs)} &	& 
} 
\startdata
970111 & -- & 39.18$\pm$0.08  & 10 & 8 \\
980329 & 3.5$^a$ & $37.53\pm0.07$ & 8 & 6  \\
990123 & 1.60 &	$205.12\pm0.03$ & 26 & 19 \\
990510 & 1.619 & $15.80\pm0.07$ & 15 & 8  \\
\enddata
\tablenotetext{a}{photometric redshift of its host galaxy \citep{Jaunsen03}.}
\tablecomments{ 
For each GRB we report the redshift, the fluence in the 2-2000 keV energy band, the 
number of time intervals in which we subdivided the time profile and those used in the time-resolved spectral analysis. The uncertainty in the fluence is only statistical.
}
\label{t:grb-intervals}
\end{deluxetable}		
%
%
%
\begin{deluxetable}{lcccccr}
\tabletypesize{\small}
\tablewidth{0pt}
\tablecaption{GRB light curve intervals in which the {\sc bf$+$bb} was tested.}
\tablehead{
\colhead{GRB} & \colhead{Interval No.} & \colhead{$\chi^2/{\rm dof}$} &
\colhead{$\chi^2/{\rm dof}$} & \colhead{F--test} & \colhead{F--test/add } & \colhead{$kT_{bb}$} \\
 & &  \colhead{{\sc bf}} & \colhead{{\sc bf$+$bb}} & \colhead{$NHP$}	& \colhead{$NHP$} & \colhead{(keV)} \\
} 
\startdata
970111 & 7 & 29.3/13 & 11.8/11 & 0.11 & 0.0067 & $8.9\pm 1.4$ \\
970111 & 8 &  32.9/13 & 19.9/11 & 0.30  & 0.07 &  $9.4\pm 1.1$ \\
980329 & 6 & 97.7/77 & 90.0/75 & 0.40 & 0.046 & $11\pm 2$ \\
990123 & 9 & 192/166 & 187/164 & 0.46  & 0.11 & $57\pm 14$  \\
990123 & 12 & 136.5/118 & 128/116 & 0.40  & 0.024 & $68\pm 12$      
 \\
990510 & 1 & 74/48 & 60.7/46 & 0.30 & 0.01 & $6.0\pm 2.0$   \\
\enddata
\tablecomments{ 
For each time interval of the GRB light curve, in which the {\sc bf} does not provide a good fit of the corresponding spectrum,  we report, in addition to the $\chi^2/{\rm dof}$ by using {\sc bf} alone and {\sc bf$+$bb}, the fit improvement level ($\chi^2/{\rm dof}$) by using {\sc bf$+$bb}.
}
\label{t:bb-plus-bf}.
\end{deluxetable}		

\clearpage
%
%
\pagestyle{empty}
\changepage{1.5in}{1in}{-0.75in}{-0.75in}{}{}{}{}{}
\begin{deluxetable}{lcccccccccccccl}
\tabletypesize{\footnotesize}
\rotate
\tablewidth{0pt}
\tablecaption{Best--fit parameters of the {\sc grbcomp} and {\sc bb$+$pl} models in  the time intervals
in which each GRB light curve was subdivided.}
\tablehead{
\colhead{}    & \colhead{}  &  \colhead{}       &  \multicolumn{5}{c}{BB $+$ PL model} & \multicolumn{7}{c}{GRBCOMP model} \\ 
\colhead{GRB} & \colhead{Interval} & \colhead{Start ($\Delta$t)} & \colhead{$kT_{bb}$} & \colhead{$N_{bb}$}  & \colhead{$\Gamma$} & \colhead{$N_{pl}$}& 
\colhead{$\chi^2/{\rm dof}$} & \colhead{$kT_{s,i}$} & \colhead{$kT_{e,i}$}& \colhead{$\tau$} & \colhead{$\delta$} &  \colhead{$\alpha_b$} &
\colhead{N} &  \colhead{$\chi^2/{\rm dof}$}\\
    &   &	 & (keV) & 	&	&	&	& (keV) & (keV) &	&	&	& ($\times 10^{-3}$)	&	\\    
} 
\startdata
970111	 & 3 & 35045.0 (3)  &  $47.1_{-0.3}^{+0.6}$ &  $15.6_{-0.12}^{+0.80}$  & $1.32_{-0.04}^{+0.03}$ & $1.12_{-0.32}^{+0.18}$ & 219.8/97 &
[3] &  $77.3_{-2.8}^{+1.4}$  & 22.1$^{+2.3}_{-2.3}$ & $0.079_{-0.009}^{+0.008}$  & $3.94_{-0.16}^{+0.22}$ & 5.80$_{-0.05}^{+0.25}$ & 126.5/97 \\
	 & 4 & 35048.0 (6)  &  $36.3_{-0.1}^{+0.3}$ &  $12.5_{-0.07}^{+0.47}$  & $1.42_{-0.04}^{+0.02}$ & $4.13_{-0.33}^{+0.02}$ & 653.4/75 &
13.7$^{+1.7}_{-1.8}$   & 86.4$^{+3.7}_{-3.6}$ & 7.9$^{+0.3}_{-  0.4}$ & $0.19_{-0.01}^{+0.01}$  & 4.7$^{+0.4}_{-0.2}$ &    $0.39^{+0.30}_{-0.10}$ & 79.8/74 \\
	 & 5 & 35054.0 (3)  &  $35.0_{-0.3}^{+0.2}$ &  $17.5_{-0.25}^{+0.20}$  & $1.50_{-0.06}^{+0.07}$ & $6.82_{-0.03}^{+0.17}$ & 677.6/73 &
11.7$^{+2.2}_{-1.5}$  &  85$^{+5}_{-5}$ & 7.3$^{+0.6}_{-0.4}$ & $0.22_{-0.02}^{+0.02}$  & 4.7$^{+0.5}_{-0.4}$ & 1.0$^{+0.7}_{-0.3}$ & 77.6/72 \\
	 & 6 & 35057.0 (5)  &  $32.7_{-0.2}^{+0.2}$ &  $16.4_{-0.1}^{+0.2}$  & $1.59_{-0.01}^{+0.01}$ & $13.30_{-0.51}^{+0.01}$ & 1030.7/73 &
9.6$^{+0.6}_{-1.1}$  &  $88_{-4}^{+4}$  & 6.5$^{+0.4}_{-0.3}$ & $0.23_{-0.01}^{+0.02}$  & 5.5$^{+0.7}_{-0.6}$ & 2.2$^{+  1.0}_{-0.4}$ & 75.7/72 \\
	& 7 & 35062.0 (3)  &  $20.6_{-0.22}^{+0.24}$ &  $6.22_{-0.10}^{+0.10}$  & $1.64_{-0.02}^{+0.02}$ & $11.2_{-1.0}^{+1.1}$ & 374.8/13 &
10.8$^{+1.8}_{-3.4}$  &   57$^{+5}_{-6}$ &  6.7$^{+0.5}_{-0.4}$  & $0.27_{-0.03}^{+0.03}$  & 4.4$^{+0.7}_{-0.4}$ & 1.1$^{+2.1}_{-0.3}$ & 18.6/12 \\
	& 8 & 35065.0 (4)  &  $16.2_{-0.1}^{+0.5}$ &  $4.16_{-0.12}^{+0.72}$  & $1.70_{-0.02}^{+0.02}$ & $7.31_{-0.04}^{+0.71}$ & 163.0/13 &
 8.5$^{+1.7}_{-2.0}$ & 37$^{+9}_{-6}$  & 8.3$^{+2.9}_{-1.4}$  & $0.42_{-0.09}^{+0.18}$  & 4.3$^{+1.5}_{-0.4}$ & 2.1$^{+  2.3}_{-1.0}$ & 16.7/12 \\
	& 9  & 35069.0 (4)  &  $14.8_{-0.2}^{+0.2}$ &  $3.00_{-0.14}^{+0.14}$  & $1.77_{-0.02}^{+0.03}$ & $11.0_{-1.0}^{+1.5}$ & 129.3/12 &
3.2$^{+0.9}_{-0.9}$ &  22$^{+3}_{-4}$  & 14.2$^{+6.2}_{-2.9}$  & $0.42_{-0.11}^{+0.19}$  & 3.3$^{+0.2}_{-0.1}$ & 34$^{+62}_{-17}$ & 12.7/11 \\
	& 10 & 35073.0 (13)  &  $15.0_{-0.3}^{+0.5}$ &  $1.34_{-0.02}^{+0.29}$  & $1.81_{-0.08}^{+0.02}$ & $9.2_{-1.2}^{+0.2}$ & 163.8/13 &
3.1$^{+1.1}_{-1.0}$&  21$^{+13}_{-4}$ & 14.3$^{+5.3}_{-6.3}$  & $0.44_{-0.21}^{+0.32}$  & 2.9$^{+0.3}_{-0.1}$ & $27_{-18}^{+73}$ & 18.3/12 \\
 &	&	  &	 &	     &	&	&	&	 &	&		&	 &	   &				\\
980329 & 2 & 13477.0 (4)  &  $42.7_{-1.4}^{+1.2}$ &  $3.92_{-0.23}^{+0.21}$  & $1.48_{-0.02}^{+0.02}$ & $7.9_{-0.8}^{+0.8}$ & 161.1/83 &
6.3$^{+3.5}_{-2.5}$  &  183$^{+21}_{-18}$  & 5.2$^{+0.6}_{-0.4}$ & $0.14_{-0.02}^{+0.02}$  &  2.23$^{+0.07}_{-0.1}$ & 7.2$^{+8.6}_{-5.3}$ & 105.5/82 \\
 	 & 3 & 13481.0 (2)  &  $42.5_{-0.5}^{+0.7}$ &  $12.61_{-0.32}^{+0.33}$  & $1.51_{-0.01}^{+0.01}$ & $20.4_{-1.3}^{+1.5}$ & 403.3/90 & 
12.0$^{+4.9}_{-2.3}$ &  190$^{+11}_{-5}$  &  5.1$^{+0.3}_{-0.3}$  & $0.140_{-0.009}^{+0.011}$  &  2.50$^{+0.09}_{-0.07}$ & 2.2$^{+2.6}_{-1.8}$ & 102.5/89 \\
	 & 4 & 13483.0 (3)  &  $42.0_{-0.4}^{+0.5}$ &  $20.3_{-0.4}^{+0.4}$  & $1.52_{-0.03}^{+0.02}$ & $30.31_{-0.11}^{+0.00}$ & 584.5/97 &
  11.1$^{+2.3}_{-1.5}$    &  201$^{+11}_{-10}$ &  4.5$^{+0.2}_{-0.3}$  & $0.15_{-0.01}^{+0.01}$  & 2.57$^{+0.08}_{-0.07}$ & 4.9$^{+3.0}_{-2.2}$ & 105.0/96 \\
	 & 5 & 13486.0 (4)  &  $40.5_{-0.3}^{+0.5}$ &  $14.3_{-0.4}^{+0.2}$  & $1.510_{-0.007}^{+0.004}$ & $32.6_{-1.5}^{+0.7}$ & 977.3/105 &
 11.9$^{+2.4}_{-1.1}$ & 202$^{+9}_{-9}$ &  4.17$^{+0.17}_{-0.15}$  & $0.161_{-0.009}^{+0.010}$  & 2.31$^{+0.04}_{-0.04}$ & 3.6$^{+1.5}_{-1.5}$& 130.3/104 \\
	 & 6 & 13490.0 (4)  &  $26.6_{-0.6}^{+1.2}$ &  $1.81_{-0.10}^{+0.14}$  & $1.65_{-0.01}^{+0.02}$ & $17.7_{-1.5}^{+1.9}$ & 213.4/77 &
 6.4$^{+3.6}_{-1.0}$  &  196$^{+46}_{-41}$ &  3.2$^{+0.5}_{-0.4}$ & $0.22_{-0.05}^{+0.06}$  &  2.34$^{+0.17}_{-0.12}$ & 11.0$^{+14.0}_{-5.4}$  & 91/76 \\
	 & 7 & 13494.0 (19)  &  $19_{-3}^{+3}$ &  $0.21_{-0.04}^{+0.04}$  & $1.75_{-0.03}^{+0.03}$ & $5.05_{-0.71}^{+0.79}$ & 73.8/47 &
4.4$^{+2.1}_{-2.3}$  & 242$^{+87}_{-83}$ &  2.55$^{+0.47}_{-0.46}$  & $0.22_{-0.08}^{+0.09}$  &  2.8$^{+1.0}_{-0.5}$ & 10$^{+112}_{-7}$ & 53.8/46 \\
 &	&	  &	 &	     &	&	&	&	 &	&		&	 &	   &				\\
990123 	 & 2 & 35221.9 (6)  &  $27.2_{-2.9}^{+3.4}$ &  $1.7_{-0.7}^{+0.2}$  & $1.34_{-0.09}^{+0.14}$ & $0.38_{-0.12}^{+0.17}$ & 50.5/35 &
[3]   &  43$^{+13}_{-13}$ &  22$^{+ 16}_{-5}$  & $0.141_{-0.053}^{+0.110}$  & 2.5$^{+0.7}_{-0.3}$ & 6.9$^{+2.2}_{-0.8}$  & 41.5/35 \\
	 & 3 & 35227.9 (6)  &  $34.0_{-3.3}^{+3.8}$ &  $2.7_{-1.1}^{+0.3}$  & $1.22_{-0.06}^{+0.11}$ & $0.37_{-0.09}^{+0.12}$ & 44.2/44 &
[3]    &  49$^{+14}_{-8}$ & 28$^{+12}_{-11}$  & $0.099_{-0.043}^{+0.052}$  &  2.2$^{+0.3}_{-0.2}$ & 12$^{+1}_{-3}$ & 29.2/44 \\
	 & 4 & 35233.9 (2)  &  $55_{-9}^{+8}$ &  $5.7_{-1.1}^{+1.5}$  & $1.25_{-0.06}^{+0.05}$ & $1.25_{-0.26}^{+0.22}$ & 75.8/57 &
[3] & 63$^{+39}_{-11}$ & 23.5$^{+7.1}_{-11.2}$   &  $0.090_{-0.046}^{+0.062}$  &  1.9$^{+0.3}_{-0.1}$ & 905$^{+32}_{-242}$ & 61.5/57 \\
	 & 5 & 35235.9 (2)  &  $71_{-5}^{+5}$ &  $12.1_{-1.3}^{+1.4}$  & $1.16_{-0.03}^{+0.02}$ & $1.90_{-0.27}^{+0.22}$ & 119.6/105 &
1.7$^{+2.1}_{-1.3}$  &  148$^{+20}_{-25}$ & 9.9$^{+2.6}_{-0.9}$  &  $0.092_{-0.018}^{+0.027}$  &  2.0$^{+0.2}_{-0.2}$   &  228$^{+831}_{-196}$ & 105/104 \\
	 & 6 & 35237.9 (2)  &  $77_{-4}^{+5}$ &  $23.1_{-2.0}^{+2.2}$  & $1.05_{-0.02}^{+0.02}$ & $2.51_{-0.21}^{+0.27}$ & 181.5/151 &
4.5$^{+2.1}_{-1.7}$  &  173$^{+29}_{-25}$  &  8.3$^{+1.5}_{-1.2}$   & $0.094_{-0.019}^{+0.023}$  & 1.7$^{+0.1}_{-  0.1}$ &  31$^{+77}_{-20}$ & 153.4/150 \\
	 & 7 & 35239.9 (2)  &  $96_{-3}^{+5}$ &  $48.5_{-2.3}^{+3.6}$  & $1.09_{-0.01}^{+0.01}$ & $4.06_{-0.27}^{+0.32}$ & 275.6/182 &
5.2$^{+1.8}_{-1.4}$   & 207$^{+23}_{-20}$ &   7.8$^{+0.9}_{-0.8}$  &  $0.083_{-0.012}^{+0.013}$  & 1.78$^{+0.08}_{-0.07}$ & 31$^{+43}_{-15}$ & 189.0/181 \\
	 & 8 & 35241.9 (2)  &  $100_{-5}^{+4}$ &  $52.4_{-2.3}^{+2.4}$  & $1.07_{-0.10}^{+0.09}$ & $3.7_{-0.1}^{+0.1}$ & 294/183 &
6.2$^{+1.6}_{-1.8}$ & 214$^{+30}_{-16}$  &  7.8$^{+0.6}_{-1.0}$  &   $0.081_{-0.012}^{+0.013}$  &  1.74$^{+0.09}_{-0.06}$ & 21$^{+33}_{-8}$  & 177.7/182 \\
	 & 9  & 35243.9 (2)  &  $76_{-3}^{+3}$ &  $28.2_{-1.8}^{+1.7}$  & $1.15_{-0.02}^{+0.01}$ & $3.6_{-0.3}^{+0.3}$ & 236.4/166 &
1.8$^{+1.3}_{-0.9}$& 134$^{+16}_{-14}$ &  11.6$^{+1.8}_{-1.5}$  & $0.087_{-0.015}^{+0.017}$  & 1.81$^{+0.08}_{-0.07}$   & 450$^{+  4152}_{-450}$ & 195.6/165 \\
	 & 10 & 35245.9 (2 &  $69_{-3}^{+2}$ &  $21.2_{-1.3}^{+1.0}$  & $1.33_{-0.03}^{+0.02}$ & $4.8_{-0.5}^{+0.3}$ & 196.0/140 &
1.5$^{+1.0}_{-1.0}$  & 154$^{+14}_{-13}$  & 8.9$^{+0.9}_{-0.8}$   & $0.098_{-0.012}^{+0.013}$  & 2.6$^{+0.2}_{-0.2}$ & 505$^{+  6920}_{-394}$ & 145.8/139 \\
	 & 11 & 35247.9 (2)  &  $40_{-3}^{+3}$ &  $9.7_{-3.2}^{+1.3}$  & $1.34_{-0.04}^{+0.08}$ & $3.9_{-0.5}^{+0.7}$ & 124.5/104 &
2.7$^{+1.2}_{-1.1}$  &  89$^{+16}_{-11}$  &  11.3$^{+2.0}_{-1.8}$  & $0.134_{-0.028}^{+0.034}$  & 2.5$^{+0.3}_{-0.2}$ &  72$^{+285}_{-49}$ & 95.9/103 \\
	 & 12 & 35249.9 (2)  &  $67_{-3}^{+3}$ &  $18.6_{-1.3}^{+1.9}$  & $1.35_{-0.04}^{+0.03}$ & $3.71_{-0.42}^{+0.45}$ & 147.8/118 &
[3]    &  145$^{+17}_{-21}$  & 10.0$^{+2.8}_{-0.7}$   & $0.093_{-0.015}^{+0.028}$  & 2.8$^{+0.4}_{-0.3}$ & 469$^{+5100}_{-361}$ & 126.6/118 \\
	 & 13 & 35251.9 (2)  &  $78_{-3}^{+2}$ &  $36.1_{-2.0}^{+1.7}$  & $1.18_{-0.02}^{+0.01}$ & $3.9_{-0.4}^{+0.2}$ & 273.0/188 &
1.8$^{+1.7}_{-1.3}$  & 136$^{+ 12}_{-11}$  & 12.9$^{+1.9}_{-  1.4}$  & $0.077_{-0.010}^{+0.013}$   & 1.95$^{+0.08}_{-0.07}$ & 459$^{+1517}_{-459}$ & 209.8/187 \\
	 & 14 & 35253.9 (2)  &  $87_{-4}^{+3}$ &  $36.5_{-2.6}^{+2.2}$  & $1.19_{-0.02}^{+0.01}$ & $5.0_{-0.4}^{+0.4}$ & 238.3/180 &
 1.6$^{+1.1}_{-0.8}$ &  176$^{+14}_{-18}$ & 8.8$^{+1.1}_{-0.5}$   & $0.087_{-0.010}^{+0.013}$  & 2.0$^{+0.1}_{-0.1}$ & 756$^{+472}_{-594}$ & 201.4/179 \\
	 & 15 & 35255.9 (2)  &  $74_{-3}^{+3}$ &  $26.2_{-1.4}^{+1.5}$  & $1.23_{-0.02}^{+0.01}$ & $5.0_{-0.4}^{+0.4}$ & 239.6/182 &
3.3$^{+1.0}_{-0.9}$ &  148$^{+15}_{-13}$  & 9.3$^{+0.5}_{-0.9}$   & $0.098_{-0.013}^{+0.011}$  & 2.03$^{+0.10}_{-0.08}$   & 83$^{+127}_{-44}$ & 170.1/181 \\
	 & 16 & 35257.9 (2)  &  $62_{-4}^{+4}$ &  $16.5_{-1.5}^{+1.7}$  & $1.32_{-0.02}^{+0.02}$ & $6.0_{-0.5}^{+0.6}$ & 188.3/125 &
2.6$^{+1.1}_{-1.1}$  &  120$^{+27}_{-12}$  & 9.7$^{+1.0}_{-1.8}$   & $0.117_{-0.025}^{+0.029}$  & 2.1$^{+0.2}_{-0.1}$   & 151$^{+593}_{-65}$ & 147.0/124 \\
	 & 17 & 35259.9 (5)  &  $49_{-3}^{+2}$ &  $7.5_{-0.6}^{+0.5}$  & $1.43_{-0.02}^{+0.01}$ & $8.64_{-0.54}^{+0.36}$ & 272.7/170 &
1.8$^{+0.5}_{-0.4}$   & 118$^{+19}_{-15}$ & 7.4$^{+1.0}_{-0.9}$	& $0.155_{-0.027}^{+0.032}$  &  2.2$^{+0.2}_{-0.1}$  & 346$^{+405}_{-188}$ & 171.6/169 \\
	 & 18 & 35264.9 (5)  &  $29_{-2}^{+2}$ &  $5.2_{-0.8}^{+0.4}$  & $1.38_{-0.01}^{+0.03}$ & $8.8_{-0.4}^{+0.5}$ & 220.0/147 &
1.6$^{+0.5}_{-0.4}$   &  74$^{+17}_{-13}$  &   8.5$^{+1.3}_{-1.3}$  & $0.215_{-0.050}^{+0.059}$  & 1.96$^{+0.12}_{-0.09}$ & 641$^{+  1247}_{-374}$   & 163.9/146 \\
 	 & 19 & 35269.9 (5)  &  $41_{-1}^{+2}$ &  $3.0_{-0.2}^{+0.2}$  & $1.46_{-0.01}^{+0.01}$ & $9.6_{-0.2}^{+0.3}$ & 185.9/147 &
2.2$^{+0.4}_{-0.5}$  &  146$^{+31}_{-29}$  & 5.3$^{+0.9}_{-0.6}$   & $0.174_{-0.041}^{+0.047}$  & 2.5$^{+0.5}_{-0.3}$ & 211$^{+263}_{-97}$   & 135.8/146 \\
	 & 20 & 35274.9 (5)  &  $44_{-5}^{+6}$ &  $3.5_{-0.7}^{+0.7}$  & $1.49_{-0.03}^{+0.04}$ & $9.97_{-0.63}^{+0.75}$ & 157.0/134 &
$1.39_{-0.43}^{+0.50}$ &  $134_{-22}^{+26}$  & $5.85_{-0.65}^{+0.75}$  & 0.172$_{-0.034}^{+0.040}$  & $2.41_{-0.24}^{+0.38}$ & $0.84_{-0.48}^{+1.90}$ & 111.9/133 \\
&	&	  &	 &	     &	&	&	&	 &	&		&	 &	   &				\\
990510 & 1 & 31745.9 (5)  &  $19.6_{-0.4}^{+0.7}$ &  $1.95_{-0.06}^{+0.16}$  & $1.60_{-0.02}^{+0.05}$ & $3.9_{-0.5}^{+1.0}$ & 151.4/48 &
9.2$^{+1.4}_{-1.7}$   &  83$^{+20}_{-21}$  &  4.7$^{+1.4}_{-0.3}$  & $0.343_{-0.092}^{+0.132}$  & 3.9$^{+0.9}_{-0.6}$ & 1.0$^{+1.2}_{-0.4}$ & 61.0/47 \\
 	 & 2 & 31750.9 (5)  &  $20.5_{-0.4}^{+0.4}$ &  $2.04_{-0.06}^{+0.08}$  & $1.75_{-0.02}^{+0.02}$ & $9.7_{-1.0}^{+1.0}$ & 179.6/53 &
5.7$^{+1.4}_{-1.0}$  &  79$^{+13}_{-13}$  & 5.3$^{+0.9}_{-0.6}$  & $0.320_{-0.064}^{+0.078}$  & 4.2$^{+0.6}_{-0.6}$ & 6.3$^{+5.1}_{-2.8}$ & 46.6/52 \\
	 & 7 & 31785.9 (5) &  $37.6_{-0.5}^{+1.0}$ &  $3.2_{-0.1}^{+0.1}$  & $1.53_{-0.01}^{+0.01}$ & $9.7_{-0.7}^{+0.7}$ & 267.4/81 & 
5.5$^{+0.6}_{-0.5}$ &  123$^{+14}_{-13}$  & 5.5$^{+0.6}_{-0.5}$   & $0.201_{-0.027}^{+0.031}$  & 2.5$^{+0.2}_{-0.1}$ & 12$^{+29}_{-8}$ & 93.0/80 \\
	 & 8  & 31790.9 (3)  &  $32.6_{-0.4}^{+0.6}$ &  $5.8_{-0.2}^{+0.2}$  & $1.55_{-0.01}^{+0.01}$ & $14.4_{-1.2}^{+0.5}$ & 313.6/80  &
5.2$^{+1.2}_{-1.2}$ & 102$^{+9}_{-9}$  &  6.1$^{+0.5}_{-0.4}$  & $0.220_{-0.025}^{+0.027}$  & 2.8$^{+0.2}_{-0.1}$ & 15$^{+15}_{-7}$ & 91.1/79 \\
	 & 9 & 31793.9 (4)  &  $27.1_{-0.8}^{+0.8}$ &  $1.85_{-0.10}^{+0.10}$  & $1.69_{-0.02}^{+0.02}$ & $12.0_{-1.1}^{+1.3}$ & 165.6/62 &
3.3$^{+0.9}_{-0.9}$  &  105$^{+19}_{-15}$  &  5.1$^{+0.4}_{-0.6}$   & $0.253_{-0.048}^{+0.047}$  &  3.3$^{+0.3}_{-0.3}$ & 36$^{+61}_{-20}$ & 72.5/61 \\
	 & 11 & 31802.9 (4)  &  $16.6_{-0.4}^{+0.5}$ &  $1.03_{-0.05}^{+0.05}$  & $1.93_{-0.03}^{+0.02}$ & $19.2_{-1.4}^{+2.7}$ & 105.1/16 &
3.5$^{+0.9}_{-0.8}$   &  50$^{+ 14}_{-6}$  & 6.6$^{+0.7}_{-1.3}$  & $0.407_{-0.091}^{+0.121}$  & 3.0$^{+0.3}_{-0.1}$ & 37$^{+44}_{-14}$ & 15.2/15 \\
 	 & 12 & 31806.9 (10)   &  $19_{-2}^{+2}$ &  $0.20_{-0.04}^{+0.06}$  & $2.07_{-0.06}^{+0.06}$ & $10.2_{-1.3}^{+1.5}$ & 48.8/34 &
[1] &  71$^{+14}_{-10}$  & 5.0$^{+0.6}_{-0.6}$  & $0.349_{-0.064}^{+0.077}$  & [3.4] & 973$^{+93}_{-92}$ & 40.8/36 \\
 	& 13 & 31816.9 (5)  &  $12.7_{-2.6}^{+2.4}$ &  $0.11_{-0.05}^{+0.03}$  & $2.23_{-0.13}^{+0.24}$ & $5.3_{-1.2}^{+1.6}$ & 30.2/14 &
2.1$^{+0.8}_{-0.9}$   &  23$^{+4}_{-8}$  & [9.3]   & $0.65_{-0.24}^{+0.12}$  & 2.20$^{+0.09}_{-0.09}$ & 109$^{+603}_{-83}$ & 12.7/13  \\

\enddata
\tablecomments{The {\sc grbcomp} parameters are given in the rest frame. In the case of GRB\,970111, a redshift $z$ equal to 1 was assumed. In the case of $\delta$, it was estimated from its dependence on $kT_{e,i}$ and $\tau$ (see
text). Uncertainties are 1$\sigma$ errors. In square parenthesis,  parameters frozen in the fits.}
\label{t:results}
\end{deluxetable}

\clearpage

\pagestyle{plaintop}
%
%
\begin{deluxetable}{lcccccl}
\tablewidth{0pt}
\tablecaption{Correlation analysis results between $E_{p,i}$ and $kT_{e,i}$.}
\tablehead{
\colhead{GRB} & \colhead{$\rho$} & \colhead{$NHP$} 
& \colhead{$K$ (keV)} & \colhead{$m$} & \colhead{$kT_{e,i}^0$ (keV)} & \colhead{$\chi_r^2$ (dof)}  \\  
} 
\startdata
970111	& 0.76 & 0.028  & $99.7 \pm 17.5$ & $1.72 \pm 0.28$ & 39.9  &  10.8 (6)	\\
980329	& $-0.60$ & 0.21  & $612\pm 314$ & $-4.1\pm 9.7$ & 225.7  & 0.33 (4)	\\
990123	& 0.88 & $4.66 \times 10^{-7}$  & $673.6\pm 65.9 $ & $1.44 \pm 0.18$ &  85.6 & 0.72 (17)	\\
990510	& 0.93 & $8.6 \times 10^{-4}$  & $98.7\pm 21.4$ & $1.89 \pm 0.34$ & 44.2 & 0.91 (6)	\\
\enddata
\tablecomments{A power--law relation between the two parameters is assumed:
$E_{p,i} = K (\frac{kT_{e,i}}{kT_{e,i}^0})^m$. In addition to the best--fit parameters $K$,  $m$, and the median value $kT_{e,i}^0$,
the correlation coefficient $\rho$, $NHP$,
and the best--fit reduced $\chi^2$ with dof are reported.}
\label{t:ep-vs-kte}
\end{deluxetable}

%
%
\begin{deluxetable}{lcccccl}
\tabletypesize{\small}
\tablewidth{0pt}
\tablecaption{Correlation analysis results between $kT_{e,o}$  and 2--2000 keV flux.}
\tablehead{
\colhead{GRB} & \colhead{$\rho$} & \colhead{$NHP$} 
& \colhead{$K$~(keV)} & \colhead{$m$} & 
\colhead{flux$_0$~(cgs)}  & \colhead{$\chi_r^2$ (dof)}  \\  
} 
\startdata
970111	& 0.95	& $2.6 \times 10^{-4}$  & $24.0 \pm 1.8$ & $0.65 \pm 0.09$ & $8.6\times 10^{-7}$  & 2.5 (6)	\\
980329	& $-0.085$ & 0.87  & $39.7\pm 3.3$ & $0.06 \pm 0.07$ & $8.03\times 10^{-7}$  & 0.49 (4)	\\
990123	& 0.82 	& $1.89 \times 10^{-5}$  & $31.2\pm 2.2$ & $0.37 \pm 0.04$ & $1.25\times 10^{-6}$  & 1.3 (17)	\\
990510	& 0.76 	& $2.8\times 10^{-2}$  & $20.1\pm 2.3$ & $0.36 \pm 0.08$ & $1.7\times 10^{-7}$ &  2.8 (6)	\\
\enddata
\tablecomments{A power--law relation between the two parameters is assumed: 
$kT_{e,o} = K (\frac{{\rm flux}}{{\rm flux_0}})^m$.
In addition to the best--fit parameters $k$, $m$ and the
median value flux$_0$, the correlation coefficient $\rho$, 
$NHP$ and the best--fit reduced $\chi^2$ with dof are reported.}
\label{t:kte-vs-flux}
\end{deluxetable}

%
\begin{deluxetable}{lcccccl}
\tablewidth{0pt}
\tablecaption{Correlation analysis results between $E_{p,i}$ and $\delta$.}
\tablehead{
\colhead{GRB} & \colhead{$\rho$} & \colhead{$NHP$} 
& \colhead{$K$~(keV)} & \colhead{$m$} & \colhead{$\delta_0$} & \colhead{$\chi_r^2$ (dof)}  \\  
} 
\startdata
970111	& $-0.93$ & $8.6\times 10^{-4}$  & $263\pm 12$ & $-0.95 \pm 0.13$ & 0.233 & 7.6 (6)	\\
980329	& $-0.71$ & 0.11  & $803\pm 204$ &  $-1.07\pm 0.75$ & 0.197   & 0.44 (4)	\\
990123	& $-0.73$ & $4.1 \times 10^{-4}$  & $1234\pm 142$ & $-1.98 \pm 0.46$ & 0.110 & 0.90 (17)	\\
990510	& $-0.97$ & $3.3 \times 10^{-5}$  & $154\pm 40$ & $-2.38 \pm 0.60$  & 0.366 & 0.10 (6)	\\
\enddata
\tablecomments{A power--law relation between the two parameters is assumed:
$E_{p,i} = K (\frac{\delta}{\delta_0})^m$. In addition to the best--fit parameters $K$,  $m$, and the median value $\delta_0$, 
the correlation coefficient $\rho$, $NHP$,
and the best--fit reduced $\chi^2$ with the corresponding dof are reported.}
\label{t:ep-vs-delta}
\end{deluxetable}

%
%
\begin{deluxetable}{lccccl}
\tablewidth{0pt}
\tablecaption{Correlation analysis results between seed photon temperature $kT_{s,o}$ 
and 2--2000 keV flux.}
\tablehead{
\colhead{GRB} & \colhead{$\rho$} & \colhead{$NHP$} 
& \colhead{$a$} & \colhead{$m$} & \colhead{$\chi^2$ (dof)}  \\  
} 
\startdata
970111	& 0.71 & $7.1\times 10^{-2}$    & $3.7\pm 0.6$  & $0.52 \pm 0.11$  & 2.8 (5)	\\
980329	& 0.71 & $1.1\times 10^{-1}$  & $301\pm 219$    & 
$0.38 \pm 0.10$  & 0.34 (4)	\\
990123	& 0.46 & $8.6\times 10^{-2}$  & $2.7\pm 0.6$   & $0.51 \pm 0.12 $  & 0.86 (13)	\\
990510	& 0.38 & $3.5\times 10^{-1}$  & $5.1\pm 1.0$    & $0.79 \pm 0.15$ & 2.8 (6)	\\
\enddata
\tablecomments{A power--law relation between the two parameters is assumed, i.e.
$kT_{s,o} =  a {\rm (flux)}^m$. 
In addition to the best--fit parameters $a$ and $m$, 
the correlation coefficient $\rho$, $NHP$,
and the best--fit reduced $\chi_r^2$ with the corresponding dof are reported.}
\label{t:kts-vs-flux}
\end{deluxetable}
%

%
%
\begin{deluxetable}{lcccccl}
\tablewidth{0pt}
\tablecaption{Correlation analysis results between seed photon temperature $kT_{s,i}$ 
and photospheric radius $R_{ph}$ for each of the GRB with known $z$ and for all of them together.}
\tablehead{
\colhead{GRB} & \colhead{$\rho$} & \colhead{$NHP$} 
& \colhead{$K$~(keV)} & \colhead{$m$} & \colhead{$R_{ph}^0$ (cm)} & \colhead{$\chi^2$ (dof)}  \\  
} 
\startdata
970111	& -0.93 & $2.5\times 10^{-3}$    & $6.0\pm 2.8$  & $-0.79 \pm 0.47$ & $4.33\times 10^{13}$ & 0.07 (5)	\\
980329	& -0.82 & $4.1\times 10^{-2}$  & $11.7\pm 6.8$    & 
$-1.58 \pm 2.96$  &  $1.35\times 10^{14}$ & 0.10 (4)	\\
990123	& -0.96 & $2.3\times 10^{-8}$  & $2.12\pm 1.18$   & $-1.01 \pm 0.67 $ & $4.03\times 10^{14}$ & 0.05 (13)	\\
990510	& -0.95 & $2.6\times 10^{-4}$  & $3.7\pm 0.8$    & $-0.65 \pm 0.13$ & $1.49\times 10^{14}$ & 0.04 (6)	\\
980329$+$990123$+$990510	& $-0.88$	& $1.7\times 10^{-10}$	& $3.76 \pm 0.72$	& $-0.92\pm 0.17$  &	$2.32 \times 10^{14}$	& 0.21 (29)  \\
\enddata
\tablecomments{A power--law relation between the two parameters is assumed, i.e.
$kT_{s,i} = K (R_{ph}/R_{ph}^0)^m$. 
In addition to the best--fit parameters $K$, $m$  and the median value  $R_{ph}^0$,
the correlation coefficient $\rho$, $NHP$,
and the best--fit reduced $\chi_r^2$ with the corresponding dof are reported.}
\label{t:kts-vs-rph}
\end{deluxetable}
%
%
\begin{deluxetable}{lcccccl}
\tablewidth{0pt}
\tablecaption{Correlation analysis results between the low-energy photon index of the
Band function $-\alpha_{bf}$ and  the bulk parameter $\delta$.}
\tablehead{
\colhead{GRB} & \colhead{$\rho$} & \colhead{$NHP$} 
& \colhead{$K$} & \colhead{$m$} & \colhead{$\delta_0$} & \colhead{$\chi_r^2$ (dof)}  \\  
} 
\startdata
970111	& 0.96 & $1.8\times 10^{-4}$  & $1.60\pm 0.17$ & $6.33 \pm 0.67$ & 0.42 & 0.9 (6)	\\
980329	& 0.94 & $4.8\times 10^{-3}$  & $1.31\pm 0.45$  & $9.0 \pm 5.0$ & 0.217  & 0.06 (4)	\\
990123	& 0.62 & $4.1\times 10^{-3}$  & $0.96\pm 0.10$  & $6.03 \pm 1.44$ & 0.16 & 0.36 (17)	\\
990510	& 0.57 & 0.13  & $2.06\pm 0.58$  & $4.87 \pm 2.00$ & 0.47 & 0.47 (6)	\\
\enddata
\tablecomments{A linear relation between the two parameters is assumed: $-\alpha_{bf} = K + m (\delta -\delta_0)$. 
In addition to the best--fit parameters $K$, $m$ and the median value $\delta_0$, 
the correlation coefficient $\rho$, $NHP$, and the best--fit reduced $\chi_r^2$ with  the corresponding dof are reported.}
\label{t:alpha-vs-delta}
\end{deluxetable}

%
%
\begin{deluxetable}{lcccl}
\tablewidth{0pt}
\tablecaption{Correlation analysis results between the intrinsic (redshift corrected) electron temperature $kT_{e,i}$ and the GRB luminosity (in units of $10^{52}$), for each of the 3 GRBs with known redshift, and for the sum of the two events (990123 and 990510) that show a variable $kT_{e,i}$ during the prompt emission.}
\tablehead{
\colhead{Parameter} & \colhead{980329} & \colhead{990123} & \colhead{990510} & \colhead{990123$+$990510}     
} 
\startdata
$\rho$   & $-0.08$ & 0.82			& 0.76  		& 0.87  \\
$NHP$	    & $0.87$ & $1.8\times 10^{-5}$	& $2.8\times 10^{-2}$  & $5.04\times 10^{-9}$ \\
$K$	& $173\pm19$	& $67.1\pm 5.5$		& $79.4\pm 4.5$ & $74.04\pm 3.47$  \\
$m$	& $0.06\pm 0.07$ 	& $0.28\pm 0.03$		& $0.34\pm 0.08$ & $0.25\pm 0.02$   \\
$L_{iso}^0$ & 1.068  	&1.068			&  1.068		& 1.068   \\
$\chi_r^2$	& 0.46	& 1.07			&	2.8 &  2.08  \\ 
dof	& 4	& 17				&     6   &  25 		\\
\enddata
\tablecomments{A power--law relation between the two parameters is assumed: $kT_{e,i} = K (\frac{L_{iso}}{L_{iso}^0})^m$. 
In addition to the best--fit parameters $k$,  $m$ and the median value $L_{iso}^0$,  
the correlation coefficient $\rho$, $NHP$, and the best-fit reduced $\chi_r^2$ with its dof are reported. The relatively high $\chi_r^2$ found for 990510, and for 990123 and 990510 by merging their data together, is due to an intrinsic  spread of the data points around the best--fit power--law.}
\label{t:kTei-vs-L}
\end{deluxetable}

\clearpage

%
%

\clearpage

%

\clearpage
%
%

\clearpage

%

\clearpage

%
%

\clearpage

%
%

\clearpage

%
%
\clearpage

%
%
\clearpage

%
%
\clearpage

%
%

\clearpage
%
%

\clearpage
%
%

\clearpage
%
%
\clearpage

%
%

\clearpage
%
%

\clearpage
%
%

\clearpage
%
%

\clearpage
%
%

\end{document}